\documentclass[useAMS, usenatbib, usegraphicx]{mn2e}
\usepackage{lscape}
\usepackage{times}
\usepackage{graphicx}

\newcommand{\He}{{\rm He}}
\newcommand{\cm}{{\rm cm}}
\newcommand{\yr}{{\rm yr}} 

\newcommand{\kms}{{\rm km}\,{\rm s}^{-1}} 
\newcommand{\K}{{\rm K}}
\newcommand{\pc}{{\rm pc}} 
\newcommand{\kpc}{{\rm kpc}}

\newcommand{\mpch}{h^{-1}\,{\rm Mpc}} 
\newcommand{\erg}{{\rm erg}}
\newcommand{\Msun}{{{\rm M}_\odot}}
\newcommand{\hMpc}{h^{-1}\,{\rm Mpc}}
\newcommand{\hkpc}{h^{-1}\,{\rm kpc}}
\newcommand{\hMsun}{{h^{-1}\,{\rm M}_\odot}}

\begin{document}

\title[Chemical enrichment in cosmological, SPH simulations]{Chemical enrichment in cosmological, smoothed particle hydrodynamics
  simulations}

\author[R. P. C. Wiersma et al.]{
Robert P. C. Wiersma,$^1$\thanks{E-mail: wiersma@strw.leidenuniv.nl} 
Joop Schaye,$^1$ Tom Theuns,$^{2,3}$ Claudio Dalla Vecchia,$^1$
\newauthor and 
Luca Tornatore$^{4,5,6}$\\ 
$^1$Leiden Observatory, Leiden University, P.O. Box 9513, 2300 RA Leiden, the Netherlands \\
$^2$Institute for Computational Cosmology, Department of Physics, University of Durham, South Road, Durham, DH1 3LE, UK\\
$^3$Department of Physics, University of Antwerp, Groenenborgerlaan 171, B-2020 Antwerpen, Belgium\\
$^4$INAF - Osservatorio Astronomico di Trieste, Via G.B. Tiepolo 11, I-34131 Trieste, Italy\\
$^5$INFN/National Institute for Nuclear Physics, Via Valerio 2, I-34127 Trieste, Italy\\
$^6$Dipartimento di Astronomia dell'Universit{\`a} di Trieste, Via G.B. Tiepolo 11, I-34131 Trieste, Italy
}

\maketitle
\begin{abstract}
We present an implementation of stellar evolution and chemical
feedback for smoothed particle hydrodynamics (SPH) simulations. We
consider the timed release of individual elements by both massive
(Type II supernovae and stellar winds) and intermediate mass stars
(Type Ia supernovae and asymptotic giant branch stars). We illustrate
the results of our method using 
a suite of cosmological simulations that include 
new prescriptions for radiative cooling, star formation, and
galactic winds. Radiative cooling is 
implemented element-by-element, in the presence of an ionizing
radiation background, and we track all 11 elements that contribute
significantly to the radiative cooling. 

While all simulations presented here use a single set of physical
parameters, we take specific care to investigate 
the robustness of the predictions of chemodynamical
simulations with respect to the ingredients, the methods, and the
numerical convergence. A comparison
of nucleosynthetic yields taken from the literature indicates that
relative abundance ratios may only be reliable at the factor of two
level, even for a fixed initial mass function. Abundances relative
to iron are even more uncertain because the rate of
supernovae Ia is not well known. We contrast two
reasonable definitions of the metallicity of a resolution element and
find that while they agree 
for high metallicities, there are large differences at low
metallicities. We argue the discrepancy is indicative of the lack of metal
mixing caused by the fact that metals are stuck to particles. We argue
that since this is a (numerical) sampling problem, solving it using a
poorly constrained physical process such as diffusion could have undesired
consequences. We demonstrate 
that the two metallicity definitions result in redshift $z=0$ stellar
masses that can differ by up to a factor of two, because of the
sensitivity of the cooling rates to the elemental abundances. 

Finally, we use several $512^3$ particle simulations to investigate the
evolution of the distribution of heavy elements, which we find to be
in reasonably good agreement with available observational constraints.
We find that by $z=0$
most of the metals are locked up in stars. The gaseous metals are
distributed over a very wide range of gas densities and
temperatures. The shock-heated warm-hot intergalactic medium has a
relatively high metallicity of $\sim
10^{-1}\,Z_\odot$ that evolves only weakly and is therefore
an important reservoir of metals. Any census aiming to account for
most of the metal mass will have to take a wide variety of objects and
structures into account.
\end{abstract}


\section{Introduction}

Nucleosynthetic processes within stars and supernovae (SNe) change the
abundances of elements in the Universe over time. As stars release
these elements back into the diffuse interstellar medium (ISM) that
surrounds them, subsequent generations of stars are born with
different chemical compositions. On larger scales, winds driven out of
galaxies by the energy released by dying stars and/or active galactic nuclei
modify the composition of the intergalactic medium (IGM) and hence
that of future galaxies. The IGM within groups and clusters of
galaxies exerts a ram-pressure on their member galaxies as they move
along their orbits. This head-wind can be sufficiently strong
to strip the 
ISM from the galaxies, thus mixing the elements that were present in
the ISM into the IGM. These and other processes make the (re-)cycling of
elements over time through stars, galaxies and diffuse gas a
complicated problem that is well-suited for studies using
hydrodynamical simulations. 

Simulating the 
exchange of elements between stars, galaxies, and their environments
is the subject of this paper. Specifically, this paper describes and
tests a method to follow the timed release and subsequent spreading of
elements by stellar populations formed in cosmological, hydrodynamical 
simulations. 

The elemental abundances of stars, galaxies, and diffuse gas are of
great interest for a variety of reasons. First, the dissipation of
binding energy by the emission of ``cooling'' radiation is a
fundamental ingredient of models of the formation of both stars and
galaxies. Because radiative cooling rates are very sensitive to abundance
variations, the rate at which gas can collapse into galaxies is to a
large extent determined by its chemical composition. Similarly,
the initial mass function (IMF) of stars may well depend on the
chemical composition of the gas from which they formed. 

Second, knowledge of elemental abundances can give great
insight into 
a variety of key astrophysical processes. For example, the
distribution of elements in the IGM provides stringent constraints on models of
galactic winds and ram-pressure stripping in clusters. The relative
abundances of elements can tell us about the IMF of the stars that
produced them. The abundance of alpha elements - which are
released on short time-scales through core collapse SNe - relative to
iron - most of which is 
released much later when intermediate mass stars explode as type Ia
SNe - tells us about the time-scale on which stars have been
formed. On the other hand, the absolute abundances of metals provide
us with a measure 
of the integral of past star formation.

Third, emission and absorption lines of individual
elements constitute key diagnostics of physical conditions, such as
gas densities, temperatures, radiation fields and dust
content. The strengths of said lines depend on a number of factors, 
but invariably the abundances play an important role.

Fourth, individual lines are the only observable
signatures for most of the diffuse gas in the Universe. For example,
oxygen lines are the most promising tools to detect the warm-hot IGM
that may host a large fraction of the cosmic baryons. To verify this
prediction of structure formation in a cold dark matter universe, it 
is necessary to know the abundance of oxygen in this elusive gas phase.

Clearly, following the timed release of the elements by stars and
their subsequent dispersal through space, a collection of processes
sometimes referred to as ``chemodynamics'', is a critical ingredient of
any realistic simulation of the formation and evolution of galaxies. 
The implementation of chemodynamics into cosmological simulations is a
challenging problem, mainly because such models lack the resolution to
resolve many of the relevant physical processes and hence require
``sub-grid'' recipes. 

The first study to implement metal enrichment
into a Smoothed Particle Hydrodynamics (SPH) code did not distinguish
between different elements and only 
considered enrichment by core collapse SNe in the instantaneous
recycling approximation \citep{Steinmetz1994}. Since then many authors
have implemented more sophisticated recipes for chemodynamics into SPH
codes
\citep[e.g.][]{RVN96,Berczik1999,Mosconi2001,K01a,LPC02,KG03,Valdarnini2003,Kobayashi2004,Sommer-larsen2005,Tornatore2007,Oppenheimer2008},
ranging from recipes that distinguish only between SN type Ia and SN
type II elements to models that follow large numbers of individual
elements released by AGB stars, SNe Ia, SNe II, and the winds from
their progenitors. Three-dimensional chemodynamical simulations have
so far been used most often for the study of individual objects such
as galaxies
\citep[e.g.][]{TBH92, RVN96,Berczik1999,Reta01,K01a,KG03,Kobayashi2004,MU06a, Geta07}
and clusters of galaxies
(e.g.\ \citealt{LPC02, Valdarnini2003, Toeta04, Sommer-larsen2005,
  Reta06, Tornatore2007}; see \citealt{Borgani2008} for a review),
but in recent years chemodynamical cosmological simulations have also
become more common \citep[e.g.][]{Mosconi2001, Seta05,
  Kobayashi2007,Oppenheimer2008}. Note that several recent
cosmological works \citep{Seta05, 
  Kobayashi2007,Oppenheimer2008} use our hydrodynamical code of
choice, namely the SPH code \textsc{gadget} \citep{Springel2005}, although 
they use different implementations of chemodynamics, star formation,
galactic winds, and radiative cooling. 

Here we present a new implementation of chemodynamics that follows the
timed release -- by AGB stars, SNe Ia, SNe II, and the winds from
their progenitors -- of the 11 elements that \cite{Wiersma2009} found to
contribute significantly to the 
radiative cooling at $T > 10^4~\K$. We present a subset of
high-resolution cosmological, hydrodynamical simulations
taken from the OverWhelmingly Large Simulations (OWLS) project (Schaye et al., in prep.)
and use them to test the robustness of our prescription to numerical
parameters and to make some predictions regarding the large-scale
distribution of heavy elements. These simulations include
element-by-element cooling, a first for cosmological chemodynamical
simulations, and they take the effect of photo-ionization
by the UV/X-ray background radiation on the cooling rates into
account, which is normally either ignored or included only for
hydrogen and helium. While this paper
focuses on a single implementation of the relevant physics, we will
present the effects of 
varying the physical parameters (e.g., cosmology,
star formation, cooling, feedback from star formation) in a future
paper that will make use of a larger fraction of the OWLS data set.

This paper is organized as follows. The simulations and the
prescriptions used for star formation, feedback from star formation,
and radiative cooling are summarized in \S\ref{sec:sim}. In
\S\ref{sec:stellarevol} we discuss in some detail the ingredients that
we take from models of stellar evolution or observations, including
the IMF, stellar 
lifetimes, stellar yields, and SN Ia rates. In \S\ref{sec:ejmass} we
discuss how we combine all these ingredients to predict the mass
released by a simple stellar population as a function of its age and
present some results. Our implementation of chemodynamics into SPH is
discussed in \S\ref{sec:sph_implementation}. This section also
contrasts two possible definitions of elemental abundances in SPH and
demonstrates that the choice of definition can be extremely important due
to limitations intrinsic to SPH. In \S\ref{sec:results} we
use our simulations to address the question ``Where are the metals?''
and we show how the answer is influenced by the definition of
metallicity  that is used. Finally, we provide a detailed summary in
\S\ref{sec:conclusions} and investigate convergence with the size of
the simulation box and with resolution in
appendices~\ref{sec:boxsize} and \ref{sec:resolution}, respectively.  

For the solar abundance we use the metal mass fraction $Z_\odot =
0.0127$, corresponding to the value obtained using the default
abundance set of \textsc{cloudy} (version 07.02; last described by
\citealt{Feta98}). This abundance set (see Table \ref{tab-abund})
combines the abundances of \cite{Allende2001,Allende2002} for C
and O, \cite{Holweger2001} for N, Ne, Mg, Si, and Fe, and assumes $n_{\rm
  He}/n_{\rm H} = 0.1$ which is a typical value for nebular with
near-solar compositions. Note that much of the literature assumes
$Z_\odot = 0.02$, which is 0.2~dex higher than the value used here.

\begin{table}
\caption{Adopted solar abundances.}
\label{tab-abund}
\centering
\begin{tabular}{ccc}
\hline\hline Element & $n_i/n_{\rm H}$ & Mass Fraction\\ 
\hline 
H & 1 & 0.7065 \\ 
He & 0.1 & 0.2806 \\ 
C & $2.46\times 10^{-4}$ & $2.07\times 10^{-3}$ \\ 
N & $8.51\times 10^{-5}$ & $8.36\times 10^{-4}$ \\ 
O & $4.90\times 10^{-4}$ & $5.49\times 10^{-3}$ \\ 
Ne & $1.00\times 10^{-4}$ & $1.41\times 10^{-3}$ \\ 
Mg & $3.47\times 10^{-5}$ & $5.91\times 10^{-4}$ \\ 
Si & $3.47\times 10^{-5}$ & $6.83\times 10^{-4}$ \\ 
S & $1.86\times 10^{-5}$ & $4.09\times 10^{-4}$ \\ 
Ca & $2.29\times 10^{-6}$ & $6.44\times 10^{-5}$ \\ 
Fe & $2.82\times 10^{-5}$ & $1.10\times 10^{-3}$ \\ 
\hline
\end{tabular}
\end{table}

\section{Simulations}
\label{sec:sim}

We performed cosmological, gas-dynamical simulations using a
modified version of the $N$-body Tree-PM, SPH code \textsc{gadget
  iii}, which is a modified  
version of the code \textsc{gadget ii} \citep{Springel2005}. Our
prescriptions for star formation, for feedback from core collapse
supernovae and for radiative cooling and heating are summarized
below. Our prescription for stellar evolution, which is the subject
of this paper, is described in detail in the next section.

\begin{table*} 
\begin{center}
\caption{List of simulations used. The columns give the comoving size
  of the box $L$, the total number of particles per component $N$ (dark
  matter and baryons), the initial baryonic particle mass $m_{\rm b}$, the dark matter
  particle mass $m_{\rm dm}$, the (Plummer-equivalent) comoving softening length
  $\epsilon_{\rm com}$, the maximum (Plummer-equivalent) proper
  softening length $\epsilon_{\rm prop}$, and the redshift at which the simulation 
  was stopped $z_{{\rm end}}$.}
\label{tbl:sims}
\begin{tabular}{lrcllrrr}
\hline
Simulation & $L$ & $N$ & $m_{\rm b}$ & $m_{\rm dm}$ &
$\epsilon_{\rm com}$ & $\epsilon_{\rm prop}$  & $z_{\rm end}$\\  
& $(\hMpc)$ & & $(\hMsun)$ & $(\hMsun)$ & $(\hkpc)$ &
$(\hkpc)$ &  \\
\hline 
L006N128 &   6.25 & $128^3$ & $1.4 \times 10^6$
& $ 6.3 \times 10^6$ & 1.95 & 0.5 & 2\\
L012N256 &  12.50 & $256^3$ & $1.4 \times 10^6$
& $ 6.3 \times 10^6$ & 1.95 & 0.5 & 2 \\
L025N128 &  25.00 & $128^3$ & $8.7 \times 10^7$
& $ 4.1 \times 10^8$ & 7.81 & 2.0 & 0 \\
L025N256 &  25.00 & $256^3$ & $1.1 \times 10^7$
& $ 5.1 \times 10^7$ & 3.91 & 1.00 & 2 \\
L025N512 &  25.00 & $512^3$ & $1.4 \times 10^6$
& $ 6.3 \times 10^6$ & 1.95 & 0.5 & 1 \\
L050N256 &  50.00 & $256^3$ & $8.7 \times 10^7$
& $ 4.1 \times 10^8$ & 7.81 & 2.0 & 0 \\
L050N512 &  50.00 & $512^3$ & $1.1 \times 10^7$
& $ 5.1 \times 10^7$ & 3.91 & 1.0 & 0 \\
L100N128 & 100.00 & $128^3$ & $5.5 \times 10^9$
& $ 2.6 \times 10^{10}$ & 31.25 & 8.0 & 0 \\
L100N256 & 100.00 & $256^3$ & $6.9 \times 10^8$
& $ 3.2 \times 10^9$ & 15.62 & 4.0 & 0 \\
L100N512 & 100.00 & $512^3$ & $8.7 \times 10^7$
& $ 4.1 \times 10^8$ & 7.81 & 2.0 & 0 \\
\hline
\end{tabular}
\end{center}
\end{table*}

We use the suite of simulations of varying box sizes and particle
numbers listed in Table~\ref{tbl:sims} together with the corresponding
particle masses and gravitational force softening scales. We use
fixed comoving softening scales down to $z=2.91$ below which we switch
to a fixed proper scale. Our largest
simulations use $2\times 512^3$ particles in boxes of comoving size
$L=25$, $50$, and $100~\hMpc$. 

The initial particle positions and velocities are obtained
from glass-like \citep{White1994} initial conditions using \textsc{cmbfast}
(version 4.1; \citealt{Seljak1996}) and employing the Zeldovich approximation
to linearly evolve the particles down to redshift
$z = 127$. We assume a flat $\Lambda$CDM universe and employ the
set of cosmological parameters
$[\Omega_m,\Omega_b,\Omega_\Lambda,\sigma_8, n_s, h] = [0.238, 0.0418, 
0.762,  0.74, 0.951, 0.73]$, as determined from the WMAP 3-year data
and consistent\footnote{Our value of $\sigma_8$ is $1.6\,\sigma$ lower
than allowed by the WMAP 5-year data.} with the WMAP 5-year data
\citep{Komatsu2008}. In addition, the primordial helium mass fraction was 
set to $X_{\rm H} = 0.248$. 

We employ the star formation recipe of \cite{Schaye2008}, to which we
refer the reader for details. Briefly, 
gas with densities exceeding the critical density for the
onset of the thermo-gravitational instability ($n_{\rm H} \sim 10^{-2}
- 10^{-1}~\cm^{-3}$) is
expected to be multiphase and star-forming
\citep{Schaye2004}. We therefore impose an
effective equation of state with pressure $P\propto \rho_g^{\gamma_{\rm
    eff}}$ for densities exceeding $n_{\rm H} =
0.1~\cm^{-3}$, normalized to $P/k = 10^3~\cm^{-3}\,\K$ for an atomic gas at the
threshold density. We use $\gamma_{\rm eff} = 
4/3$  for which both the Jeans mass and the ratio 
of the Jeans length and the SPH kernel are independent of the density,
thus preventing spurious fragmentation due to a lack of numerical resolution. 

The local Kennicutt-Schmidt star formation law is
analytically converted 
and implemented as a pressure law. As we demonstrated in
\cite{Schaye2008}, our method allows us to reproduce
arbitrary input star formation laws for any equation of state without
tuning any parameters. We use the observed \cite{Kennicutt1998} law,
\begin{equation}
\dot{\Sigma}_\ast = 1.5 \times 10^{-4}\, \Msun \yr^{-1} \kpc^{-2} \left
    ({\Sigma_{\rm g} 
    \over 1 ~\Msun \pc^{-2}}\right )^{1.4},
\label{eq:KS}
\end{equation}
where we divided Kennicutt's normalization by a factor 1.65 to account for the
fact that it assumes a Salpeter IMF whereas we are using a \cite{C03}
IMF.

Galactic winds are implemented as
described in \cite{Dallavecchia2008}. Briefly, after a short
delay of $t_{\rm SN} =
3\times 10^7~\yr$, corresponding to the maximum lifetime of stars that
end their lives as core-collapse supernovae, newly-formed star particles
inject kinetic energy into their surroundings by
kicking a fraction of their neighbours in a random direction. The
simulations presented 
here use the default parameters of \cite{Dallavecchia2008}, which
means that each SPH neighbour $i$ of a newly-formed star particle $j$
has a probability of $\eta m_j/\sum_{i=1}^{N_{\rm ngb}}m_i$ of
receiving a kick with a velocity $v_{\rm w}$. We choose $\eta = 2$ and
$v_{\rm w} = 600~\kms$  
(i.e., if all baryonic particles had equal
mass, each newly formed star particle would kick, on average, two of
its neighbours). Assuming that each star with
initial mass in the range $6-100~\Msun$ injects
$10^{51}~\erg$ of kinetic energy, these parameters imply that the
total wind energy accounts for 40 per cent of the available kinetic
energy for a Chabrier IMF and a stellar mass range
$0.1-100~\Msun$ (if we consider only stars in the mass range $8-100~\Msun$ for 
type II SNe, this works out to be 60 per cent). The value $\eta=2$ was
chosen to 
roughly reproduce the peak in the cosmic star formation rate. Note
that contrary to the widely-used kinetic 
feedback recipe of \cite{Springel2003}, the kinetic energy is injected
\emph{locally} and the wind particles are \emph{not} decoupled
hydrodynamically. As 
discussed by \cite{Dallavecchia2008}, these differences have important
consequences. 

Radiative cooling was implemented according to
\cite{Wiersma2009}\footnote{We used their equation (3) rather than (4)
  and \textsc{cloudy} version 05.07 rather than 07.02.}. In brief, net
radiative cooling 
rates are computed element-by-element in the presence of the cosmic
microwave background (CMB) and a \citet[][hereafter HM01]{HM01} model
for the UV/X-ray background radiation from quasars and galaxies. The
contributions of the eleven elements hydrogen, helium, carbon, nitrogen,
oxygen, neon, magnesium, silicon, sulphur, calcium, and iron are
interpolated as a function of density, temperature, and redshift from
tables that have been pre-computed using the publicly available
photo-ionization package \textsc{cloudy}, last described by
\cite{Feta98}, assuming the gas to be optically thin and in
(photo-)ionization equilibrium.

Hydrogen reionization is implemented by turning on the evolving,
uniform ionizing 
background at redshift $z=9$. Prior to this redshift the cooling rates
are computed using the CMB and a photo-dissociating background which
we obtain by cutting off the $z=9$ HM01 spectrum at 1~Ryd. Note that
the presence of a photo-dissociating background suppresses H$_2$
cooling at all redshifts. 

\begin{figure} 
\includegraphics[width=84mm]{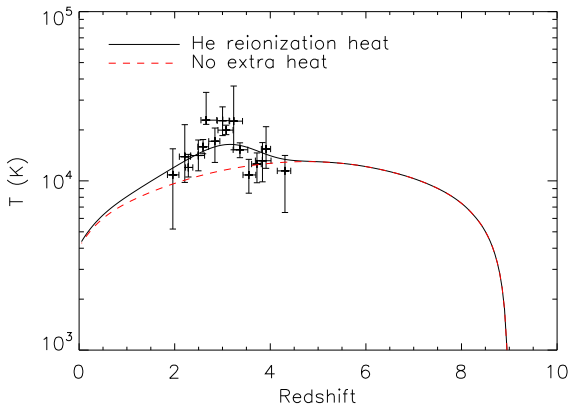}
\caption{A comparison between the simulated and observed evolution of
  the temperature of gas with density equal to the cosmic mean. The
  dashed curve indicates the temperature evolution for a gas parcel
  undergoing Hubble expansion, computed
  using the same radiative cooling and heating rates as are used in our simulations for
  gas of primordial composition. These net
  cooling rates include the effects of the CMB and, below $z=9$, of the
  evolving UV/X-ray background (UVB) from galaxies and quasars as modeled
  by HM01, assuming the gas to be optically thin and in ionization
  equilibrium. The redshift of hydrogen reionization was set to $z=9$
  by turning on the UVB at this redshift, resulting in a steep
  increase of the temperature to $T\sim 10^4~\K$. 
  A comparison with with the data points of \protect\cite{Seta00}, which were
  derived from the widths of Ly$\alpha$ absorption
  lines, shows that the simulation underestimates the temperature
  at $z\approx 3$, the redshift around which the reionization of helium is
  thought to have ended. To mimic the expected extra heat input, relative to
  the optically thin limit, during helium reionization, the solid
  curve shows the result of injecting an extra
  $2~{\rm eV}$ per proton, smoothed with a $\sigma(z)=0.5$ Gaussian
  centered at $z = 3.5$. The reduced $\chi^2$ for the solid and
  dashed curves is 1.1 and 4.5, respectively. \label{fig-themhist}} 
\end{figure}

Our assumption that the gas is optically thin may lead to an
underestimate of the temperature of the IGM shortly after
reionization \cite[e.g.][]{Abel1999}. Moreover, it is well known that without
extra heat input, 
hydrodynamical simulations underestimate the temperature of the IGM at
$z\ga 3$, the redshift around which helium reionization is thought
to have ended
\cite[e.g.][]{Theuns1998,Theuns1999,Bryan1999,Seta00,Ricotti2000}. We therefore
inject $2~{\rm eV}$ per proton, 
smoothed with a $\sigma(z)=0.5$ Gaussian centered at $z = 3.5$. Figure
\ref{fig-themhist} shows that with this extra energy, the predicted
temperature at the mean density agrees well with the measurements of
\cite{Seta00}.

\section{Ingredients from stellar evolution}
\label{sec:stellarevol}

Cosmological simulations cannot at present resolve
individual stars. Instead, a single particle is taken to represent a
population of stars of a single age, a so-called simple stellar
population (SSP), with some assumed stellar initial mass function
(IMF). The task of stellar evolution modules in cosmological
hydrodynamical simulations is to implement 
the timed release of kinetic energy and mass by star particles. The
feedback processes originating from stellar evolution that we will consider
are winds from asymptotic giant branch (AGB) stars, type Ia supernovae
(SNe), type II (i.e.\ core collapse) SNe\footnote{Type Ib and Ic SN are also core collapse 
supernova, but we follow other authors and use type II to refer to the 
entire class of core collapse SN.} and the winds from their
progenitors. In this section, we briefly present the ingredients that 
go into our model, while deferring a discussion of alternatives to appendix \ref{app:stellarevol}.

Although many formulations of the stellar initial mass function (IMF) 
exist in the literature, there is consensus that a simple 
power law predicts too many low mass stars below $1~\Msun$. We have 
decided on a \citet{C03} IMF - because it is an example of an IMF with a low mass 
turnover that fits the observations within the uncertainties. For the 
parameters we assume $M_{\rm c} = 0.079$, $\sigma = 0.69$ (see appendix \ref{sec:IMF}) 
with mass limits of $0.1~\Msun$ and $100~\Msun$. While this 
is a sensible choice, uncertainties in the IMF can have a significant 
effect on the global metal production.

For stellar yields, we choose the complete set of \cite{M01} and
\cite{Peta98}, along with the type Ia SN yields of the W7 model of
\cite{Teta03}. This covers mass loss from intermediate mass stars as
they pass through the AGB phase \citep{M01}, which with this yield set
has an upper limit of $5~\Msun$, as well as mass loss and explosive
nucleosynthesis that occur in high mass stars up to $100~\Msun$
\citep{Peta98}\footnote{The high mass stellar yields were modified
  following the advice of L. Portinari - see appendix
  \ref{sec-yields} for more details.}.  The yields of \cite{Teta03} cover the
regime of intermediate mass stars in close binaries which end their
life as type Ia SNe. These two modes of SNe inject energy
into the ISM. We inject energy from core collapse SNe in kinetic form
as discussed in section \ref{sec:sim}
and in more detail in \cite{Dallavecchia2008}. The energy from SN Ia
is injected thermally among the SPH neighbors of the star particles.

For a more detailed discussion of the yields we refer the reader to
appendix~\ref{app:stellarevol}, where we show, by comparing different
sets of 
nucleosynthetic yields taken from the literature, that they represent
an important source of uncertainty. In particular, for
AGB stars, as well as for  
the elements heavier than nitrogen in massive stars, the yields vary
by factors of a few. We make the choice 
that most suits our needs, while keeping in mind the resulting uncertainty.

As in most recent chemodynamical implementations, we relax the instantaneous 
recycling approximation. This means we apply a finite lifetime for each star 
as a function of mass. In consistency with our yield sets, we use 
the metallicity-dependant lifetimes of \cite{Peta98}. As 
shown in appendix~\ref{sec:lifetimes} however, there is little difference 
between published lifetimes.

Since type Ia SNe are thought to be a product of binary evolution, 
there is no simple `lifetime' to assign to their progenitors. Instead, 
theoretically or empirically derived rates must be used. We apply a
modification  
of a e-folding delay function that is normalised to the cosmic 
type Ia SNe rate. In appendix \ref{sec:SNIa_rates} we show that 
this is yet another source of uncertainty in chemical evolution 
models. Since the precise progenitors are still unknown, and 
since the observations of type Ia SNe are inconsistent, there is 
freedom to change the total number of type Ia SNe by factors 
of a few.

\section{The mass ejected by a simple stellar population}
\label{sec:ejmass}

The mass that is ejected during a
single time step $(t,t+\Delta t)$ by a star
particle of initial mass $m_{\ast,0}$ and metallicity $Z$ that was
created at time $t_\ast<t$ is given by 
\begin{eqnarray}
\Delta m_\ast &=& m_\ast(t) - m_\ast(t+\Delta t) \\
&=& m_{\ast,0}
\int_{M_Z(t-t_\ast+\Delta t)}^{M_Z(t-t_\ast)} \Phi(M) m_{\rm ej}(M,Z)\,dM,
\end{eqnarray}
where $M_Z(\tau)$ is the inverse of the lifetime function $\tau_Z(M)$,
$m_{\rm ej}(M,Z)$ is the mass ejected by a single star and
$\Phi(M)$ is the IMF, normalized such that $\int M\Phi(M)dM = 1$.
It is straightforward to generalize the above equations to give $\Delta
m_{\ast,j}$, the mass ejected in the form of element $j$. The
functions $\tau_Z(M)$ and $m_{j,{\rm ej}}(M,Z)$ need to be taken from stellar
evolution and nucleosynthesis calculations provided in the literature.

\subsection{Implementation}
\label{sec:implementation}

Since stellar evolution and nucleosynthesis depend on initial
composition, a correct treatment of chemodynamics would require
yields (i.e.\ the ejected mass of various elements) to be tabulated in
a space with dimensionality equal to the number of elements. Most
yields are, however, given only as a 
function of metallicity, where metallicity is the total mass fraction
in metals. The models used to compute the yields assume implicitly
that the relative abundances of the heavy elements are initially
solar. 

One way to implement the release of elements is to interpolate the
ejected masses, which are given as a function of metallicity, to the
metallicity of the star particle and to ignore possible differences
between the relative abundances assumed by the yield
calculation and the actual relative abundances of the star
particle. However, this can have some very undesirable
consequences. For example, because iron does not 
participate in the nucleosynthesis of intermediate mass stars, the
iron that these stars eject was already present when the star was
born. If an intermediate mass star has a high iron
abundance for its metallicity, then simple interpolation of the yield
tables with respect to 
metallicity would lead to a large underestimate of the ejected iron
mass. That is, it would result in the spurious destruction of iron.

A more accurate way to implement chemodynamics would thus be to 
distinguish metals that were produced/destroyed from those that pass
through the star without participating in the nucleosynthesis. If we
knew, for each ejected element, what fraction of the ejected mass
resulted from net nucleosynthetic production (i.e., production minus
destruction) and what fraction simply passed through the star (the two
fractions should add up to one), then it would make sense to assume
that the mass that passed through the star is proportional to the
star's initial abundance. In that case we would still neglect the
effect of relative abundances on the nucleosynthesis itself (this can
only be done by repeating the nucleosynthesis calculations or using the 
somewhat cumbersome Q-matrix formalism, e.g.\ \citealt{Peta98}), 
but we would take into account the consequences of varying relative
abundances on metals that do not participate in the nucleosynthesis.

We have adopted this strategy - as is often done 
\cite[e.g.][]{Reta01, LPC02, Sommer-larsen2005, Reta06, Tornatore2007} - 
and have implemented it as follows\footnote{Note that this strategy applies directly 
for our chosen yields since they are presented as what is produced minus what was 
initially in the star. Some yields \cite[e.g.][]{WW95} would have to be modified 
to implement this approach.}. For a star of
given initial mass and metallicity, we can write the mass of element $j$ that
is ejected into the surrounding medium, $m_{j,{\rm ej}}$, as the sum
of two parts:  
\begin{equation}
m_{j,{\rm ej}} = m_{j, 0:{\rm ej}} + m_{j, {\rm p:ej}}.
\label{eq:m_j,ej}
\end{equation}
Here $m_{j,0:{\rm ej}}$ is the mass in element $j$ that would
have been ejected if no nucleosynthesis had taken place and if the
star were well-mixed, 
\begin{equation}
m_{j, 0:{\rm ej}} = \left(\frac{m_{j,0:{\rm star}}}{m_{\rm tot,
    0:star}}\right)_{\rm sim} \left (m_{\rm tot, ej}\right )_{\rm tbl},
\label{eq:m_j,0:ej}
\end{equation}
where $\left(m_{j,0:{\rm star}}/m_{{\rm tot, 0:star}}\right)_{\rm sim}$ is the
initial mass fraction of element $j$ in the star \textit{as tracked by
  the simulation} and $\left (m_{\rm tot, ej}\right )_{\rm tbl}$ is the total 
ejected mass according to the yield table. Note that we use the
subscript ``sim'' to indicate values predicted by the simulation and
the subscript ``tbl'' to indicate values that are published by
authors presenting a particular set of stellar yields. The second term
of equation 
(\ref{eq:m_j,ej}) represents the  
ejected mass in element $j$ that was produced minus destroyed. 
Assuming
  that the star is well-mixed, we can write it as
\begin{equation}
m_{j,{\rm p:ej}} = \left (m_{j,{\rm ej}}\right )_{\rm tbl} -
\left(\frac{m_{j,0:{\rm star}}}{m_{\rm tot, 0:star}}\right)_{\rm tbl}
\left (m_{\rm tot,ej}\right )_{\rm tbl},
\label{eq:m_j,p:ej}
\end{equation}
where $\left(m_{j,0:{\rm star}}/m_{\rm tot, 0:star}\right)_{\rm tbl}$ is
the initial mass fraction of element $j$ that was used for the
nucleosynthesis calculation. This term, which is often referred to as
the yield of element $j$, will be negative if an element
effectively gets destroyed (this is for example the case for hydrogen).

Combining equations (\ref{eq:m_j,ej}),
(\ref{eq:m_j,0:ej}), and (\ref{eq:m_j,p:ej}), we obtain:
\begin{equation}
\begin{array}{ll} m_{j, {\rm ej}} = (m_{j, {\rm ej}})_{\rm tbl} ~+ 
\\ ~~~~\left[\left(\frac{m_{j,0:{\rm star}}}{m_{\rm tot,0:star}}\right)_{\rm sim}
  - \left(\frac{m_{j,0:{\rm star}}}{m_{\rm tot,
      0:star}}\right)_{\rm tbl}\right] (m_{\rm tot, ej})_{\rm tbl}.\end{array}
\label{eq:m_j,ej2}
\end{equation}
Note that this expression behaves as expected in the limit that the
simulated abundances agree with the ones assumed by the yield tables (i.e.\
the second term vanishes). Note also that although the abundances
appearing in equation (\ref{eq:m_j,ej2}) are absolute, it is the
abundance relative to other heavy elements that is important because
we interpolate the yield tables to the metallicity (i.e.\ $\sum_j m_{j,0}
/m_{\rm tot,0}$ where the sum is over all elements heavier than helium) of
the simulation star particle.

The ejected mass predicted by equation (\ref{eq:m_j,ej2}) can in
principle be negative if the simulated relative abundance of some
element is much lower than the relative abundance assumed in the yield
calculation and if the element is not produced in significant
amounts. To prevent stellar particles from obtaining negative total
abundances, we therefore impose that for each element $j$, the \emph{total}
ejected mass, which consists of the sum of the mass ejected by
intermediate mass stars and SNe of types Ia and II, is
non-negative at each time step. If it happens to be negative, we set
it to zero. In practice we
have, however, not observed this to occur in any of our simulations.

If some $m_{j,{\rm ej}}$ has been set to zero, as in the scenario described
in the previous paragraph, or if the simulated abundances $\left
(m_{j,0:{\rm star}}/m_{\rm tot,0:star}\right )_{\rm sim}$ do not add up to
unity (which can happen when using smoothed abundances; see
section~\ref{sec:zsm}), then the sum of the ejected 
masses, $\sum_j m_{j,{\rm ej}}$, may differ from the total
ejected mass, $(m_{\rm tot, ej})_{\rm tbl}$, predicted by the yield
tables. To ensure that we release the correct 
amount of mass, we therefore normalise the mass released from each element,
including the total metal mass (see section \ref{sec-totmet}), such
that the total mass released agrees with the value found in the yield
tables.

\subsubsection{Tracking the total metal mass} 
\label{sec-totmet}

Of the many elements tracked by stellar evolution calculations, only a
fraction contribute significantly to the radiative cooling
rates. These elements tend to be the most 
abundant and thus the most observed. We therefore only track 9 elements
individually: H, He, C, N, O, Ne, Mg, Si, and Fe (plus we
take Ca and S, whose contributions to the cooling rates can be
significant, see \citealt{Wiersma2009}, to be proportional to
Si). Because we only track a finite number of elements, we cannot
define the metallicity to be $Z = \sum_j m_j/m_{\rm tot}$ where the sum
is over all elements heavier than He that are tracked by the
simulation. Using $Z = 1 - (m_{\rm H} + m_{\rm He})/m_{\rm tot}$ is
also problematic since this definition is susceptible to round-off
errors. 

We therefore treat the total metal mass in the same manner as we treat
individual elements, i.e., we include it as an additional array, as
has also been done by others \citep[e.g.][]{LPC02, KG03,
  Valdarnini2003, Sommer-larsen2005, Reta06}. There 
are a number of ways to determine the yield of the total metals
released by a stellar population, $Y_Z \equiv m_{Z,{\rm p:ej}}$. Since the
sum of all $Y_j \equiv m_{j,{\rm p:ej}}$ (including 
hydrogen and helium) should be zero (see
eq.\ \ref{eq:m_j,p:ej}), one could simply define 
$Y_Z = - (Y_{\rm H} + Y_\He)$. 
Unfortunately, yield tables often give  $\sum_j Y_j < 0$, which
implies that more  
mass is destroyed than produced. This is mainly due to non-conservation of 
nucleons and round-off errors (L.~Portinari, private communication). We 
choose therefore to define $Y_{\rm Z}$ as the sum of the yields of all
elements heavier than helium and 
redefine the hydrogen yield as:
\begin{equation}
Y_{\rm H} \equiv - (Y_\He + Y_{\rm Z}).
\end{equation}
Note that $Y_{\rm Z}$ will differ from the sum of the metal yields of
all elements tracked by the simulation if the yield tables contain
more elements than are tracked by the simulation. The differences and round-off 
errors may not be significant in light of the uncertainties in our method (e.g., 
different yield sets, the definition of metallicity, etc.), but we choose to try 
to minimize the uncertainties at each step.

\subsection{Results}

\begin{figure}
\includegraphics[width=84mm]{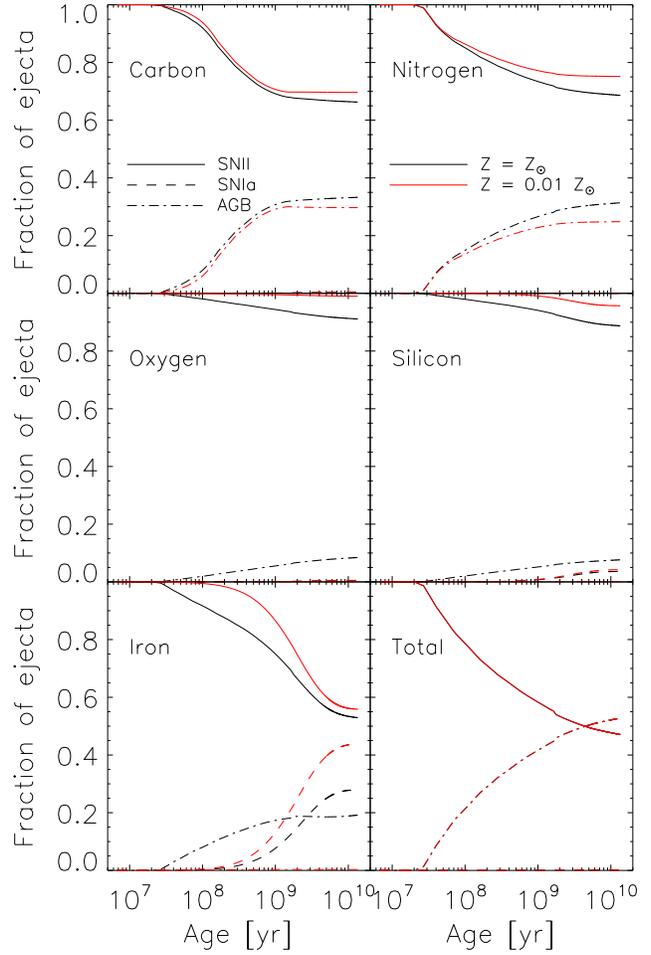}
\caption{The fraction of the integrated mass ejected by an SSP that is
  due to SNII (solid), SNIa (dashed), and AGB (dot-dashed) as a
  function of the age of the SSP for two initial stellar
  metallicities: solar (black) and one per cent of solar (red). All
  curves assume a Chabrier IMF. The
  bottom-right panel corresponds to the total ejected mass, whereas
  the other panels correspond to individual elements, as indicated in
  the plot. While nearly all the ejecta come initially from SNII, for
  ages $\tau \ga 10^8~\yr$ the contributions from AGB stars become significant
  for carbon, nitrogen and, if the metallicity is high, for other
  elements as well. The contribution from SNIa very important for iron
  for $\tau \ga 10^9~\yr$. Note that the predictions are uncertain at
  the factor of two level, mainly due to the freedom in the
  normalization of the SNIa rate and the uncertainty in the
  yields.\label{fig:ejfrac}} 
\end{figure}

\begin{figure*}
\includegraphics[width=84mm]{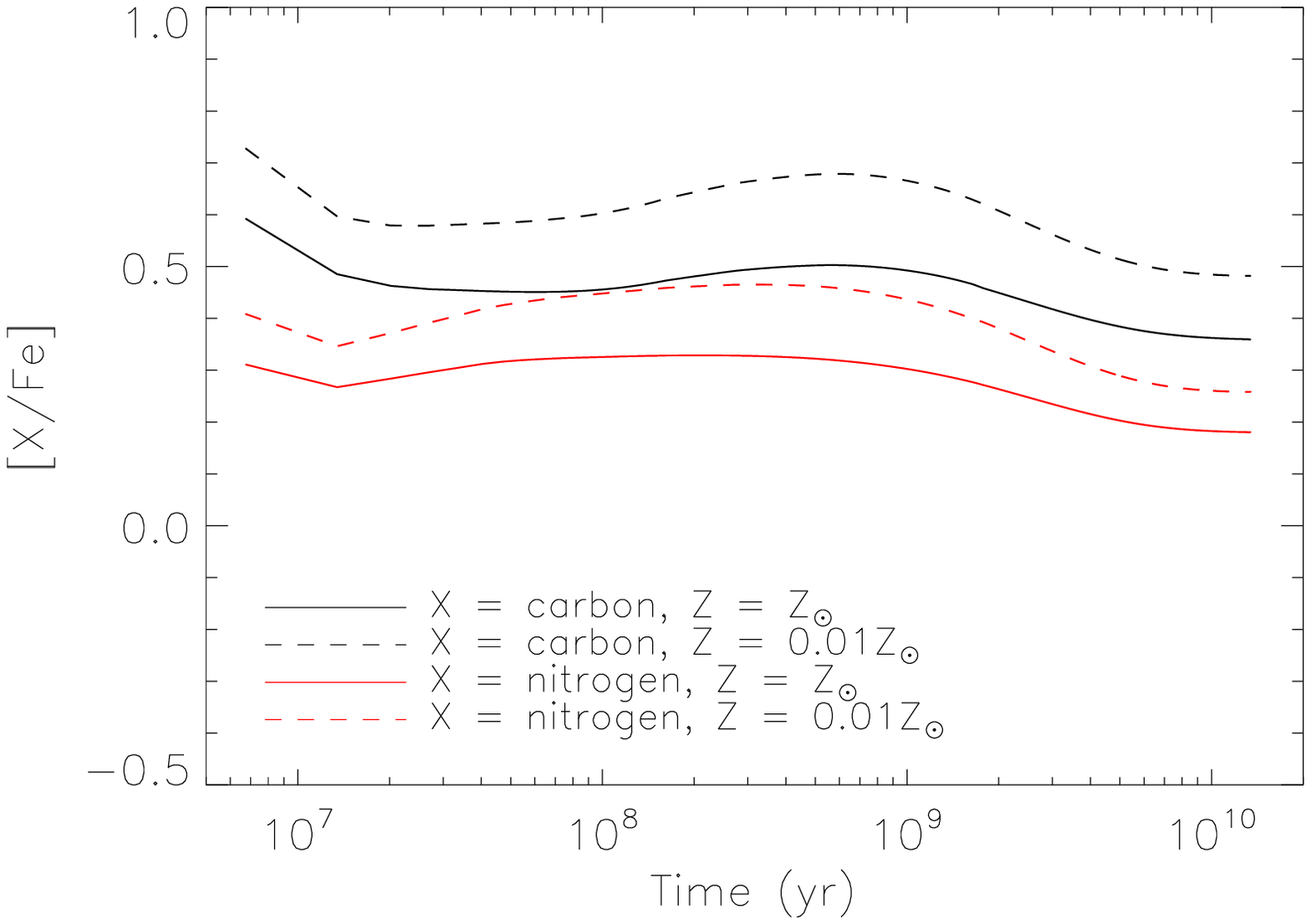}
\includegraphics[width=84mm]{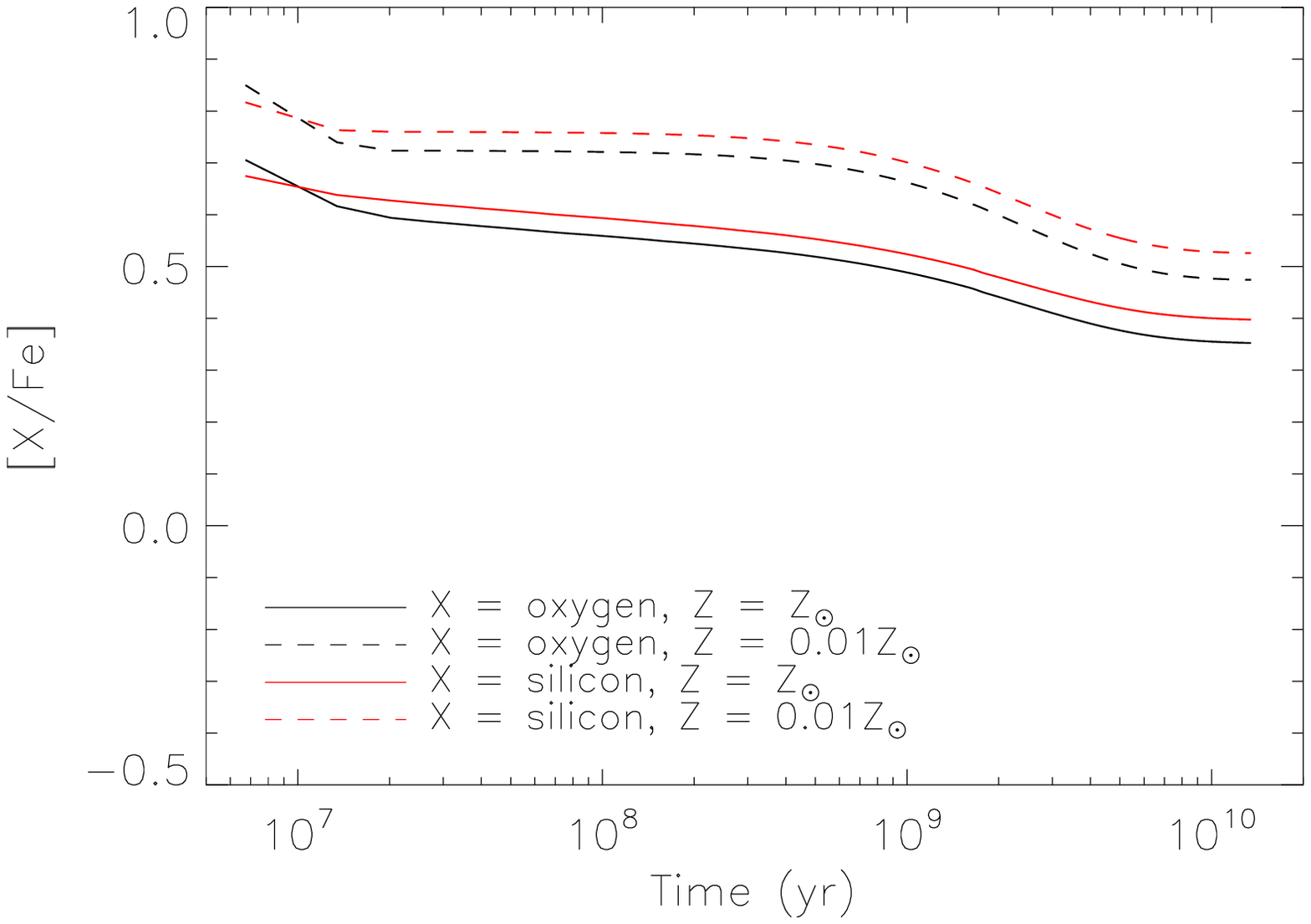}
\caption{The abundances of various elements relative to iron in the
  integrated ejecta of an SSP as a function of its age for two
  metallicities: solar (solid) and one per cent of solar (dashed). All
  curves assume a Chabrier IMF. The abundances of all elements shown
  are initially enhanced relative to iron, particularly for lower
  stellar metallicities. The integrated abundances decrease relative
  to that of iron for $\tau \ga 10^9~\yr$ when large amounts of iron
  are ejected by SNIa. Note that the predictions are uncertain at
  the factor of two level, mainly due to the freedom in the
  normalization of the SNIa rate and the uncertainty in the
  yields. \label{fig:relabund}}
\end{figure*}

Figures~\ref{fig:ejfrac} and \ref{fig:relabund} show the result of
combining all the ingredients described in
appendix~\ref{app:stellarevol} using the implementation described in
this section. Both figures provide information about the composition
of the integrated (i.e., cumulative in time) ejecta of an SSP as a
function of its age. While figure~\ref{fig:ejfrac} shows the fraction of
the mass due to SNII, AGB and SNIa, figure~\ref{fig:relabund} plots the
abundance of a few elements relative to iron. 

At early times, $\tau \la 10^8~\yr$, the integrated ejecta are totally
dominated by SNII and consequently the abundances relative to iron of
elements such as carbon, nitrogen, oxygen, and silicon are highly
supersolar. As the stellar population ages, the contributions of AGB stars
become very important for carbon and nitrogen and those of SNIa for
iron. Higher initial stellar metallicities tend to result in larger
abundances relative to iron, although the effect becomes smaller at
late times because the yields of SNIa are independent of metallicity. 

Note that these figures
show the properties of the mass released integrated over time rather
than at a particular time. For ages $\tau > 10^8~\yr$ all mass loss
comes from AGB stars and SNIa and the elemental abundances relative to iron
of the instantaneous ejecta are therefore much lower than those in the
integrated ejecta that are shown here.

We emphasize that the predictions shown in these figures are very
uncertain, even for a fixed IMF. As we
discuss in appendix~\ref{sec-yields}, the yields are uncertain at the
factor of two level. For example, we rescaled the SNII yields of C,
Mg, and Fe by factors of 0.5, 2, and 0.5, respectively (note that
these factors were included in Figs.~\ref{fig:ejfrac} and
\ref{fig:relabund}, but not in any of the figures appearing in
appendix~\ref{sec-yields}). The relative contribution of SNIa is proportional
to the normalization of the SNIa rate, which, as discussed in
appendix~\ref{sec:SNIa_rates}, is also highly uncertain. As shown in
Fig.~\ref{fig:ejfrac}, this is particularly important for iron.

We assumed a Chabrier IMF, but the results presented in this section
would be similar for other IMFs as long as they have similar slopes
for stellar masses above a few solar masses, as is for example the
case for the Salpeter IMF.

\section{Implementation into SPH}
\label{sec:sph_implementation}

\subsection{Enrichment scheme}

Having determined the ejected masses $m_{j,{\rm ej}}$ for each stellar
  particle during a given time step, we still need to define how much
  of this mass should go to each of the surrounding gas particles. We  
would like the mass transfer to be isotropic, but it is not obvious
how to accomplish this. As discussed in detail in appendix~A of
\cite{Pawlik2008}, the transport of mass\footnote{\cite{Pawlik2008} considered
photon transport, but the analysis is equivalent for mass transport.} from a
source particle to its SPH neighbours is in general highly biased
towards the direction of the centre of mass of the neighbours. In
essence this is a consequence of the fact that particle-to-particle
transport is only possible in directions where there are particles.

\cite{Pawlik2008} demonstrate that using an SPH kernel transport
scheme, as for example laid out by \citet{Mosconi2001}, results in a
statistical bias that is much smaller than alternative schemes such as
equal weighting or solid angle weighting. It should be
noted, however, that while SPH weighting results in only a small
statistical bias, the transport from individual source particles is
still highly anisotropic. As in previous works 
\cite[e.g.][]{Steinmetz1994, K01a, KG03, Toeta04, Seta05, Seta06a, Kobayashi2007, Tornatore2007}, 
we choose to employ SPH weights and
transfer to each SPH neighbour $k$ of a given star particle the
fraction of the ejected mass given by 
\begin{equation}
w_k = \frac{\frac{m_k}{\rho_k}W(r_k,h)}{\sum_i \frac{m_i}{\rho_i}W(r_i,h)},
\end{equation}
where $h$ is the smoothing length of the star particle (which we determine
in the same manner as for gas particles), $r_i$ is the distance from
the star particle to neighbour $i$, $W$ is the SPH kernel, and the
sum in the denominator is over all SPH neighbours of the star
particle. \cite{Tornatore2007} have investigated changing the number of
neighbours and kernel length used for distribution of heavy
elements. They found that using more neighbours gives somewhat higher
star formation rates, presumably because more particles are affected
by metal-line cooling. We choose to use 48 neighbours, the same number
as is used in the other SPH calculations. 

Finally, we note that we do not change the entropy or, equivalently,
the internal energy per unit mass of the receiving gas particles. The
total internal energy of a receiving particle therefore
increases when its mass increases. In reality the thermal energy of
the gas surrounding a star undergoing mass loss will also change as a
result of the mass transfer, but
not necessarily by the amount that we implicitly assume. In fact, in
the case of SN explosions we inject (part of) this energy explicitly
(see \S\ref{sec:sim}). On the other hand, many of the receiving gas
particles may be star-forming, in which case we impose an effective
equation of state (see \S\ref{sec:sim}). Assuming all star particles
lose 100 per cent of 
their mass and that all this mass is transferred to non-star-forming
particles with $T=10^4\,\K$, the total energy per unit stellar mass
that we implicitly use to heat the transferred mass to this
temperature is only $(c_{\rm s}/v_{\rm w})^2/\eta \sim 10^{-4}$ of the
kinetic energy released by 
core collapse SNe, where $c_s$ is the sound speed of the ambient
medium. Thus, the energy that we implicitly use in the mass 
transfer is negligible. 

Similarly, we also neglect the change in the momentum of star
particles and their gaseous neighbours as a result of the change in
the particle masses. We made this choice partly because we also do
not attempt to follow the actual momentum of the ejecta except for core collapse SNe, in
which case we inject the kinetic energy explicitly. Neglecting
momentum transfer from AGB stars and SNIa can be justified 
for the star particles if the mass
transfer is isotropic, which it is in our case. For the receiving gas
particles this argument does not apply, even when averaged over many
mass transfer events, if the stars and gas have systematically
different bulk flows. However, even in that case the energy
associated with this momentum transfer is negligible compared with the
kinetic energy injected by core collapse SNe unless the bulk flow
velocities of the stars differ systematically from those of the gas
by $\ga 10^2\,\kms$.

\subsection{Smoothed metallicities}
\label{sec:zsm}

\begin{figure*}
\includegraphics[width=56mm]{samplingproblem1.eps}
\includegraphics[width=56mm]{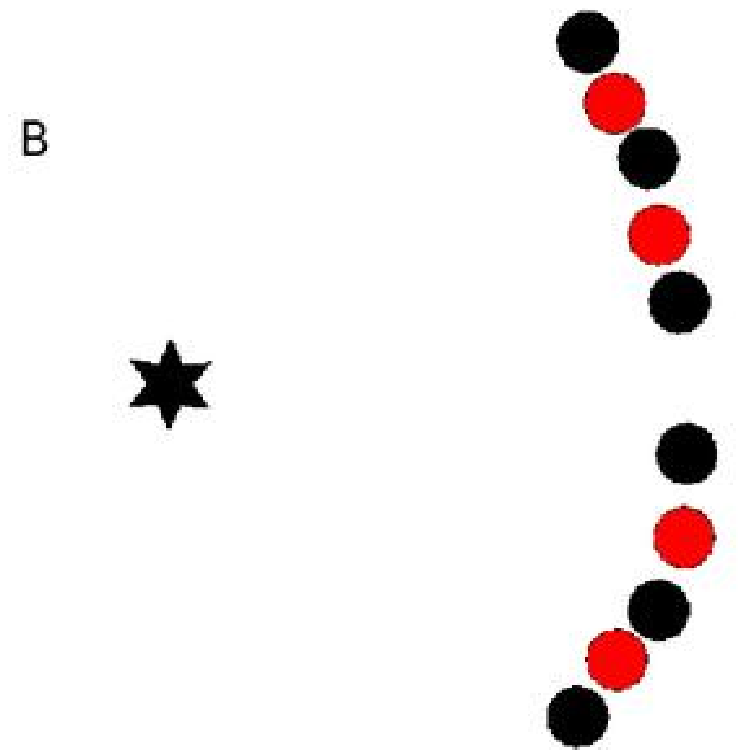}
\includegraphics[width=56mm]{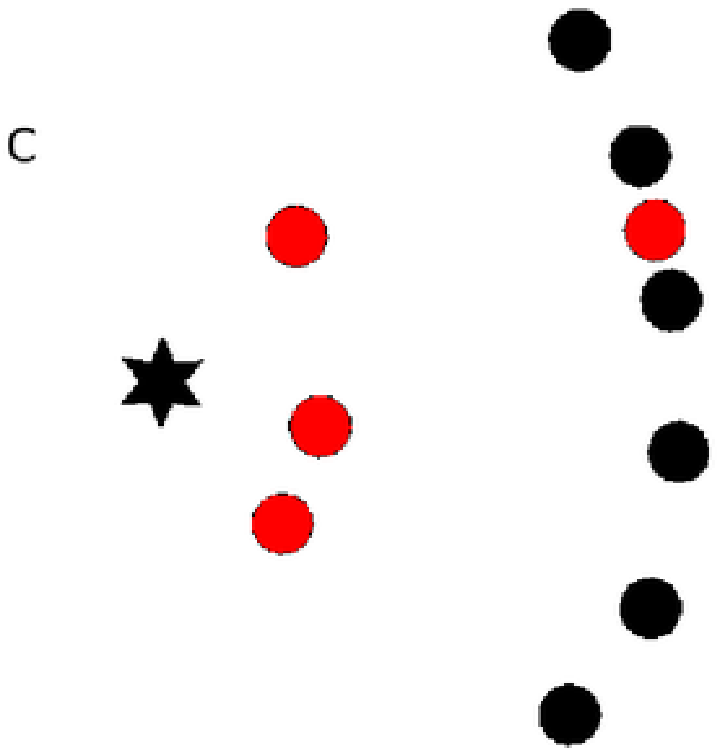}
\caption{The enrichment sampling problem. \textit{A}: A star particle
  enriches its neighbouring gas particles (red). \textit{B}: The
  energy released by massive stars within the 
  star particle drives its neighbours away. Because metals are stuck to
  particle the local metallicity in the shell fluctuates. \textit{C}:
  Using kinetic feedback the problem 
  is worse because only a small fraction of the neighbours are
  kicked.\label{fig-samp}} 
\end{figure*}

The elemental abundances of a particle are not only of interest in the
analysis of the simulations, they are also required during the
simulation itself because gas cooling rates depend on them. Furthermore, the
lifetimes and yields of star particles depend on their metallicities. 

Having tracked and transferred the total metal mass released by star
particles during all time steps (see section \ref{sec-totmet}), we
could simply define the metallicity of a particle to be 
\begin{equation}
Z_{\rm part} \equiv {m_Z \over m}.
\label{eq:Zpart}
\end{equation}
We will refer to metallicities computed using the above expression as
``particle metallicities''.
Alternatively, we could define the metallicity as the ratio of the SPH
smoothed metal mass density and the SPH smoothed gas mass density (as in \citealt{Okamoto2005, Tornatore2007}),
\begin{equation}
Z_{\rm sm} \equiv {\rho_{\rm Z} \over \rho},
\label{eq:Zsm}
\end{equation}
where the gas and metal mass densities of particle $i$ are given by
the standard SPH expressions
\begin{eqnarray}
\rho_i &=& \sum_j m_j W(r_i-r_j,h_i),\\
\rho_{{\rm Z},i} &=& \sum_j m_{{\rm Z},j} W(r_i-r_j,h_i),
\label{eq-smrho}
\end{eqnarray}
where the sum is over all SPH neighbours $j$ and
$h_i$ is the smoothing length of particle $i$. We will refer to
metallicities computed using (\ref{eq:Zsm}) as ``smoothed
metallicities''. Similarly, we will make use of particle and smoothed
abundances of individual elements. Note that while particle abundances
can only change when a particle gains or loses mass, smoothed
abundances will generally vary from time step to time step because
they depend on the distances between particles and their SPH
neighbours. Although most studies do not explicitly define
what they mean by ``metallicity'', particle metallicities are
generally used.   

The use of smoothed abundances has the advantage that it is most
consistent with the SPH formalism. For example, in the absence of
metals gas cooling rates depend on $\rho$ and it therefore makes sense
for the metallic cooling rates to depend on $\rho_Z \equiv Z_{\rm
  sm}\rho$, computed in the same manner as $\rho$. Another
advantage of smoothed 
abundances is that they partially counter the lack of metal mixing
that is inherent to SPH. In particular, there will be many less
particles with zero metallicity if smoothed abundances are used
because every neighbour of a particle with a non-zero particle
metallicity will have a non-zero smoothed metallicity.

The reason SPH underestimates metal mixing is that, in the absence of
some implementation of diffusion, metals are stuck to particles. Even
in the absence of microscopic diffusion processes, this will result in
a lack of metal mixing due to a straightforward sampling problem. This
sampling problem is most easily illustrated by imagining a shell of
constant thickness swept
up by a spherically symmetric explosion driven by a single stellar
particle in a uniform, initially primordial gas, just after it has
completed the transfer of metals produced by massive stars to its
SPH neighbours (see Fig.~\ref{fig-samp}). As the metal-rich bubble
expands, the local  
metal density should decrease as $1/r^2$. However, because the metals
are stuck to a fixed number of particles, the local metal density will
instead fluctuate strongly within the swept-up shell
(Fig.~\ref{fig-samp}, middle panel). The situation may actually be
even worse in practice because we are using kinetic feedback which
implies that only a subset of the SPH neighbours of the star particle
are kicked (Fig.~\ref{fig-samp}, right panel). Note that this
sampling problem does not become smaller when the resolution is
increased. 

The metal sampling problem of SPH can be reduced by implementing
diffusion (and tuning the diffusion coefficient). We have, however,
decided not to do this because diffusion 
is a physical process whereas the sampling problem is in essence a
numerical problem. Attempting to solve it with a poorly constrained
physical mechanism may have undesired consequences. Depending on the
severity of the sampling problem, it may for example require an
unphysical amount of diffusion. The addition of diffusion may also
lead to the introduction of new physical scales (through the 
choice of a particular value for the diffusion coefficient). 

We therefore consider the problem of (turbulent) diffusion separate
from the sampling problem described here. While diffusion processes
may well be important and have a significant
impact on our results \cite[e.g.][]{deAvillez2007}, sub-grid recipes
are required to implement them as they relevant physical processes are
unresolved in cosmological simulations. 

Note that a static gas will not
suffer from the sampling problem but diffusion will still alter its
metallicity distribution. A possible way out may therefore be to
implement a recipe for (turbulent) diffusion that depends explicitly on
the relative velocities of neighbouring particles
\citep{Greif2008}. Although this will  
still implicitly use physical processes (e.g.\ hydrodynamical instabilities) to
get around a numerical problem, it does have the important advantage that
it asymptotes to the standard, no mixing, solution for the case of a
static gas. We intend to test such sub-grid mixing prescriptions in the
future. For now, we note that our use of smoothed abundances decreases the
severity of -- but does not solve -- the sampling aspect of the mixing
problem, leaving the physical aspect of the problem unaddressed. 

We decided to use smoothed abundances in the calculation of the cooling
rates, the stellar lifetimes and the yields. When a gas particle is
converted into a stellar particle its smoothed abundances are frozen.
Note that because the
smoothed abundances of gas particles are recalculated at every time
step from the metal mass fractions, we need to
store both smoothed and particle abundances and can thus compare the two.   

It is important to note that metal mass is only approximately
conserved when smoothed abundances are used using the ``gather
approach'', equation (\ref{eq-smrho}). The non-conservation is exacerbated
by the fact that the smoothed abundances of a particle are frozen
when it is converted from a gas into a star particle. Metal
conservation could be enforced by using a ``scatter''
approach, 
\begin{equation}
\rho_{{\rm Z},i} = \sum_{j} m_{{\rm Z},j} W'(r_i-r_j,h_j),
\end{equation}
normalized such that $\sum_i W'(r_i-r_j,h_j) = 1$ for every particle $j$. However, this is
computationally more expensive and would in any case only conserve
metal mass if the time steps of all particles were synchronized, which
is not the case for \textsc{gadget}. 

It is a well known problem that the SPH technique overestimates 
densities of low density gas in regions where the density gradient is
steep, which may result in overcooling \cite{Pearce1999}. By using
smoothed metal densities we  
are vulnerable to a similar problem that may result in an overestimate
of the metal mass and hence increased overcooling. Indeed, in our low
resolution L025N128 and L100N128 
runs the final total smoothed metal masses are, respectively, 3.6\% and 18\%
greater than the final total particle metal masses. However, as
expected, the difference dereases rapidly with 
increasing resolution. For the L100N256, L100N512, and L025N512 runs
the total  
smoothed metal mass is 8.3\% higher, 3.7\% higher, and 0.14\% lower
than the total particle metal mass in the respective
simulations. Thus, metal mass conservation errors are 
negligible for our highest resolution simulations. 

However, note
that we compared total smoothed 
and particle metal masses in simulations that always used smoothed
abundances to compute the cooling rates, stellar lifetimes and
yields.
Because increased metal abundances will result in increased cooling
and thus increased star formation and increased metal production, we
expect the difference 
to increase if we compare the 
total smoothed metal mass predicted by a simulation employing smoothed
abundances with the total particle metal mass predicted by a
simulation employing particle abundances. We have tested this by
comparing two L100N256 (and two L100N128) simulations. One simulation
used smoothed abundances while the other used particle abundances for
the calculation of the cooling rates. Indeed, we found that at $z=0$
the total smoothed metal mass in the L100N256 (L100N128) simulation
using smoothed abundances was 
49\% (82\%) higher than the total particle metal mass in the
corresponding simulation using particle abundances. 

\begin{figure}
\includegraphics[width=84mm]{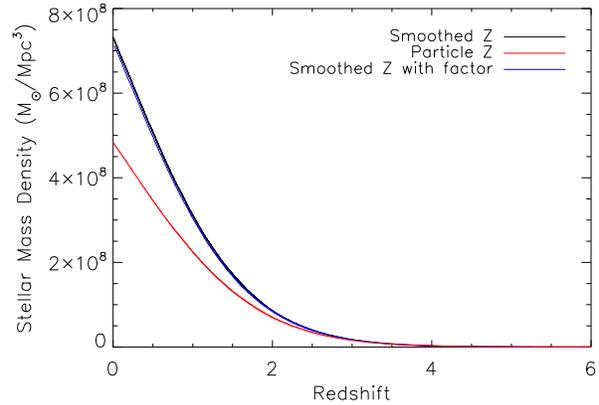}
\caption{Stellar mass density as a function of redshift for $256^3$,
  $100~\mpch$ runs. The black curve shows the total stellar mass density
  in L100N256 which employs smoothed abundances in the calculation of the
  cooling rates. The red curve indicates the total stellar
  mass in a simulation that used particle abundances for the 
  cooling rates. The use of smoothed abundances increases the
  stellar mass produced by $z=0$ by 51 \%. The blue curve used smoothed
  abundances for the cooling after multiplying them by a factor $1/1.08$
  to correct for the increase in the total metal mass, at fixed
  stellar mass, due to the use
  of smoothed abundances. The fact that the blue curve is close to the black one
  indicates that the increase in the stellar mass when using smoothed
  abundances is mostly due to the increased metal mixing rather than
  the small increase in the total metal mass.\label{fig-sfr_sm}}
\end{figure}

As can be seen from Figure~\ref{fig-sfr_sm}, the 
simulation that uses smoothed abundances (black curve) produces about
1.5 times as many stars as the simulation that used particle
abundances (red curve) for the calculation of the cooling
rates. We performed a further
test to determine if the increase in the stellar mass resulting from
our use of smoothed abundances is due to the non-conservation of metal
mass or to the increased metal mixing (i.e., metal cooling affects
more particles). To this end we ran another version of the L100N256
simulation that used smoothed 
abundances but in which the smoothed abundances were multiplied by a
factor $1/1.08$, which is the ratio between the final total particle and
smoothed metal mass in the simulation that used smoothed abundances
for the cooling, 
before passing them to the cooling routine. As can be
seen from Figure~\ref{fig-sfr_sm} (blue curve), the result is much
closer to the simulation using smoothed abundances (but without the
reduction factor) than to the simulation using particle
abundances. Hence, the increase in the total stellar mass is mostly
due to the increase in the metal mixing. 

This example demonstrates that the poor performance of SPH with
regards to metal sampling 
and mixing is an important problem that can have a very large effect
on the predicted star formation history. On the other hand, we expect
the differences to be smaller for our higher resolution
simulations. The fact that
non-conservation of metal mass is much less important than metal
mixing and that the mismatch between the total metal masses
becomes very small for our highest resolution simulations, support our
decision to use smoothed abundances during the simulations. 
In the rest of this paper we will therefore only make use of
simulations that employed smoothed abundances for the calculation of
the cooling rates.

\begin{figure}
\includegraphics[width=84mm]{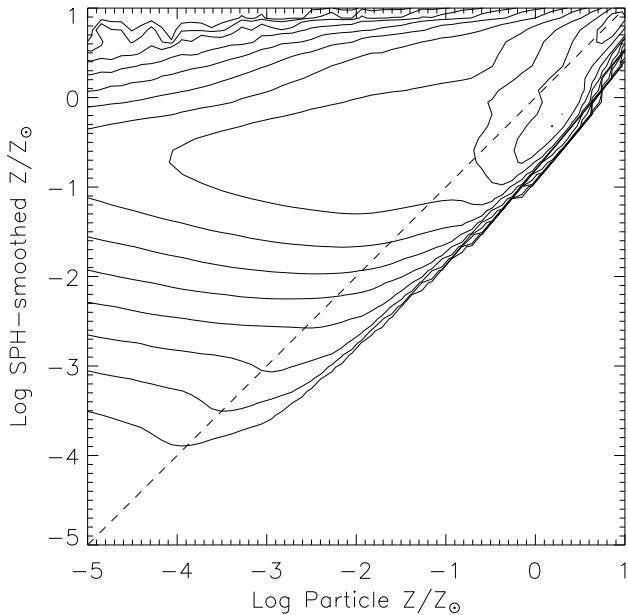}
\caption{Two-dimensional probability density function of particle 
  number density in the smoothed
  metallicity - particle metallicity plane for the L100N512 simulation
  at $z=0$. The contours are spaced logarithmically by 0.5~dex. The two
  metallicity definitions agree in highly enriched regions where the
  metals are well mixed, but at lower metallicities the smoothed
  metallicities are typically higher. The lowest possible smoothed
  metallicity for a particle with non-zero particle metallicity
  corresponds to the case when all its neighbours are metal-free and agrees
  with the lower envelope visible in the plot. \label{fig-smZZ}}
\end{figure}

In figure~\ref{fig-smZZ} we directly compare the $z=0$ smoothed and
particle metallicities in the L100N512 simulation. The contours show
the logarithm of the particle number density in the $Z_{\rm sm} -
Z_{\rm part}$
plane. The dashed line indicates 1-1 correspondence. The
two metallicities are strongly correlated for high
metallicities, although significant scatter exists. This is expected, because high-metallicity gas will
have been enriched by many star particles during many time steps and
we therefore expect the metals to be well mixed. For low particle
metallicities, however, the two metallicity definitions become
essentially uncorrelated and the smoothed metallicities are typically
higher than the particle metallicities. This is because a single
enriched particle can give its neighbours a range of
smoothed abundances depending on their distances. 

For a given particle metallicity, there is a well defined lower limit 
to the smoothed metallicity. This minimum smoothed metallicity corresponds to that of
a metal-bearing particle that is surrounded by metal-free
particles. It is visible as the lower envelope of the contours
parallel to the dashed line in 
figure~\ref{fig-smZZ}. From equations (\ref{eq:Zsm}) and (\ref{eq-smrho}) we can
see that the minimum 
smoothed metallicity is $Z_{{\rm sm},i}= m_{{\rm
    Z},i}W(0,h_i)/\rho_i$. Using the fact that for \textsc{gadget} the value
of the kernel at zero lag is $W(0,h_i) = 8/(\pi h_i^3)$ and that for
particles with $m_i\approx m$ the kernel satisfies $4\pi\rho_i h_i^3/3
\approx N_{\rm ngb} m$ (where the number of SPH neighbours $N_{\rm
  ngb}= 48$ in our simulations), it is easy to show that $Z_{{\rm sm},i}
\approx 2 Z_{{\rm part}, i}/9$. 

In summary, we use smoothed abundances during the simulation because
it is consistent with the SPH method and because it reduces, though
not eliminates, the metal
sampling problem. Smoothed and particle metallicities are very close in
high-metallicity regions but in low-metallicity regions the smoothed
abundances will typically exceed the particle ones. Using SPH
simulations to study low-metallicity gas and stars is hence
problematic because the results are sensitive to the definition of
metallicity that is used. The use of
smoothed abundances increases the number of particles for which metal
cooling is important and thus the star formation rate, particularly
when the resolution is low. 

\begin{figure*}
\includegraphics[width=168mm]{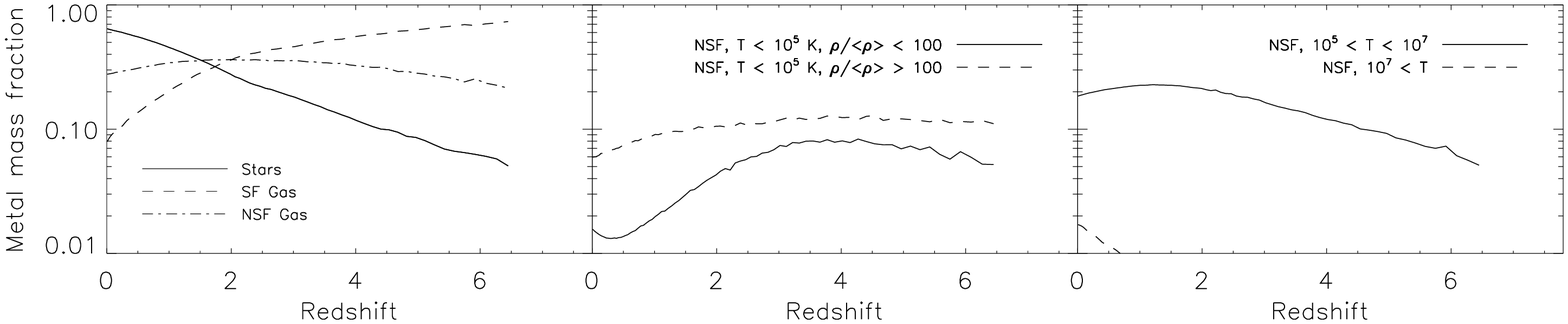}
\caption{Evolution of fractional metal mass in various components for
  the L100N512 simulation. \textit{Left:} Stars (solid), star-forming
  gas (dashed) and  
  non-star-forming gas (dot-dashed). Note that these curves add up
  to unity. \textit{Centre:} Cold-warm IGM (non-star-forming, $\rho
  < 100 \langle \rho 
  \rangle$, $T < 10^5\,\K$) (solid), and cold-warm halo gas
  (non-star-forming, $\rho > 100 \langle \rho \rangle$, $T < 10^5\,\K$)
  (dashed). \textit{Right:} WHIM
  (non-star-forming, $T > 10^5\,\K$) (solid) and ICM
  (non-star-forming, $T > 10^7\,\K$) (dashed). The parts of the curves
  corresponding to redshifts for 
  which the total metal mass is smaller than $10^{-6}$ of the total
  baryonic mass have been omitted because they become very
  noisy. While most metals initially reside in star-forming gas, by
  $z=0$ most are locked up in stars. At present most of the
  gaseous metals reside in the WHIM.\label{fig-metevol}}
\end{figure*}

\section{The predicted distribution of metals}
\label{sec:results}

In this section we will investigate the cosmic metal
distribution predicted by our simulations. We will only present results from the
simulations listed in Table~\ref{tbl:sims}, which are drawn from the
reference simulations of the OWLS set. While the simulations analysed
here differ in terms of their box sizes and particle numbers, they all
used the same simulation code and the same prescription for star
formation, galactic winds and cooling. In a future publication we will
use the full OWLS set to study the effect of changes in the
baryonic processes on the metal distribution, which turns out to be very
important. Here we will
instead focus on convergence tests and on the 
effect of different metallicity definitions.

\begin{figure*}
\includegraphics[width=168mm]{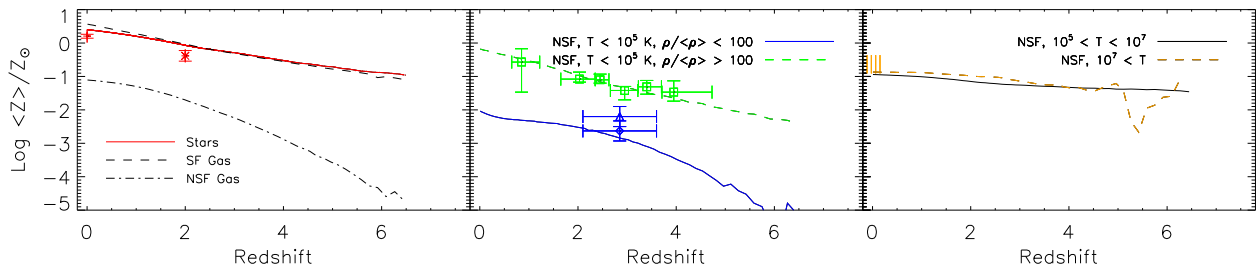}
\caption{Evolution of the mass-weighted metallicities of various
  components for the 
  L100N512 simulation. The panels and line styles are identical to
  those shown in 
  Fig.~\protect\ref{fig-metevol}. The metallicity of stars and the
  ISM evolves much more weakly than that of cold-warm diffuse gas. The
  mean metallicity of the WHIM is $\sim 10^{-1}\,Z_\odot$ at all
  times. Symbols indicate observational estimates of metallicities 
  in various phases - stars: \protect\citep{Gallazzi2008} (plus sign) and 
  \protect\citep{Halliday2008} (cross symbol); cold halo gas:
  \protect\citep{Prochaska2003} (squares);
  diffuse IGM: \protect\citep{Aguirre2008} (triangle) and
  \protect\citep{Seta03} (diamond); 
  ICM: \protect\citep{Simionescu2009} (orange region). \label{fig-Zevol}} 
\end{figure*}

\begin{figure*}
\includegraphics[width=58mm]{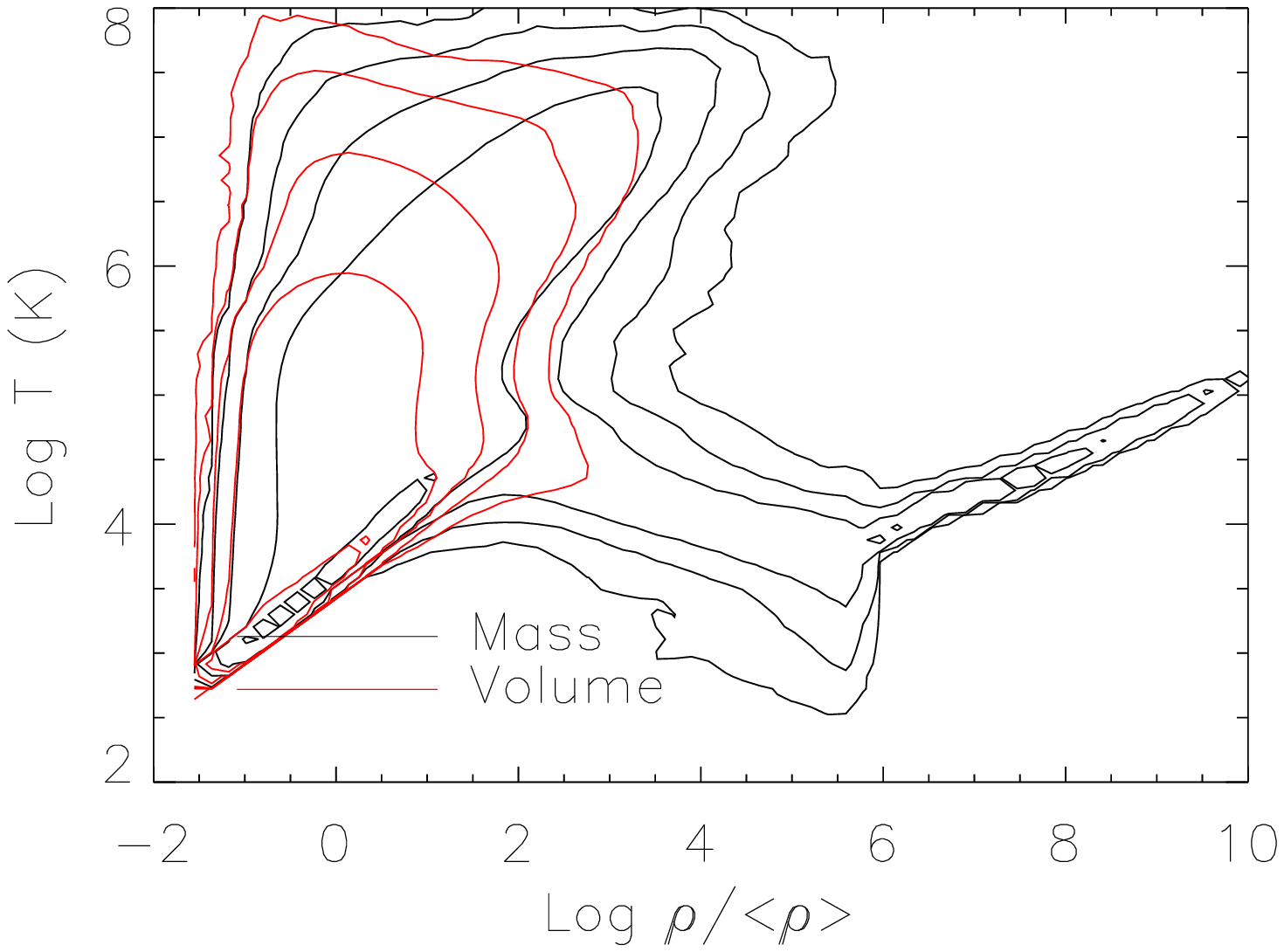}
\includegraphics[width=58mm]{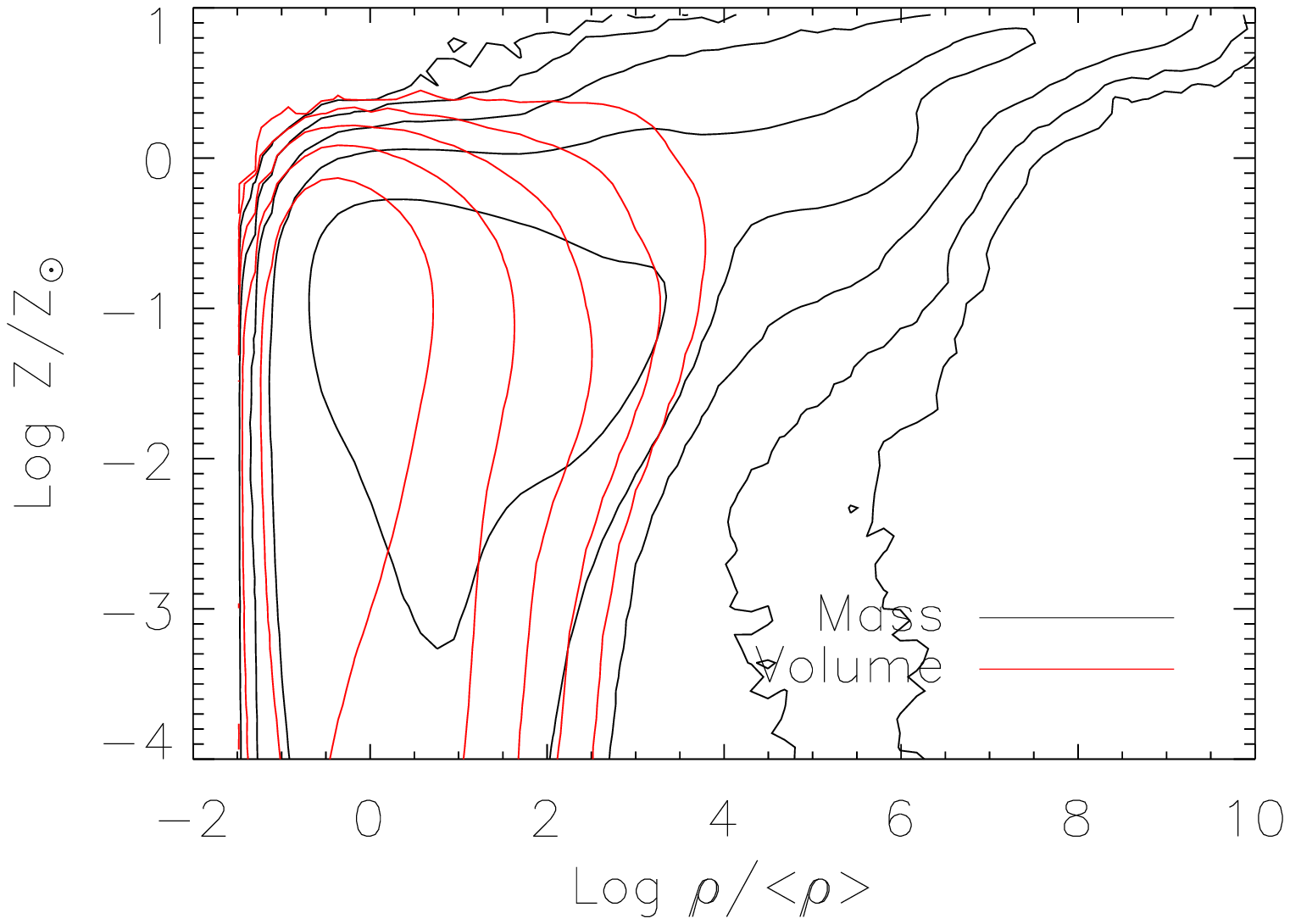}
\includegraphics[width=58mm]{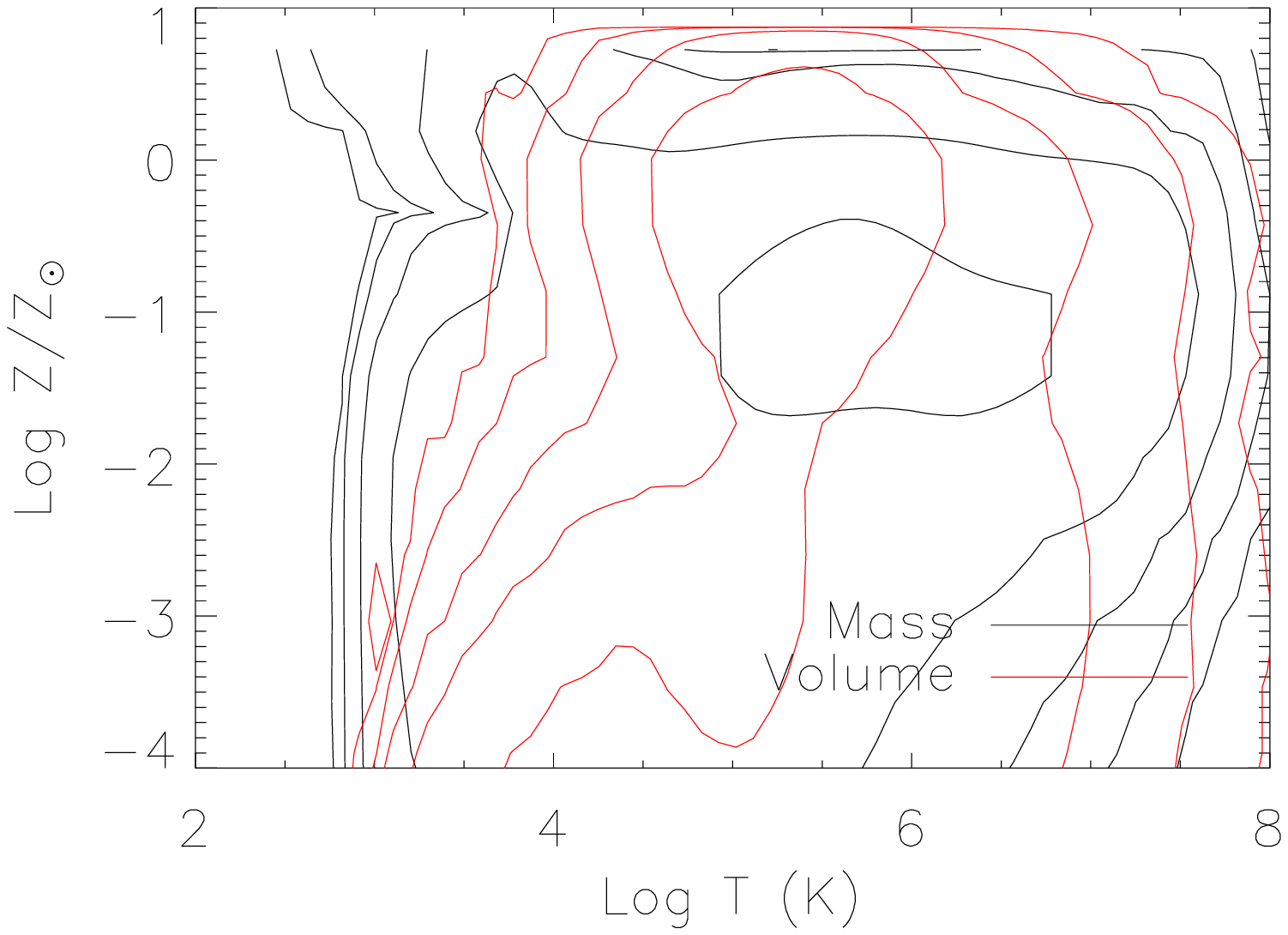}
\caption{Gas distribution weighted by mass (\textit{black}) and volume
  (\textit{red}) in the temperature - density
  (\textit{left}), metallicity - density (\textit{centre}),
  and metallicity - temperature (\textit{right}) planes at $z = 0$
  for the L100N512 simulation. Note that star-forming gas has been
  excluded from the right-hand panel. The contours are
  spaced by 1 dex. While low-density gas displays a wide range of
  metallicities, there is a strong positive gradient of metallicity
  with density that becomes narrower for higher densities. Combined
  with the fact that 
  low-density gas must account for a larger fraction of the volume
  than the mass, the metallicity-density correlation implies that
  volume-weighting favors lower metallicities than mass
  weighting. There is no well defined relation between temperature and
  metallicity. 
\label{fig-h3massweight}}
\end{figure*}

There are a number of approaches we can take to study the cosmic
metal distribution in our simulations. Before analysing the
metal distribution at $z=0$ in more detail, we will investigate the
evolution of the metal mass fractions in various components. We will
then proceed to discuss the dependence of the results on the
definition of metallicity (\S\ref{sec:smoothedvsparticle}), the size of
the simulation box (appendix \ref{sec:boxsize}) and the
numerical resolution (appendix \ref{sec:resolution}). While appendices
\ref{sec:boxsize} and \ref{sec:resolution} will make use
of the full suite of simulations listed in Table~\ref{tbl:sims},
we will only consider the L100N512 (which contains $2
\times 512^3$ particles in a box that is $100~\mpch$ on a side)
 run in the rest of this section. We note, however, that
our much higher resolution L025N512 simulation (which was,
however, stopped at $z=1$) gives qualitatively similar results.

\begin{figure*}
\includegraphics[width=58mm]{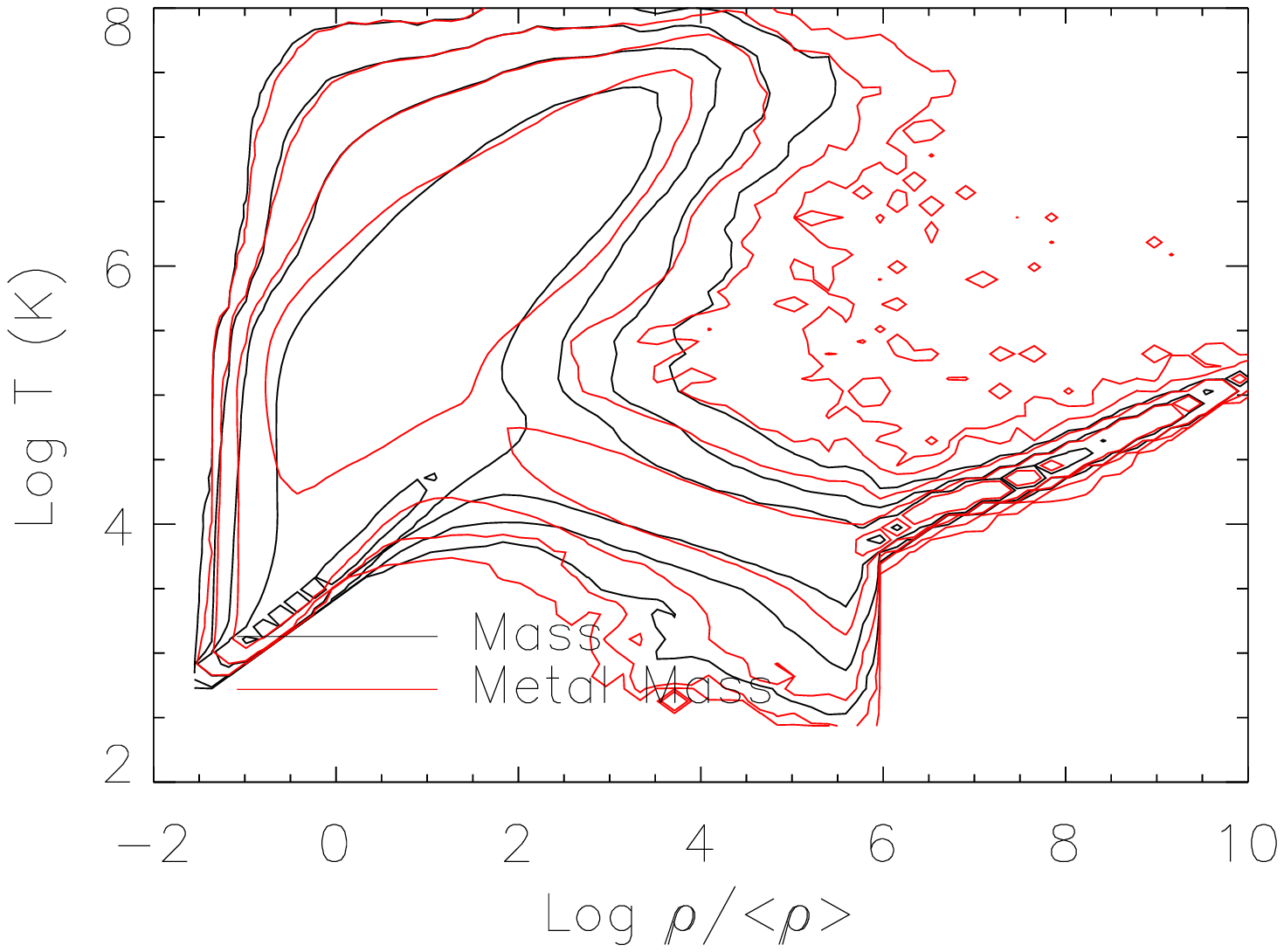}
\includegraphics[width=58mm]{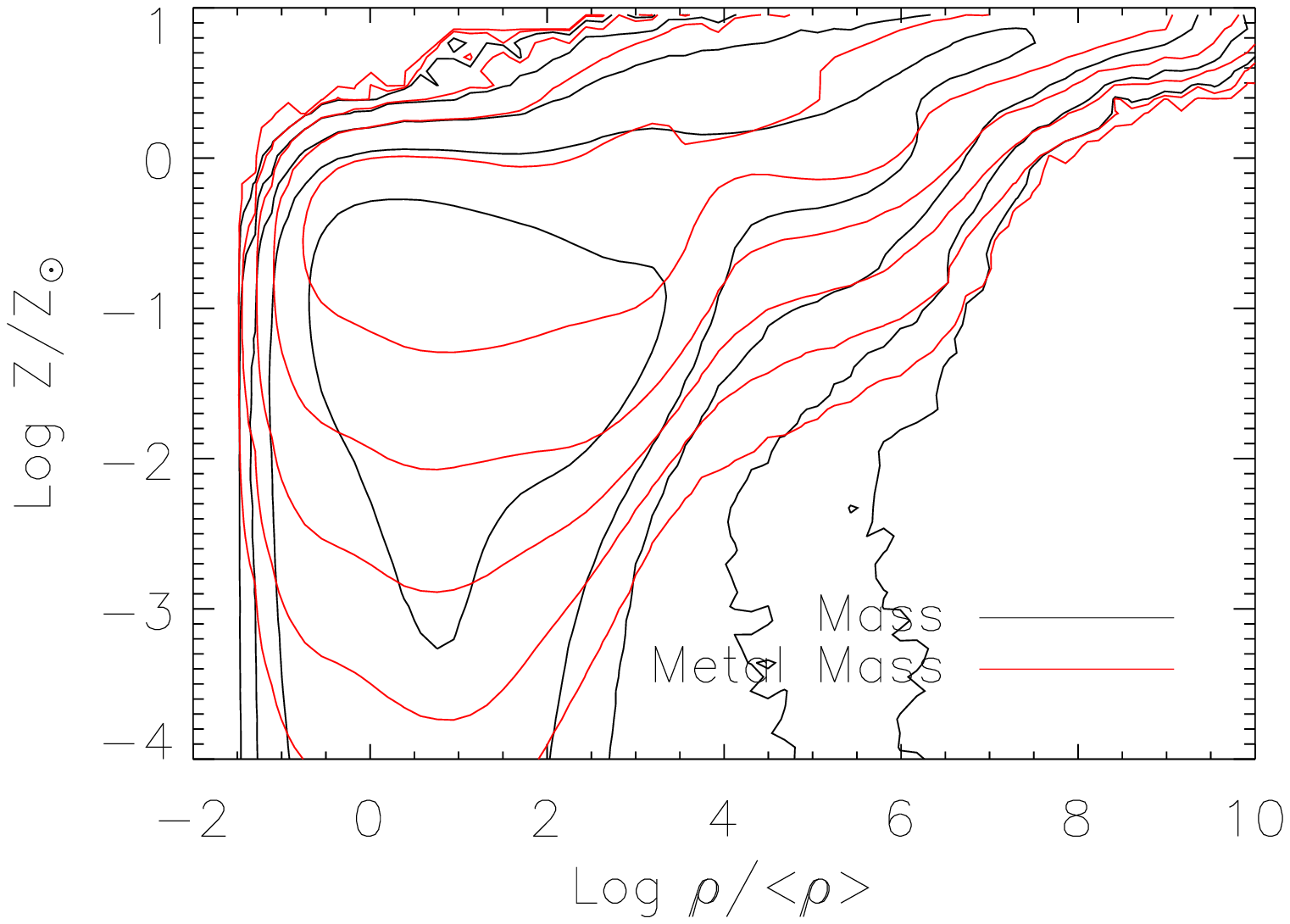}
\includegraphics[width=58mm]{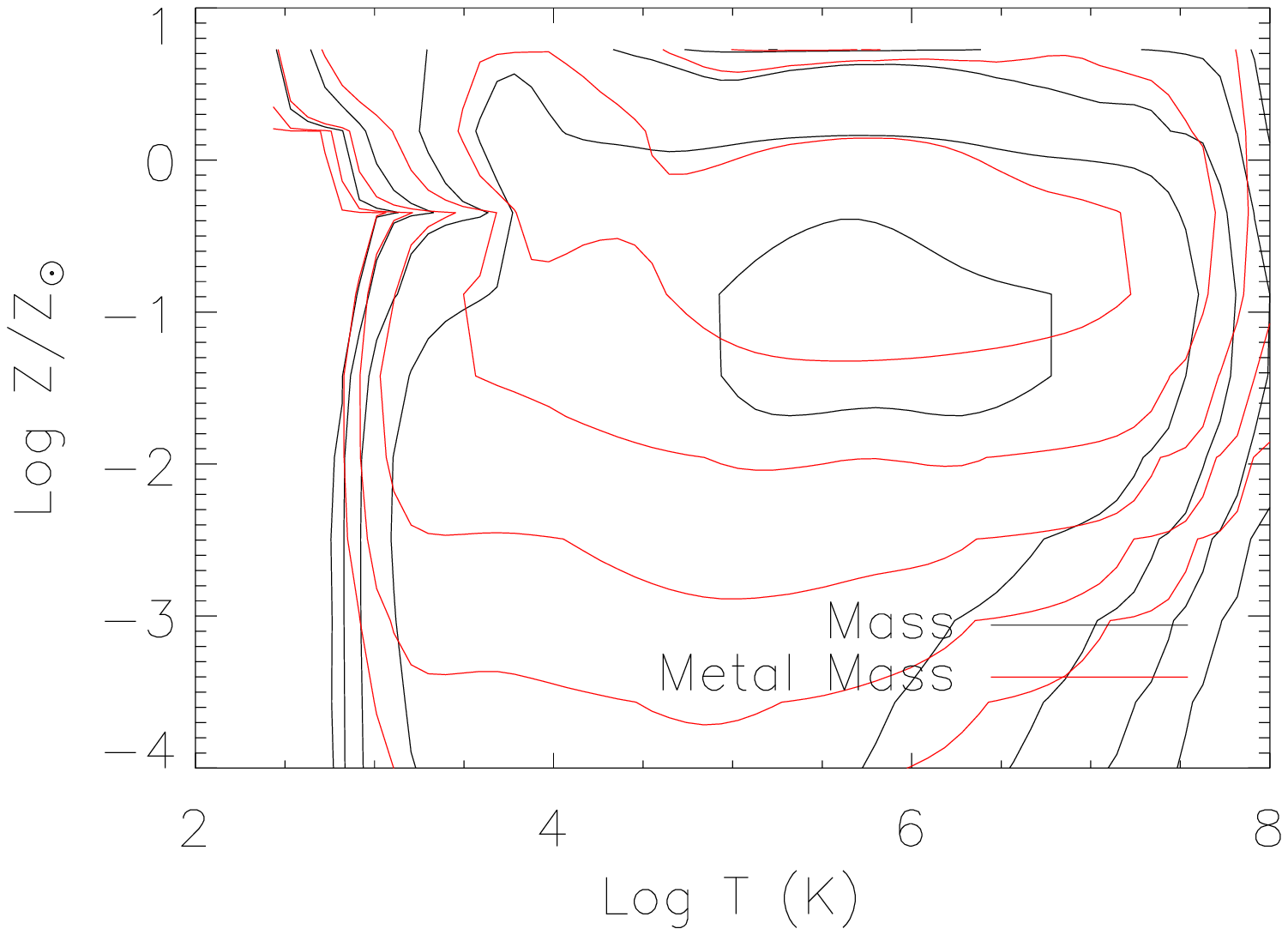}
\caption{Gas mass (\textit{black}) and gas metal mass
  (\textit{red}) distribution in the temperature - density
  (\textit{left}), metallicity - density (\textit{centre}),
  and metallicity - temperature (\textit{right}) planes at $z
  = 0$ for the L100N512 simulation. Note that star-forming gas has been
  excluded from the right-hand panel. The contours are
  spaced by 1 dex. While gaseous metals track the general gas mass
  distribution at high densities (or metallicities), at the densities
  corresponding to diffuse structures ($\rho\la 10^2 \left <\rho\right
  >$) metals reside in gas that is both hotter and more metal-rich
  than is typical for these densities.
\label{fig-metcont}}
\end{figure*}

Figure~\ref{fig-metevol} shows the evolution of the fraction of the metal mass
contained in various components. By isolating the gaseous metals into
various phases, we can develop a picture of where the metals are at a given
redshift. To that end we have plotted the total fraction of
the metal mass in stars (left, solid), star-forming gas (left,
dashed), and non-star-forming gas (left, dot-dashed). Recall (see
\S\ref{sec:sim}) that by
star-forming we mean gas particles that are allowed to form stars and
for which we impose an equation of state. These particles all have
densities $n_{\rm H} > 10^{-1}\,\cm^{-3}$ and can be thought of as
ISM gas. We have divided the non-star-forming gas into
cold-warm ($T < 10^5\,\K $; middle) and warm-hot ($T > 10^5\,\K $;
right) components. The cold-warm gas has been further subdivided into diffuse  
($\rho < 10^2 \left < \rho \right >$; middle, solid) and halo ($\rho >
10^2 \left < \rho \right >$; middle, dashed) gas and we have
subdivided the warm-hot, non-star-forming gas into the warm-hot IGM
(WHIM) ($10^5\,\K < T < 10^7\,\K $; right, solid), and the intracluster
medium (ICM) ($T > 10^7\,\K $; right, dashed). 

Focusing first on the left-hand panel of Fig.~\ref{fig-metevol}, we see that the
most prominent trend is the strong decrease in the fraction of the 
metals in star-forming gas. While this phase harbors most of the
metals at $z=8$, by $z=0$ it contains only a small fraction. The
opposite is true for the stars: while  
most of the
metals are initially in the gas phase, today the majority of the
metals are locked up in 
stars. The metal fraction contained in non-star-forming gas varies
between twenty and forty per cent. 

\cite{DO07} have recently used a cosmological SPH
simulation (using $256^3$ gas particles in a $32~\mpch$ box) to
investigate the cosmic distribution of metals. While we postpone a
detailed comparison to a future paper, we note a clear, qualitative
difference in the results. While \cite{DO07} find that initially (at
$z=6$) the diffuse IGM contains much more metal than the ISM, 
we find the opposite. We will show elsewhere that this difference
mostly reflects the higher initial mass loading of galactic winds
assumed by \cite{DO07}. This difference illustrates the importance of
varying the subgrid recipes used in the simulations. We will report on
such variations elsewhere.

The other two panels of Fig.~\ref{fig-metevol} show that while the
metal fraction in cold-warm 
gas decreases with time, the fraction contained in warm-hot and hot
gas increases. This is not unexpected, as growth of structure will
result in more and stronger accretion shocks, heating an increasing
fraction of the gas to high temperatures. While most of the metals in
non-star-forming gas are initially cold, more than half of them reside
in the WHIM for $z< 2$. The ICM, on the other hand, still contains only a
few percent of the metal mass by $z=0$, an order of magnitude less
than the WHIM.

Besides the metal mass fractions, the metallicities of the various
components are also of interest. Together with the evolution of the
baryonic mass fractions, which we will present elsewhere, the
metallicities determine the metal mass fractions. Figure~\ref{fig-Zevol} 
shows the evolution of 
the mass-weighted mean metallicities of the same components that were
shown in Fig.~\ref{fig-metevol}. 

As expected, the metallicity of the star-forming gas
roughly traces the metallicity of the stars throughout the simulation,
although the ratio of the mean stellar metallicity to the mean ISM
metallicity decreases slightly with time. The stellar and ISM
metallicities increase relatively slowly with time. At $z=3$ the mean
stellar metallicity is about 0.5~dex lower than at $z=0$. Interestingly, the
WHIM and the ICM have a metallicity $\sim 10^{-1}\,Z_\odot$ at all
times. This lack of evolution is in stark contrast to the behavior of
the cold-warm, diffuse gas whose metallicity increases rapidly with
time. Consequently, the 
differences in the metallicities of the various components become
smaller at lower redshifts. However, even by $z=0$ the metallicity of the
diffuse, cold-warm IGM is only of order 1 per cent of solar.

Although our pimary aim is not to perform a direct comparison to
observations, we have included in Fig.~\ref{fig-Zevol} a number of
observational estimates of metallicity\footnote{The observations were
  converted to our solar abundances where necessary.} in various
phases. In the left 
panel we show the estimates for the mean stellar metallicity at $z = 0$ and $z
= 2$ from \cite{Gallazzi2008} and \cite{Halliday2008}, respectively. Our
simulations fall about 0.2 dex above the redshift zero point and land
about 0.3 dex above the higher redshift point. We note however, that
\cite{Halliday2008} find their result to be low compared to that of
\cite{Eeta06}. They explained that this could represent an alpha
enhancement at high redshift (\citealt{Eeta06} measured oxygen while
\citealt{Halliday2008} measured iron). We compare our predictions for
overdense, cold, 
non-starforming gas to the observations of 
damped Ly$\alpha$ (DLA) systems of \cite{Prochaska2003} and find
excellent agreement. Our measurement of the mean metallicity of
the $z\approx 3$ IGM agrees with the measurement based on carbon of
\cite{Seta03}, but falls slightly below that based on oxygen of
\cite{Aguirre2008}. In appendix \ref{sec:resolution} we show that the
metallicity of this
phase is particularly sensitive to resolution and that higher
resolution simulations may predict higher diffuse IGM
metallicities. Finally, the metallicity of the low-redshift ICM falls along the
lower range of cluster measurements compiled by
\cite{Simionescu2009}. For this we compiled a range of all metallicity
measurements in the most outer regions of the clusters (regions which
should dominate the mass averaged metallicity). We note that since
most radial profiles flatten at outer radii, this should be a safe
assumption. From these comparisons we conclude that our simulation is
in reasonable agreement with a range of observations.

Before investigating the robustness of these results to numerical
effects, we will study the $z=0$ metal distribution in more
detail. To interpret the results, it is, however, helpful to
first consider the gas mass rather than the metal mass distribution.

The solid contours in the left-hand panel of
figure~\ref{fig-h3massweight} indicate the $z = 0$ gas mass-weighted
2-dimensional (2-D) probability density function (PDF) in the
temperature-density plane. That is, the contour values correspond to
$dM_{\rm g}/(M_{\rm g,tot} d\log{\rho \over \left <\rho\right >} d\log
T)$. Similarly, 
the red contours show the volume-weighted\footnote{Volume weighting
  was implemented as weighting by $h_i^3/\sum_j h_j^3$, where $h_i$ is the
  SPH smoothing length of particle $i$. We verified that using $m/\rho$
  instead of $h^3$ gives
  nearly identical results.} 2-D PDF, i.e., $dV/(V_{\rm
  tot}d\log{\rho \over \left <\rho\right >} d\log T)$. All contour
plots in this section show contours spaced by one dex, with each plot 
containing the same levels for both linestyles. 

There are a number of distinct features in this
temperature-density diagram. Much of the mass, and nearly all of the
volume, resides in a narrow strip ranging from about $(\log\rho/\left
<\rho\right >, \log T)\sim (-1.5,3)$ to $\sim (1.5,4.5)$,
which corresponds to the diffuse, photo-ionized IGM. The relation
between temperature and density of this component is set by the
balance between photo-heating and adiabatic cooling
\citep{Hui1997}. For gas at higher overdensities photo-heating is
balanced by radiative rather than adiabatic cooling and the
temperature slowly decreases with increasing density. The plume of
much hotter gas corresponds to gas that has been shock-heated, either
through gravitationally induced accretion shocks or galactic
winds. For gas with densities $\rho/\left <\rho\right> \ga 10^2$ there
is less mass at temperatures $T\sim 10^5\,\K$ than there is at lower
and higher temperatures because the radiative cooling rates peak at these
temperatures (due mostly to collisional excitation of oxygen, carbon,
and helium), resulting in a bimodal distribution of the temperature in
that density regime. The well-defined $T$-$\rho$ relation at the highest
densities reflects the equation of state $P\propto \rho_g^{4/3}$, which
we impose on gas with densities exceeding our star formation threshold
of $n_{\rm H}=0.1~\cm^{-3}$ ($\rho/\left <\rho\right>\sim 10^6$ at $z=0$). As
discussed in \S\ref{sec:sim}, the temperature of this gas merely
reflects the imposed effective pressure of the unresolved multiphase ISM. 

\begin{figure*}
\includegraphics[width=168mm]{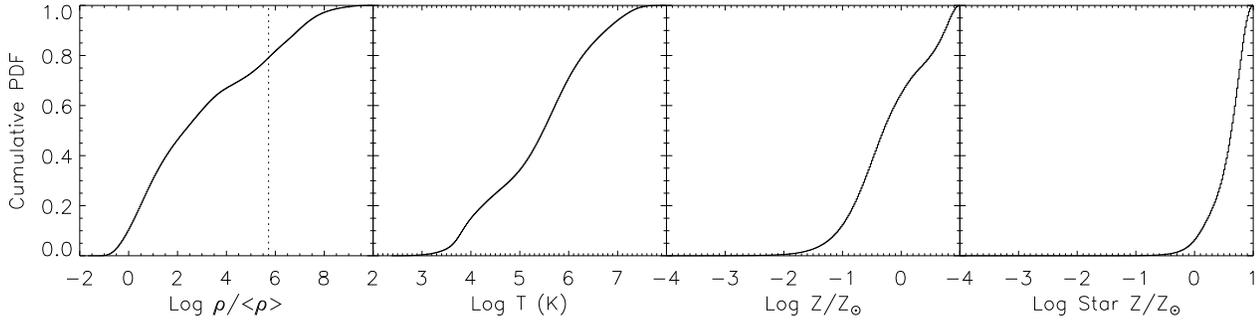}
\caption{Cumulative probability density function of the gas metal mass
  as a function of gas density (\emph{left}), temperature
  (\emph{centre-left}), and metallicity (\emph{centre-right}) at $z = 0$ for the 
  L100N512 simulation. The vertical, dotted line in the left panel
  indicates our star formation  
  threshold. Also note that star-forming gas has been excluded from
  the temperature plot. The right-hand panel shows the
  cumulative PDF of the stellar metal mass as a function of the stellar
  metallicity. While the gas metal mass is distributed over a wide range
  of densities and temperatures, it is concentrated in gas that has a
  relatively high (local) metallicity.
\label{fig-cumulative}}
\end{figure*}

The middle and right panels of Fig.~\ref{fig-h3massweight} show how the
gas mass (black) and the volume (red) are distributed in the
metallicity-density and the metallicity-temperature planes,
respectively. We have excluded star-forming gas from the right-hand
panel since its temperature would merely reflect the pressure of the
equation of state that we have imposed. We will exclude this component
from all plots investigating temperature dependencies, except for
temperature-density diagrams such as the left-hand panel of
Fig.~\ref{fig-h3massweight} (because in that case the gas density can
be used to identify the star-forming gas). 

While there is a clear positive correlation between metallicity and
density for highly overdense gas (see the black contours in the middle panel),
gas at low overdensities can have a large range of metallicities
extending all the way to zero (outside the plotted range). Note that the highest
density gas is highly supersolar. This reflects the fact that massive
galaxies in this simulation have too high metallicities because
galactic winds cannot escape their high-pressure ISM. In a future paper
we will show that this problem is not present for some other
implementations of galactic winds driven by SNe and,
particularly, for more efficient feedback mechanisms such as energy
injected by active galactic nuclei. The right-hand panel shows that
there is no well-defined  
relation between metallicity and temperature. Gas at a given
temperature displays a wide range of metallicities.

Figure~\ref{fig-metcont} uses the same panels as
Fig.~\ref{fig-h3massweight} to compare the distribution of gas mass (black)
and metal mass (red). Since we use smoothed metallicities, metal mass
is defined as $Z_{\rm sm}M_{\rm g}$. Note that the black contours are
identical to those shown in Fig.~\ref{fig-h3massweight}. Focusing first on the
left-hand panel, which shows the 
distributions in the temperature-density plane, we see that 
while metal and gas mass track each other well for high overdensities
(although the gas mass is distributed over a narrower range of
temperatures for a given density),
they differ in one important respect for $\rho/\left <\rho\right >\la
10^2$. Whereas most of the mass is concentrated at the lowest densities and
temperatures, this diffuse, photo-ionized IGM contains only a small
fraction of the metals. Instead, the vast majority of metals in
low-density gas are hot ($T\ga 10^5~\K$). This is probably caused by
the fact that 
high-velocity winds are required to transport metals to regions far from
galaxies and that such winds shock-heat the medium. This agrees with
\cite{Theuns2002a} and \cite{Aguirre05}, who also found that most of
the diffuse metals reside in hot gas,   
using different implementations of feedback from star formation
and using simulations that did \emph{not} include metal-line
cooling. \cite{Aguirre05} speculated that the inclusion of radiative
cooling by metals might be important, but our simulations show that
even when metal-line cooling is included, 
the WHIM ($5< \log T < 7$, $\rho/\left <\rho\right >\la
10^3$) is an important reservoir of cosmic metals. From the right-hand
panel of Fig.~\ref{fig-h3massweight} we can see that although some
WHIM gas has very low metallicities, much of it has $Z\ga
10^{-1}~Z_\odot$. 

The left-hand panel of Fig.~\ref{fig-cumulative} shows the cumulative
PDF for the gas metal mass as a 
function of $\log \rho/\left <\rho\right >$, i.e., ${1 \over M_{\rm
    Z_{\rm sm},tot}}
  \int_0^\rho d\rho'{ d(M_{\rm g}Z_{\rm sm}) \over d\rho'}$, where $M_{\rm
    Z_{{\rm sm},tot}}$ is the total smoothed metal mass in gas.
Similarly,
  the second and third panels from the left show the cumulative
PDF of the gas metal mass as a function of temperature and
metallicity and the right-hand panel shows the cumulative PDF of the
stellar metal mass as a function of metallicity. While the gas metal mass
is spread over remarkably large ranges of densities and temperatures, it is
concentrated in relatively high-metallicity gas. Note, however, that
  metallicity is evaluated locally, at the resolution limit of the
  simulation. High metallicity gas therefore
  includes patches of enriched gas embedded in structures whose overall
  metallicity is low. 

Considering one 
variable at a time, ninety percent of the $z=0$ 
gas metal mass resides in gas with $10^{-0.5} \la \rho/\left <\rho\right >
\la 10^{7.5}$, $10^4\,\K \la T \la 10^{6.5}\,\K$, or $10^{-1.5} \la Z_{\rm sm}/Z_\odot \la
10^{0.7}$. Half the gaseous metal mass resides in gas with $\rho < 10^2 \left
<\rho\right >$ which implies that diffuse and collapsed
structures contain similar amounts of metal mass. In terms of
temperature the midpoint lies at about $10^5~\K$, which means that
half the gas metal mass resides in shock-heated gas. Clearly, any
census aiming to account for most of the metal mass in gas will have
to take a wide variety of objects and structures into account.

In the next section we will investigate the metal 
distribution and its evolution in more detail, including the
sensitivity of the results to the definition of metallicity. The
convergence of the predictions with respect to the size of the 
simulation box and the numerical resolution is studied in appendices
\ref{sec:boxsize} and \ref{sec:resolution}, respectively. 

\begin{figure*}
\includegraphics[width=58mm]{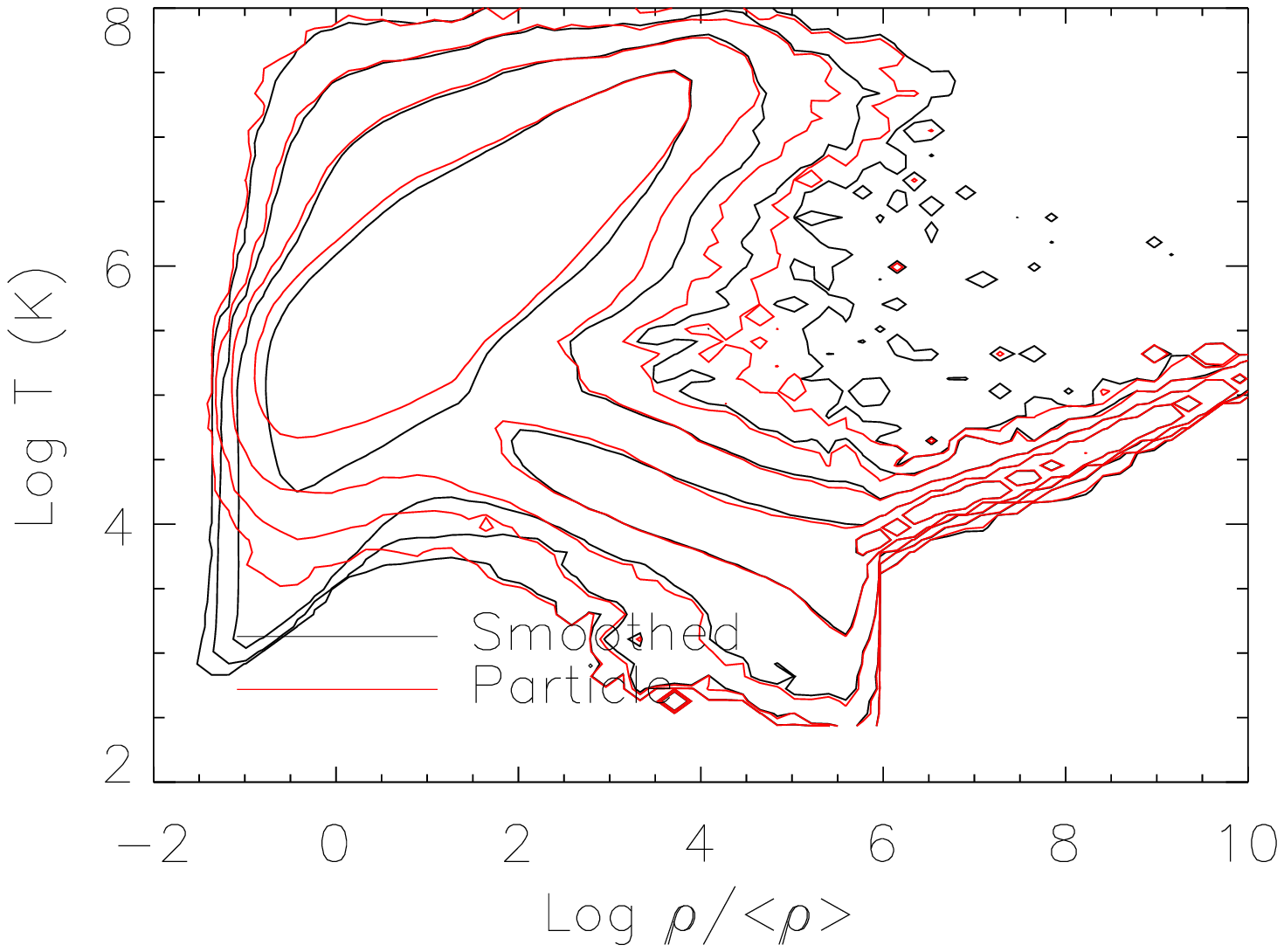}
\includegraphics[width=58mm]{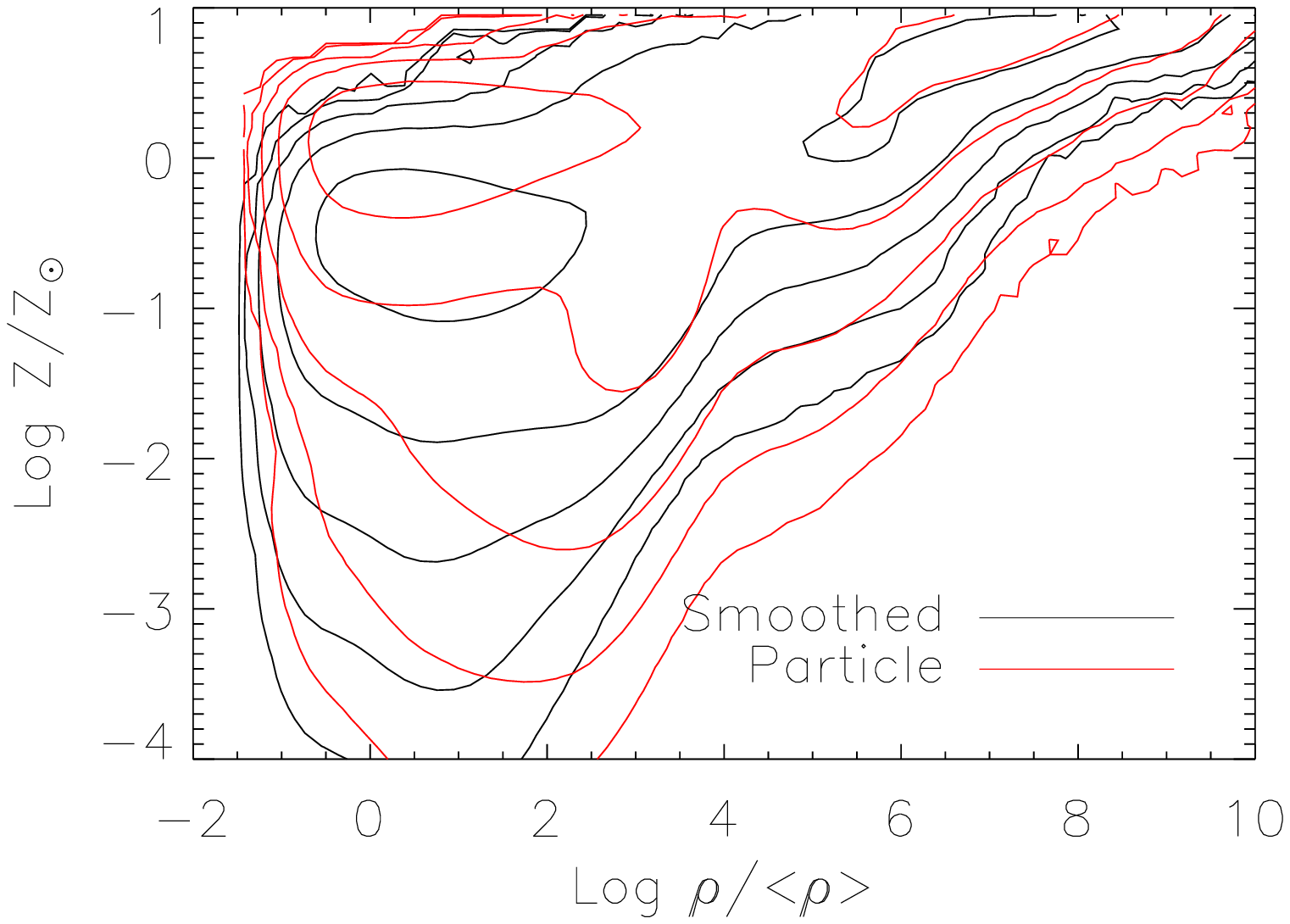}
\includegraphics[width=58mm]{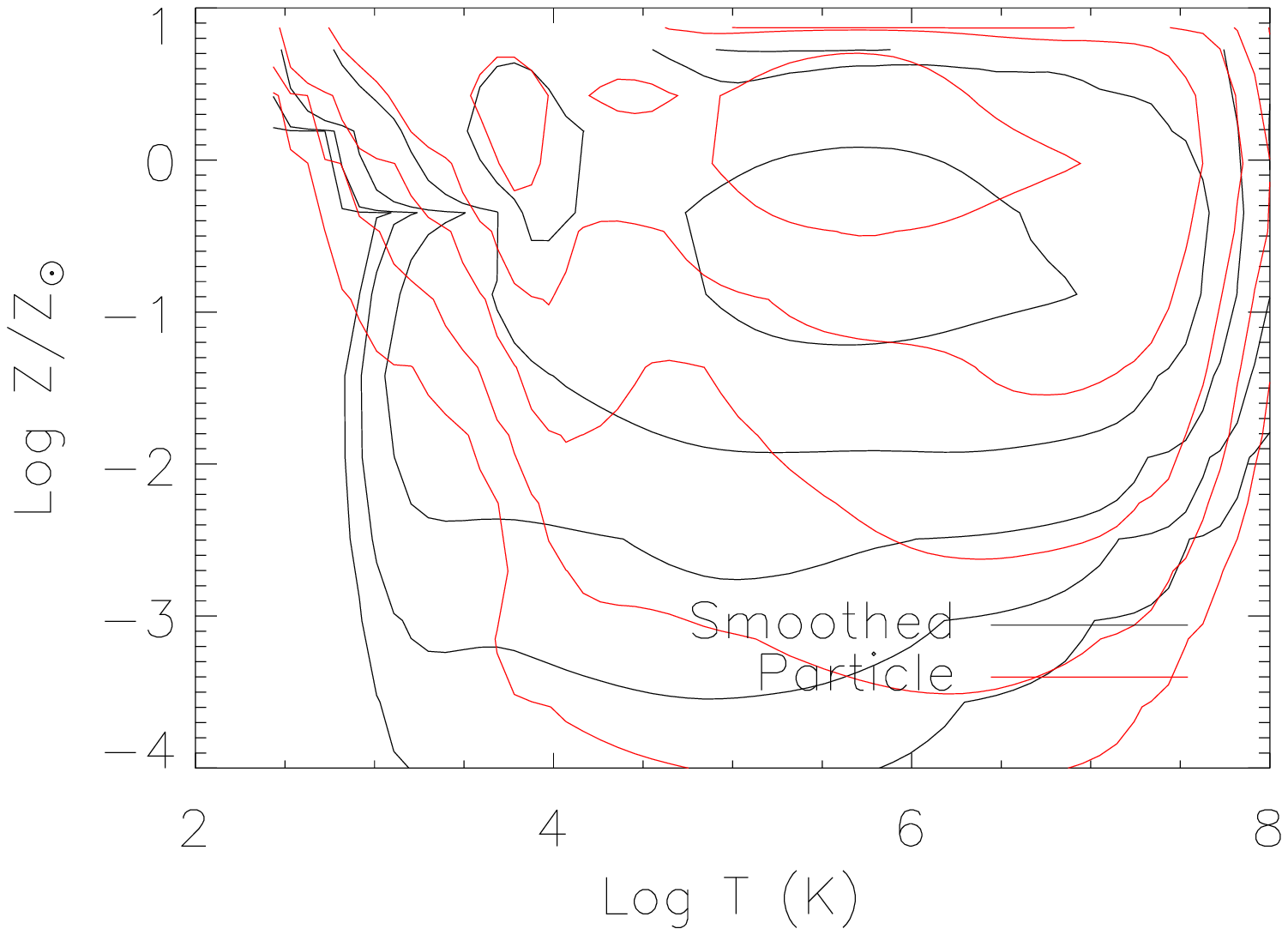}
\caption{Gas smoothed metal mass (\textit{black}) and gas particle metal mass
  (\textit{red}) distribution in the temperature - density
  (\textit{left}), metallicity - density (\textit{centre}),
  and metallicity - temperature (\textit{right}) planes at $z
  = 0$ for the L100N512 simulation. Note that star-forming gas has been
  excluded from the right-hand panel. The contours are
  spaced by 1 dex. Except for low-metallicity gas, which mostly has
  low densities and temperatures, smoothed and particle
  metallicities yield similar results.
\label{fig-metZsmZcont}}
\end{figure*}

\begin{figure*}
\includegraphics[width=168mm]{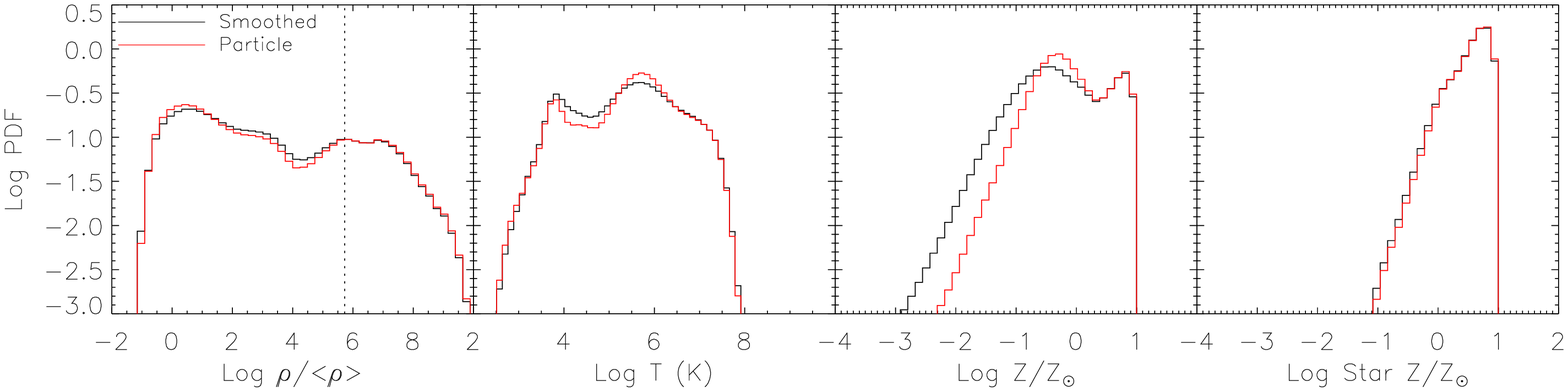}
\caption{Probability density function of the gas density
  (\textit{left}), temperature 
  (\textit{centre-left}) and smoothed/particle metallicity
  (\textit{centre-right}) weighted by smoothed
  (\textit{black}) and particle (\textit{red}) metal mass. The black
  (red) histogram in the right-hand
  panel shows the PDF of the smoothed (particle) stellar metallicity
  weighted by smoothed (particle) metal mass. All panels correspond to
  the L100N512 simulation at $z = 0$. The vertical, dotted line in the
  left panel indicates the threshold for star formation. Note that
  star-forming gas was 
  excluded from the temperature panel. The predictions for
  smoothed and particle metallicities only differ significantly for
  low-metallicity gas.
 \label{fig-ZsmZmet1D}}
\end{figure*}

\subsection{Smoothed vs.\ particle metallicity}
\label{sec:smoothedvsparticle}
As discussed in \S\ref{sec:zsm}, there is some ambiguity in the
definition of the metallicity in SPH. We have opted to use SPH
smoothed metallicities $Z_{\rm sm} \equiv \rho_Z/\rho$ rather than
particle metallicities $Z_{\rm part} \equiv m_Z/m$, because using
smoothed abundances 
is consistent with the SPH method and because it partially counters
the lack of metal mixing that results from the fact that metals are stuck to
particles. We already demonstrated in \S\ref{sec:zsm} that the definition
of metallicity is particularly important for low metallicity gas and
that it has important consequences for gas cooling rates (and hence
for the predicted star formation histories). Here we investigate its
effect on the metal distribution by comparing the gas metal mass distributions
computed using smoothed and particle metallicities. Note, however,
that smoothed metallicities were always used during the simulation for
the calculation of the radiative cooling rates, stellar lifetimes, and
stellar yields.

Figure~\ref{fig-metZsmZcont} compares the gas metal mass distribution
using smoothed (black) and particle (red) metallicities in the
temperature-density (left), metallicity-density (middle) and
metallicity-temperature (right) planes. Note that the black contours
are identical to the red contours in
Fig.~\ref{fig-metcont}. The smoothed and particle metal distributions trace
each other reasonably well, except for the diffuse, photo-ionized IGM
($\rho/\left <\rho\right > < 10^2$, $T < 10^5\,\K$; left panel) and
generally for low-metallicity gas ($Z \ll Z_\odot$; middle and right
panels). When smoothed metallicities are used larger fractions of the
metal mass reside in low-metallicity gas, which is consistent with
Fig.~\ref{fig-smZZ}.

The left-hand panel of figure~\ref{fig-ZsmZmet1D} shows the 1-D
gaseous metal mass weighted PDF for the gas density, i.e.,   
$d(M_{\rm g}Z)/(M_{Z,{\rm tot}}d\log{\rho \over \left <\rho\right
  >})$ as a function of $\log
\rho/\left <\rho\right >$, where $Z$ is the smoothed/particle
metallicity and $M_{Z,{\rm 
    tot}}$ is the total smoothed/particle metal 
mass in gas.  Thus, if the
$y$-axis were linear (which it is not), the
fraction of the area underneath the histogram that is in a given bin
would correspond to the fraction of the gas metal mass in that bin. 
Similarly, the second and third panels from the left show the
metal-mass weighted PDF of the temperature and
metallicity, respectively, and the right-hand panel shows the metal
mass-weighted PDF of 
the stellar metallicity. The distribution of metal mass over density
is similar for smoothed and particle metallicities, but using smoothed
metallicities results in a significant shift of metal mass from $T\sim
10^6\,\K$ to $T< 10^5\,\K$. While the differences in the metal
distributions as a function of metallicity are small for stars,
presumably because they tend to form in well-mixed, high density gas,
the differences are very substantial for the gas. As expected, the use
of smoothed metallicities results in a prominent shift of gaseous metal mass
to lower metallicities. For stars, the particle metallicity PDF exceeds
that for smoothed abundances by about an order of magnitude in the
range $-4 < \log Z < -2$ (i.e., outside the plotted range).

We stress, however, that because the
metallicities are evaluated locally, i.e.\ at the spatial resolution limit,
they may differ substantially from the overall mean metallicity of the
structure containing the particles. For structures that contain many
particles the mean metallicity will be insensitive to the definition
of metallicity that is used. On the other hand, local
gas metallicities are the ones that are used for the calculation of the
cooling rates, so they are certainly important.

In summary, SPH simulations such as the ones presented here are
not well-suited to study the metal distribution in low-metallicity
gas. Because metals are stuck to particles, SPH suffers from a
sampling problem which becomes worse in regions where only a small
fraction of the particles have been enriched. Consequently, in the
low-metallicity regime the distribution of mass is sensitive to the
definition of metallicity. 
However, the bulk of the
metals reside in gas of relatively high metallicity for which the
predicted metal distribution is not strongly affected by the lack of
metal mixing. Except for the
small fraction of stars that have $Z\la 10^{-1}\,Z_\odot$, the results
are more robust for stars because they
form in high-density gas that is better mixed. This is probably
because high-density gas particles tend to be near star particles
which means that they are likely to have received metals
during many time steps.

\section{Summary}
\label{sec:conclusions}

The elemental abundances of stars, galaxies, and diffuse gas are of
great interest because they determine radiative cooling rates and the
stellar initial mass function, because they can give insight into
physical processes such as the interactions between galaxies and the
IGM, because they change the strength of lines that are used as
diagnostics of physical conditions and because they determine the
observability of lines used as tracers of various gas phases. In this
paper we have presented a method to follow the timed release of
individual elements by stars, including mass loss through stellar
winds and SNe, and their subsequent dispersal through space. 

The ingredients of our stellar evolution module include a choice of
IMF (we use Chabrier), stellar lifetimes as a function of metallicity
(which we take from \citealt{Peta98}), the rate of
type Ia SNe for an SSP as a function of time (we use an empirical time
delay distribution tuned to obtain agreement between the evolution of
the observed, cosmic rates of SNIa and star formation), and stellar
yields (we use \citealt{M01} for AGB stars, \citealt{Peta98} for core
collapse SNe, and the W7 model of \citealt{Teta03} for SNIa). 

We compared different
sets of nucleosynthetic yields taken from
the literature after integrating them over the IMF. While the 
different studies predict similar total ejected masses, the
predictions for heavy elements differ substantially. In particular, the
ejected masses of 
individual elements differ by factors of a few for AGB stars and -- for
elements heavier than nitrogen -- for core collapse
SNe. Thus, even disregarding the fact that most nucleosynthetic yield 
calculations ignore potentially important effects such as rotation,
the elemental ratios are uncertain by factors of a few for a fixed
IMF. Predictions for abundances relative to iron are even less robust
because much, perhaps even most, of the iron released by an 
SSP is produced by SNIa and the normalization of the SNIa rate is
only known to within factors of a few.

Our implementation of mass transfer implicitly splits, for each
element, the ejected mass 
into terms accounting for the mass that is produced (minus
destroyed) and the mass that is simply passing through. The latter
term is assumed to scale with the initial abundance, allowing us to
correct for relative abundance variations with respect to solar at a
fixed metallicity.\footnote{Note that this treatment still ignores the
  fact that the nucleosynthesis may itself depend on the relative
  abundances.} We have taken special care to correctly track the total
metal mass even if not all elements are tracked explicitly. 

We discussed two possible definitions of metallicity of SPH
particles: the commonly 
used ``particle metallicity'', defined as the ratio of the metal to
the total mass ($m_Z/m$), and the ``smoothed metallicity'', defined as
the ratio of the metal
mass density to the total mass density ($\rho_Z/\rho$), where each
density is computed
using the standard SPH formalism. We argued that smoothed metallicities (and,
analogously, smoothed elemental abundances) are preferable because
they are most consistent with the SPH formalism (particularly with
regards to radiative cooling rates which depend explicitly on
densities) and because they partially counter the
lack of metal mixing that is inherent to SPH. We discussed in some
detail the fact that SPH underestimates metal 
mixing because metals are stuck to particles and how this
results in a sampling problem that may not become smaller when the
resolution is increased. While the use of smoothed abundances eases
this problem  
somewhat, it by no means eliminates it entirely (while also leaving 
the potential for small scale metal mixing due to physical processes
unaddressed). 

The use of smoothed abundances (which are frozen when a gas particle
is converted into a star particle) leads to a slight non-conservation
of metal mass, but we showed this to be negligible for our
high-resolution simulations. A comparison of smoothed and particle
metallicities (in a simulation that used smoothed abundances for the
cooling rates and the stellar evolution) shows that while they are similar
at high metallicities, they differ strongly at low metallicities,
with smoothed abundances typically exceeding particle abundances by
large factors. In
particular, there are many more particles with non-zero smoothed
metallicities than with non-zero particle metallicities. Since there
is no overwhelming reason to prefer one definition of metallicity over
the other, the 
fact that the choice matters indicates that care
must be taken when interpreting the predictions of SPH simulations at
low metallicities.

A comparison between two low-resolution simulations, one
using smoothed abundances and the other using particle abundances for the
calculation of the cooling rates, revealed large differences in the
star formation rate. By redshift zero the simulation employing
smoothed abundances had formed about 1.5 times as many stars as the run
using particle metallicities. We demonstrated that this difference
could not be attributed to non-conservation of metal mass and was
instead caused by the increased mixing (i.e., metal cooling is
important for more particles when smoothed abundances are
used). Although the difference is expected to decrease with increasing
resolution, we conclude that it will be
necessary to solve the metal mixing problem before SPH simulations can
be used to make precise predictions for the cosmic star formation
rate.

We used a suite of large simulations with up to $2\times 512^3$ particles
to investigate the distribution of heavy elements. All simulations
used identical physical parameters and sub-grid modules. The
simulations made use of recently developed modules for star formation
\citep{Schaye2008} and
kinetic feedback from core collapse SNe \citep{Dallavecchia2008}. We
followed all 11 elements (H, 
He, C, N, O, Ne, Mg, Si, S, Ca, Fe) that \cite{Wiersma2009} found to
contribute significantly to the cooling of photo-ionized
plasmas. Radiative cooling (including photo-heating) was implemented
element-by-element, assuming the gas to be optically thin and in
ionization equilibrium in the presence of an evolving photo-ionizing radiation
background. We did this by making use of tables that had been
pre-computed using \textsc{cloudy} following the methods of
\cite{Wiersma2009}. Element-by-element cooling and the inclusion of
photo-ionization -- not only for H and He, but also for heavy elements --
are both novel features for cosmological simulations.

Our predictions for the metallicities of various baryonic phases are
in reasonably good agreement with available observations.
While most of the cosmic metal mass initially resides in 
gas of densities typical of the ISM ($n_{\rm H}>10^{-1}\,\cm^{-3}$),
by redshift zero stars have become the dominant reservoir of metal
mass and the ISM contains only a small fraction of the metals. Diffuse
gas ($n_{\rm H}<10^{-1}\,\cm^{-3}$) contains a significant fraction of
the metals at all times. Except at very high redshifts, most of the
diffuse metals reside in the WHIM ($10^5\,\K < T < 10^7\,\K$), but
the metal mass residing in the ICM ($T>10^7\,\K$) is always
negligible. By the present time, the gaseous metal mass is distributed
over a wide range of densities (5, 50, and 95 per cent of the metal
mass resides in gas with $\rho/\left < \rho\right > \la 10^{-0.5}$,
$10^2$, and $10^{7.5}$, respectively) and temperatures (5, 50, and 95
per cent of the metal 
mass resides in gas with $T~(\K) \la 10^4$,
$10^5$, and $10^{6.5}$, respectively), but is concentrated in gas that has a
relatively high, local metallicity (95 per cent has $Z\ga
10^{-1.5}\,Z_\odot$ averaged over the mass scale corresponding to the
resolution limit). Clearly, any census aiming to account for most of
the metal mass will have to take a wide variety of objects and
structures into account.

Although the mean stellar metallicity slowly increases with time, it
is already of order ten per cent of solar by $z=8$. The evolution of
the metallicity is much stronger for cold-warm ($T < 10^5\,\K$)
diffuse gas, particularly for $\rho/\left <\rho\right> < 10^2$. By
redshift zero the diffuse, photo-ionized IGM has a metallicity $\sim
10^{-2}\,Z_\odot$, while the cold-warm gas in halos has
$Z>10^{-1}\,Z_\odot$. Interestingly, the metallicity of the WHIM and
the ICM is $\sim 10^{-1}\,Z_\odot$ at all redshifts.

A comprehensive convergence study (see appendices \ref{sec:boxsize}
and \ref{sec:resolution}) revealed that, except for the ICM, a
$50~\mpch$ box is sufficiently large to obtain a converged result for
the cosmic metal mass fractions and metallicities down to $z=0$ and
$12.5~\mpch$ suffices for $z>2$. The convergence with respect to
numerical resolution was consistent with expectations based on a
comparison with the Jeans scales. Our use of an effective equation of
state for star-forming gas guarantees that the Jeans mass does not
fall below $f_{\rm g}^{3/2} 10^7\,\hMsun$, where $f_{\rm g}$ is the
local gas fraction, but even this scale is only marginally resolved in
our highest resolution simulations (which have $m_{\rm g}\sim
10^6\,\hMsun$). We found that it is much easier to 
obtain converged predictions for the metallicity of the different
components than for their metal mass fractions. While
the simulations presented here can provide (numerically) robust
predictions for the metallicities at $z<4$, higher resolution models may be
required for higher redshifts. 

Here we have investigated only a single set of physical parameters. We
postpone such a comparison to 
a follow-up paper that will investigate how the predictions for the metal
distribution vary if we change the physical assumptions and sub-grid
prescriptions, such as the cosmology, the star formation recipe, the
implementation and efficiency of galactic winds, the cooling rates,
and feedback from AGN. In this way we hope to isolate the processes
that drive the evolution of the cosmic metal distribution.

\section*{Acknowledgments}
We are grateful to Volker Springel for help with the
simulations and to Laura Portinari and Robert Izzard for 
discussions about stellar yields. We are also grateful to Stefano 
Borgani, Andreas Pawlik, and the anonymous referee for their careful
reading of the manuscript.  
This work was supported by Marie
Curie Excellence Grant MEXT-CT-2004-014112.

\bibliographystyle{mn2e} 
\bibliography{ms}

\appendix
\section{Uncertainties and diversity in stellar evolution choices}
\label{app:stellarevol}

In this appendix we explain in detail our implementation of stellar evolution, giving an 
indication of possible sources of uncertainties in chemodynamical 
simulations.

We need a method for recycling products of stellar evolution back into the 
ISM. We therefore take the standard assumption that a star particle represents 
a simple stellar population. The
feedback processes originating from stellar evolution that we will consider
are winds from asymptotic giant branch (AGB) stars, type Ia supernovae
(SNe), type II (i.e.\ core collapse) SNe and the winds from their
progenitors. The above stellar processes differ in a number of ways,
such as their nucleosynthetic production, their energetic output, and the 
timescale on which they act. 

Progenitors associated with AGB stars are typically
intermediate mass  ($0.8~\Msun \le M \le 8~\Msun$) stars which
have long ($10^8~\yr \la \tau \la 10^{10}~\yr$) main sequence
lifetimes \cite[see][]{Kippenhahn1994}. For AGB stars the mass loss occurs during a period that is
very short compared with both their lifetimes and the dynamical
timescales in cosmological simulations and we therefore
assume the mass loss for an AGB star of given mass to happen within a single simulation time step at
the end of the main sequence lifetime. The kinetic energy injected by
AGB stars is assumed to be unimportant, which is an excellent
approximation both because observed AGB wind velocities are not 
large compared to the velocity dispersion in the ISM and because it
takes an SSP billions of years for all the energy to be released.
AGB stars are
major producers of carbon and nitrogen, both of them being among the
few elements that can be detected in the diffuse IGM
\cite[e.g.][]{Seta03, Fechner2009}. 

Since type II SNe (SNII) only result from massive stars 
\cite[see][and references therein]{Crowther2007}, most of these 
events take place within tens of millions of years after the
birth of a stellar population. For SNII progenitors mass loss on the
main sequence can be substantial. However, since their main sequence lifetimes are
short compared with the dynamical timescales in cosmological
simulations, it is still a good approximation to release all the mass
ejected by stars of a fixed initial mass in a single time step at the 
end of their lifetimes. Note that if the time step governing a
star particle is shorter than the lifetime of the least massive SNII
progenitor, then the mass will still be ejected over multiple time
steps. 

Type II SNe explosions dump so much energy in the ISM ($\sim
10^{51}~\erg$ of kinetic energy per SN) in a short time that they may 
drive galactic-scale outflows from starbursting galaxies
\cite[e.g.][]{veilleux2005}. The implementation of these 
winds is troublesome since simulations lack the resolution
to implement them correctly
\cite[e.g.][]{Ceverino2007,Dallavecchia2008}.  In our simulations
forty percent of the kinetic energy from SNII is injected in kinetic
form as discussed in section \ref{sec:sim} and in more detail in
\cite{Dallavecchia2008}. The remainder is assumed to be lost
radiatively on scales below the resolution limit of the
simulation.

Unlike type II SNe and AGB stars, type Ia SNe are a
result of binary stellar evolution, and consequently somewhat
complicated to model. Currently there are two major theories for the
progenitors of type Ia SNe. The most common is the single
degenerate model. In this model, a white dwarf experiences enough
mass transfer via a main sequence or giant companion to bring its mass
above the Chandrasekhar limit, inducing an explosion. According to the double degenerate
model, two white dwarfs merge after a long period of binary
evolution. In order to predict the type Ia SN rate from a
stellar population, one must consider a wide range of relatively uncertain
processes (e.g.\ binary stellar evolution) and poorly constrained parameters
(e.g.\ binary fraction, binary separation, binary mass function).

Although each SNIa is thought to inject a similar amount of kinetic
energy as a type II SN, they are thought to be much less efficient at driving
galactic outflows, mainly because the energy of a stellar population
is released over billions of years rather than the tens of millions of
years over which core collapse SNe release their energy. SNIa may,
however, dominate the stellar energy feedback in galaxies with very
low specific star formation rates. We therefore do include energy feedback by
SNIa, which distribute in thermal form among the SPH neighbors of star
particles. 

These two types of supernovae also have different chemical signatures.
Type II SNe produce copious amounts of oxygen and so-called `alpha elements'
(neon, magnesium, silicon, etc.). These elements are primarily a
result of $\alpha$-capture reactions and, in addition to oxygen, make
up the bulk of the metal mass ejected from type II SNe
\citep{WW95}. What is known about type Ia SNe (SNIa) is that they are
a major source of iron in the Universe and that a large fraction of
them involve relatively old ($\ga 10^9~\yr$) stars \cite[see
][]{GR83}. Because of the time difference between the release of
$\alpha$ elements by type II SNe and iron by SN Ia, the $\alpha/$Fe
ratio can provide information about the time since the last starburst,
at least for the case of closed box models.

\subsection{Stellar initial mass function}
\label{sec:IMF}

The stellar IMF has long been a subject of debate. 
\citet{S55} made the first attempt to derive an IMF from local stellar counts and his
formulation is still widely 
used today. As star formation is still not understood well
enough to predict the IMF from first principles, a wide range of IMFs
are available in the 
literature \cite[e.g.,][]{Reta05}. Although there is consensus about
the fact that the IMF is less steep than Salpeter below $1~\Msun$,
there are many uncertainties. In addition, the IMF may depend on
redshift or on properties 
of the environment such as metallicity, gas pressure, or the local
radiation field. 

For our purposes,
the particularly interesting IMFs are the Salpeter IMF - because it
serves as a reference model and because it is still used in many
simulations - and the \citet{C03} IMF - because it is an example of an IMF with a low mass 
turnover that fits the observations much better than the Salpeter
IMF. In this work we will assume a Chabrier IMF.

The Salpeter mass function takes a very simple power-law form,
\begin{equation}
\Phi(M) \equiv {dN \over dM} = A M^{-2.35},
\end{equation}
where the normalization constant $A$ is chosen such that:
\begin{equation}
\displaystyle\int M\Phi(M) dM = 1~\textrm{M}_{\odot}.
\label{eq:imfnorm}
\end{equation}
While the Chabrier IMF has the advantage that it
does not overproduce low mass stars, it is slightly more complicated:
\begin{equation}
M\Phi(M) = \left\{\begin{array}{ll} A e^{-(\log M - \log M_{\rm c})^2/2\sigma^2}
& \textrm{if}~ M \leq 1 \Msun \\ B M^{-1.3} &
\textrm{if}~ M > 1 \Msun
\end{array}\right.
\end{equation}
where  $M_{\rm c} = 0.079$, $\sigma = 0.69$, and the coefficients $A$ and 
$B$ are set by requiring continuity at $1~\Msun$ and by the
normalization condition (\ref{eq:imfnorm}). Although the shape of the
IMF above $1~\Msun$ is very similar to 
Salpeter, the lognormal decrease at the low mass end results in a much
lower stellar mass-to-light ratio. 

\begin{figure}
\includegraphics[width=84mm]{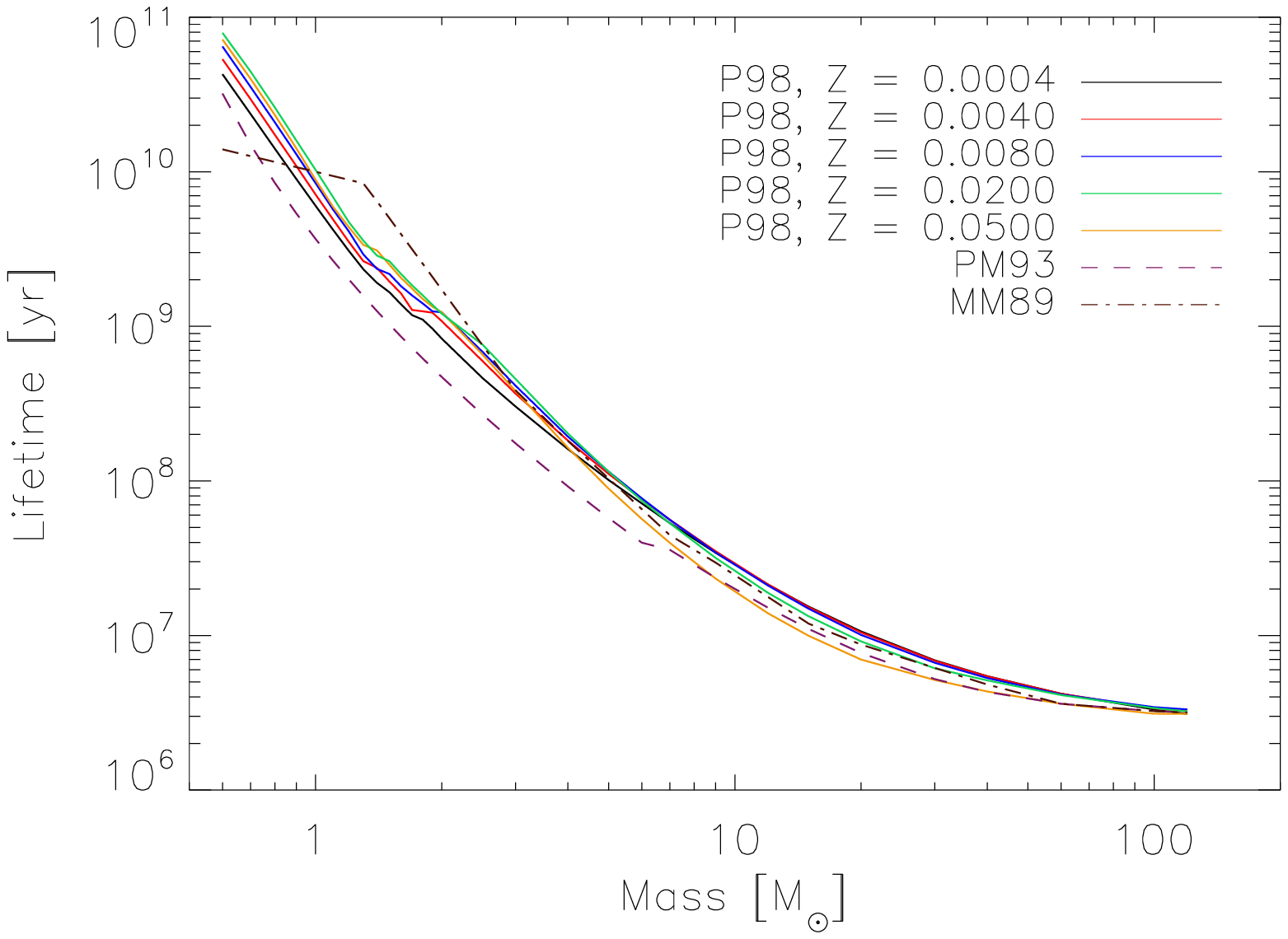}
\caption{Stellar lifetime as a function of initial mass for various
  metallicities (metal mass fractions). Solid lines show lifetimes as 
  given by \protect\citet{Peta98}, while the dot-dashed line shows 
  calculations from \protect\citet{PM93}, who compiled results from
  the literature for a unspecified metallicity, and the dashed line shows 
  the lifetimes given by \protect\citet{Maeder1989}, who assumed solar
  abundances.
  Lifetime is a strongly
  decreasing function of mass, but it is only weakly dependent on
  metallicity. \label{fig-lifetimes}}
\end{figure}

Besides the shape of the IMF, we also need to specify the mass
limits. The lower limit is defined by the hydrogen
burning limit of a star (typically cited as $0.08~\Msun$;
e.g.\ \citealt{Kippenhahn1994}), while
the upper limit is significantly more uncertain, and the value most often
found in the literature of $100~\Msun$ roughly reflects the current
upper limit of observed 
stars. For consistency with previous studies, we choose mass limits of
$0.1~\Msun$ and $100~\Msun$.

\subsection{Stellar lifetimes}
\label{sec:lifetimes}

Although a star can be active (via accretion) for an indefinite period
of time, it is useful to define its stellar `lifetime' as the time a
star takes to move from the zero age main sequence 
up the giant branch and through any subsequent giant evolution. Using
this definition, most stellar mass loss occurs at the end
of the star's lifetime during the AGB/SN phase. 

There is no easy way to determine the
lifetime function for a stellar population observationally. As a
result, the published lifetimes are functional fits to the results of
stellar evolution models (e.g., \citealt{Maeder1989} and
\citealt{PM93}; see \citealt{Reta05} for an overview).

\citet{Peta98} have published metallicity-dependent lifetimes from
their stellar evolution calculations in tabular form and we show their
results in figure~\ref{fig-lifetimes} together with the widely used
lifetimes of \cite{Maeder1989}, who assumed solar abundances, and
\cite{PM93}, who presented a compilation taken from the literature but
did not specify the assumed metallicity. The different lifetime sets
agree well at high mass, but differ by nearly an order of magnitude
for stars of a few solar masses with \cite{PM93} giving systematically
lower values.

Lifetime is a strongly decreasing function of mass and only weakly
dependent on metallicity. Although
metallicity seems to make little difference, it is
attractive to be consistent in treating stellar evolution (in that we
use yields based on the same calculations). We thus employ
the metallicity-dependent lifetime tables of \citet{Peta98}. 

\begin{table*}
\begin{center} 
\caption{AGB yield references and grids}
\label{tab-agb-param}
\begin{tabular}{lcc}
\hline 
Reference & Initial stellar mass ($\Msun$) & Metallicity (metal mass
fraction) \\  
\hline 
van den Hoek \& Groenewegen (1997) & [0.8, 0.9, 1, 1.3, 1.5, 1.7, 2,
  2.5, 3, 3.5, 4, 4.5, 5, 6, 7, 8] & [0.001, 0.004, 0.008, 0.02,
  0.04]\\ 
Marigo (2001) & (various) [0.85 - 5] & [0.004, 0.008, 0.019] \\
Izzard et al.\ (2004) & [0.5, 1, 1.5, 2, 2.5, 3, 3.5, 4, 4.5, 5, 5.5, 6,
  6.5, 7, 7.5, 8] & [0.0001, 0.004, 0.008, 0.02] \\ 
\hline
\end{tabular}
\end{center}
\end{table*}

\begin{figure*}
\includegraphics[width=84mm]{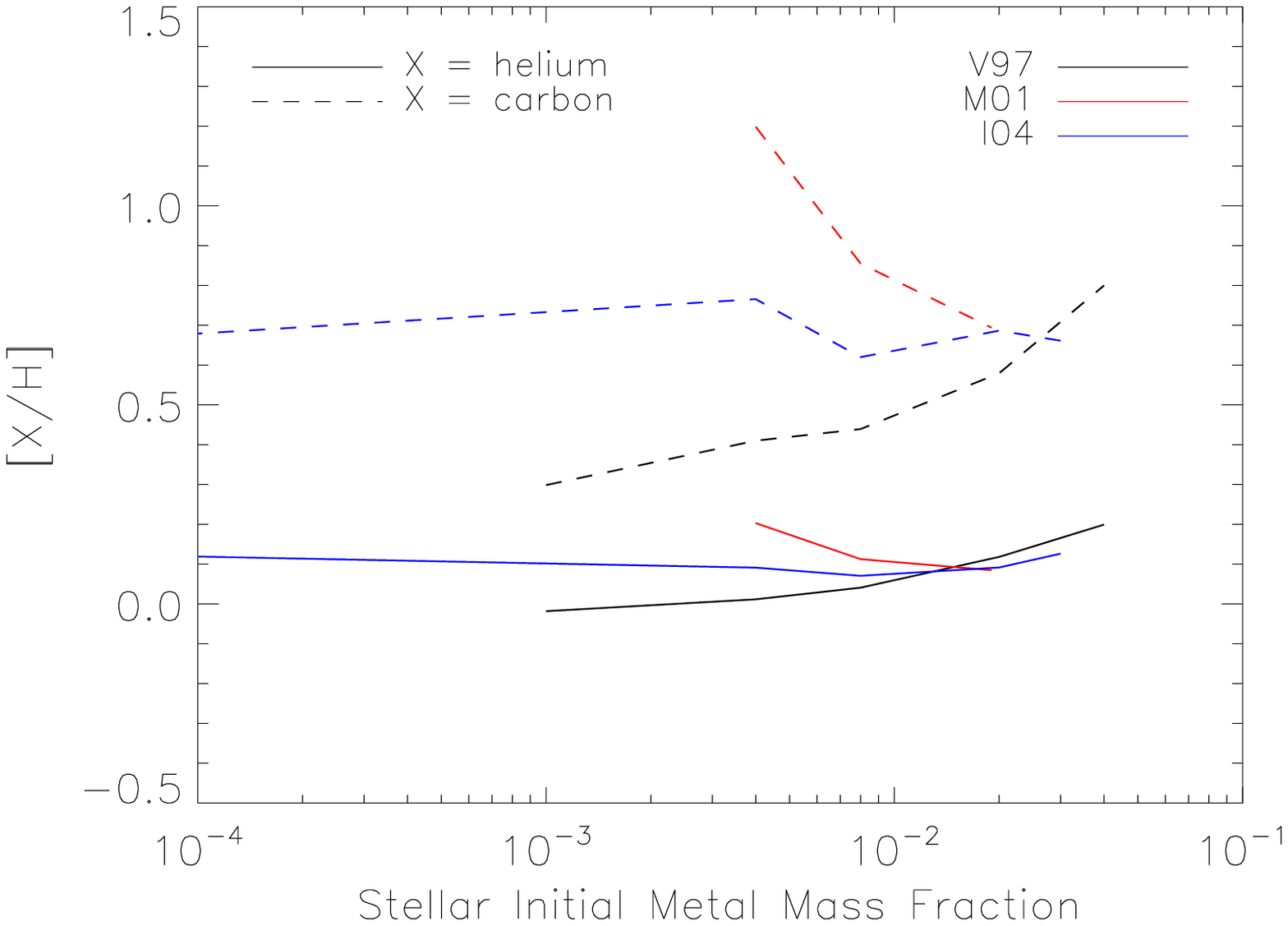}
\includegraphics[width=84mm]{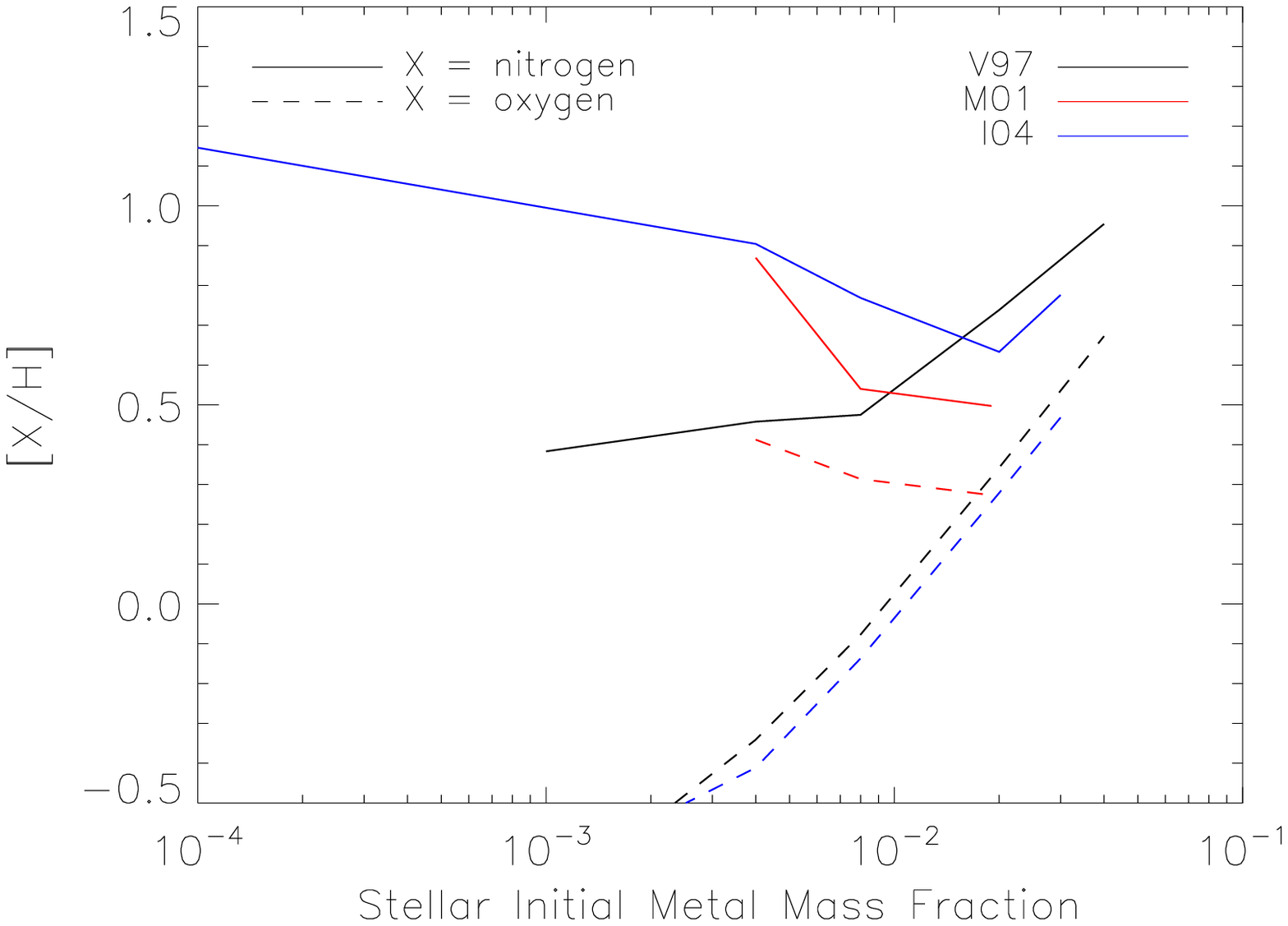}
\caption{Composition of the integrated AGB ejecta of an SSP at time
  $t=\infty$ as a 
  function of its initial stellar metal mass fraction, assuming
  the yields of 
  \protect\citet{vdHG97} (\textit{black}), \protect\citet{M01} (\textit{red}), or
  \protect\citet{Ieta04} (\textit{blue}). Shown are the abundances of
  helium (\emph{left, solid}), carbon (\emph{left, dashed}), 
  nitrogen (\emph{right, solid}) and oxygen (\emph{right, dashed}) relative to
  hydrogen in solar units. The calculations  
  assume a Chabrier IMF and integrate the yields over the stellar initial
  mass range $[0.8,6]~\Msun$. The different yield sets agree around
  solar metallicity, but differ significantly for lower
  metallicities. \label{fig-AGB}} 
\end{figure*}

\subsection{Stellar yields}
\label{sec-yields}

The mass ejected in each element must be obtained from stellar
evolution and (explosive) nucleosynthesis calculations. 
Differing treatments of physics can often lead to drastically
different results in yield calculations. For instance, is it largely 
unknown what determines
the convection boundaries and hot-bottom burning in AGB stars, leading
to discrepancies between yields sets\footnote{In this section we use
  the term yield in the generic sense, referring to stellar ejecta. We
  will define it more rigorously in appendix \ref{sec:implementation}.}, as
we will see 
below. \citet{Heta05} found that adding rotation to type II SN 
models changes the yields by factors ranging from two to an order
of magnitude. These problems are most evident in the fact that the
yields of individual elements are often rescaled in order to get
chemical evolution models to match observations. In fact, some authors
\cite[e.g.][]{Franc2004} use observations and chemical evolution 
models to derive stellar yields for a number of elements, and the
results show marked differences from yields derived from
nucleosynthetic calculations. 

Another important point to make sure of when using yields from various
processes, is that the theoretical assumptions made in the models
match. For instance, if one uses AGB yields for masses up to
$8~\Msun$, then one must take care to use type II SNe yields 
that begin higher than $8~\Msun$ \citep{M01}. The problem is more
often than not in reverse, however, since most yield sets for massive
stars are tabulated down to $\approx 10~\Msun$, but most
intermediate mass yield sets only go up to at most $8~\Msun$,
leaving it up to the user to decide what to do for the
transition masses. Ideally, one would like to use a consistent set of
yields, in the sense that the same stellar evolution model is used for
both intermediate mass stars and the progenitors of core collapse
SNe. 

\subsubsection{AGB stars}

The asymptotic giant branch phase of stellar evolution occurs in
intermediate mass stars near the very end of their lifetimes. During
this phase the envelope of the star puffs up and is eventually shed,
causing the star to lose up to 60~\% of its mass.  Prior to the AGB
phase, material in the core (where most of the heavy elements
reside) is dredged up into the envelope via convection. As a result,
the ejecta are particularly rich in carbon and nitrogen.

Building on pioneering work by \citet{IT78} and \citet{RV81}, various
groups have published AGB yields
\cite[e.g.][]{FC97,vdHG97,M01,Ceta01,Keta02,Ieta04}. Table~\ref{tab-agb-param}
outlines the extent of the resolution (in mass and metallicity) of the
AGB yields of \cite{vdHG97}, \cite{M01} and \cite{Ieta04}, which are
some of the most complete sets for our purposes. These yields are
compared in Fig.~\ref{fig-AGB}, which shows the abundance relative
to hydrogen, in solar units\footnote{We use the notation 
$[{\rm X}/{\rm H}] \equiv {\rm log}({\rm X}/{\rm H}) - {\rm log}({\rm
    X}/{\rm H})_\odot$.}, of various elements in the ejecta as a
function of stellar metallicity. These calculations are for an SSP with 
a Chabrier IMF and the mass range $[0.8,6]~\Msun$ 
at time $t = \infty$. The yields agree very well at solar metallicity, and 
for the case of helium, this agreement extends to lower
metallicities. However, for nitrogen, oxygen, and particularly carbon,
different yields sets give very different results at low
metallicities. 

\begin{figure}
\includegraphics[width=84mm]{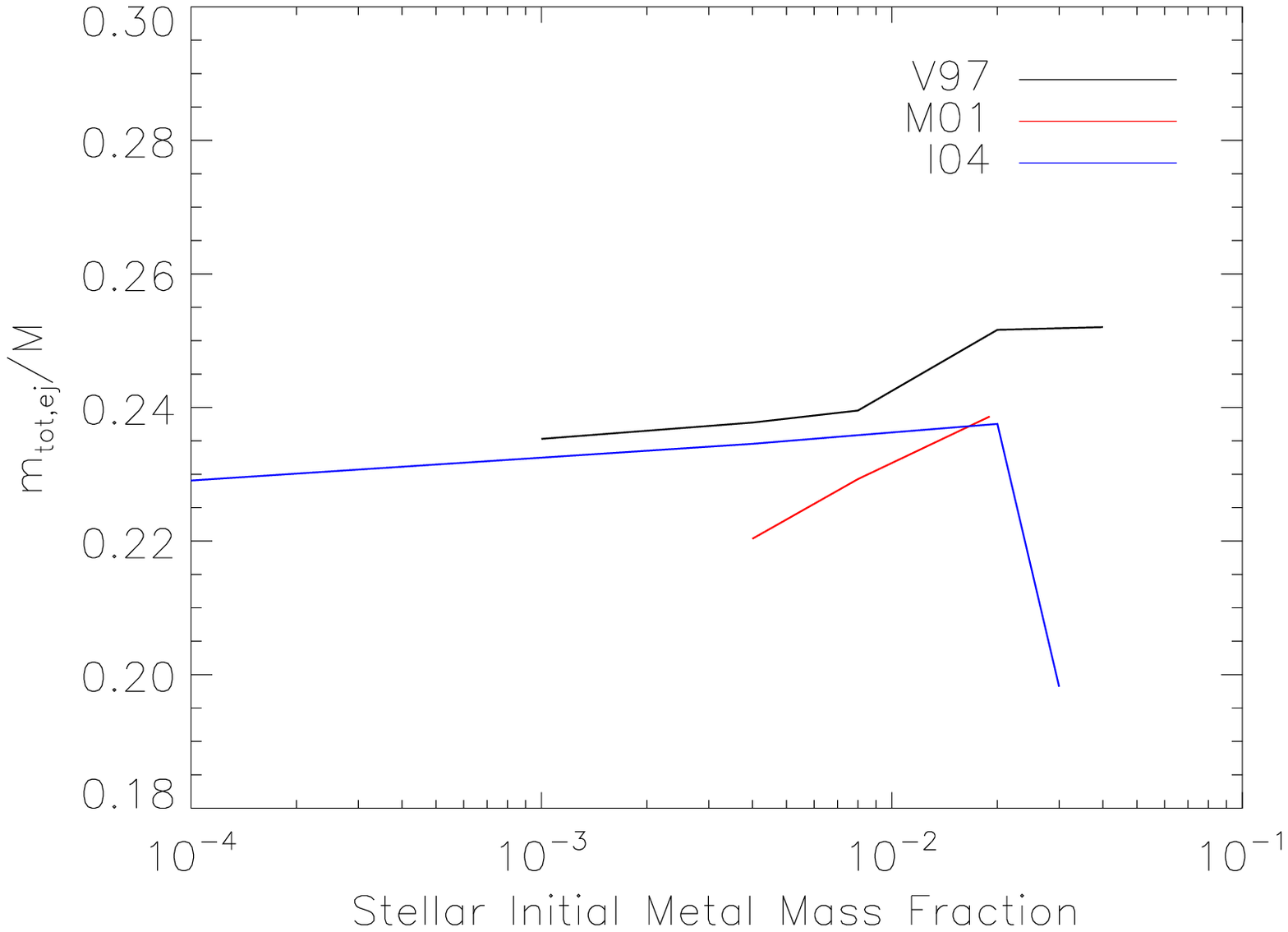}
\caption{Total fraction of the initial mass of an SSP ejected by AGB
  stars at time
  $t=\infty$ as a function of initial stellar metal mass fraction, assuming 
  the yields of \protect\citet{vdHG97} (\textit{black}), \protect\citet{M01} 
  (\textit{red}), or \protect\citet{Ieta04} (\textit{blue}). The
calculations assume 
  a Chabrier IMF and integrate the yields over a stellar initial
  mass range $[0.8,6]~\Msun$, but the fractions are normalized to the
  mass range $[0.1,100]~\Msun$. Different yield sets predict similar
  ejected mass fractions. The ejected mass fraction is insensitive to
  metallicity. \label{fig-AGBej}} 
\end{figure}

We show in Fig.~\ref{fig-AGBej}, for each yield set, the integrated
fraction of the initial SSP mass ejected by stars in the mass range
$[0.8,6]~\Msun$ at time $t=\infty$, normalized to the total initial 
stellar mass over the range $[0.1,100]~\Msun$. 
The ejected mass fractions are very similar for the different yield sets.

In this work we use the yields of \citet{M01}. Although these
only go up to $5~\Msun$, they form a complete set with the SN Type II
yields of \citet{Peta98} since they are both based on the Padova evolutionary 
tracks. Indeed, there are very few yield pairings that form a
consistent set across the full mass range.

\subsubsection{Type II Supernovae} 

Although major advances have been made in modelling type II
SNe, theorists have yet to come up with a model that
self-consistently explodes. As a result, the yields are very sensitive
to the location of the mass cut, i.e., the explosion radius. The
material interior to this radius 
is assumed to end up in the stellar remnant, while the mass exterior
to this radius is assumed to be ejected in the
explosion. Unfortunately, the mass cut is very uncertain,
thus creating another degree of freedom.  
Although uncertainties in the
explosion radius do not affect the 
oxygen yields much, the yields of heavier elements such as iron
are very sensitive to this choice. 

A large number of groups have published tables of type II SN yields
\cite[e.g.,][]{Maeder1992, WW95,Peta98,Reta02,Heta05, Nomoto2006}. Although the more recent
models tend to be more sophisticated, for example including the
effects of improved stellar and nuclear physics and rotation, the
\citet{WW95} yields are still the most widely used. \citet{Peta98}
combine models of pre-SN stellar mass loss with the nucleosynthesis 
explosion calculations of \citet{WW95}, who ignored pre-SN mass
loss. To do this they link the 
carbon-oxygen core mass predicted by their models with those of
\citet{WW95}. They find low metallicity, massive stars to have very 
inefficient mass loss for these large core masses (for 
which there are no type II SN yields), they only 
consider mass loss by winds and assume the rest of the star collapses
directly into a black hole.

\begin{table*}
\begin{center} 
\caption{SN type II yield references and grids}
\label{tab-SNII-param}
\begin{tabular}{lcc}
\hline 
Reference & Initial stellar mass ($\Msun$) & Metallicity (metal mass
fraction) \\  
\hline 
Woosley \& Weaver (1995) & [11, 12, 13, 15,
  18, 19, 20, 22, 25, 30, 35, 40] & [0.0, 0.000002, 0.0002, 0.002,
  0.02] \\ 
Portinari et al.\ (1998) & [6, 7, 9, 12, 15, 20, 30, 60,
  100, 120, 150, 200, 300, 500, 1000] & [0.0004, 0.004, 0.008, 0.02,
  0.05]\\ 
Chieffi \& Limongi (2004) & [13, 15, 20, 25, 30, 35] & [0.0,
  0.000001, 0.0001, 0.001, 0.006, 0.02] \\ 
\hline
\end{tabular}
\end{center}
\end{table*}

\begin{figure*}
\includegraphics[width=84mm]{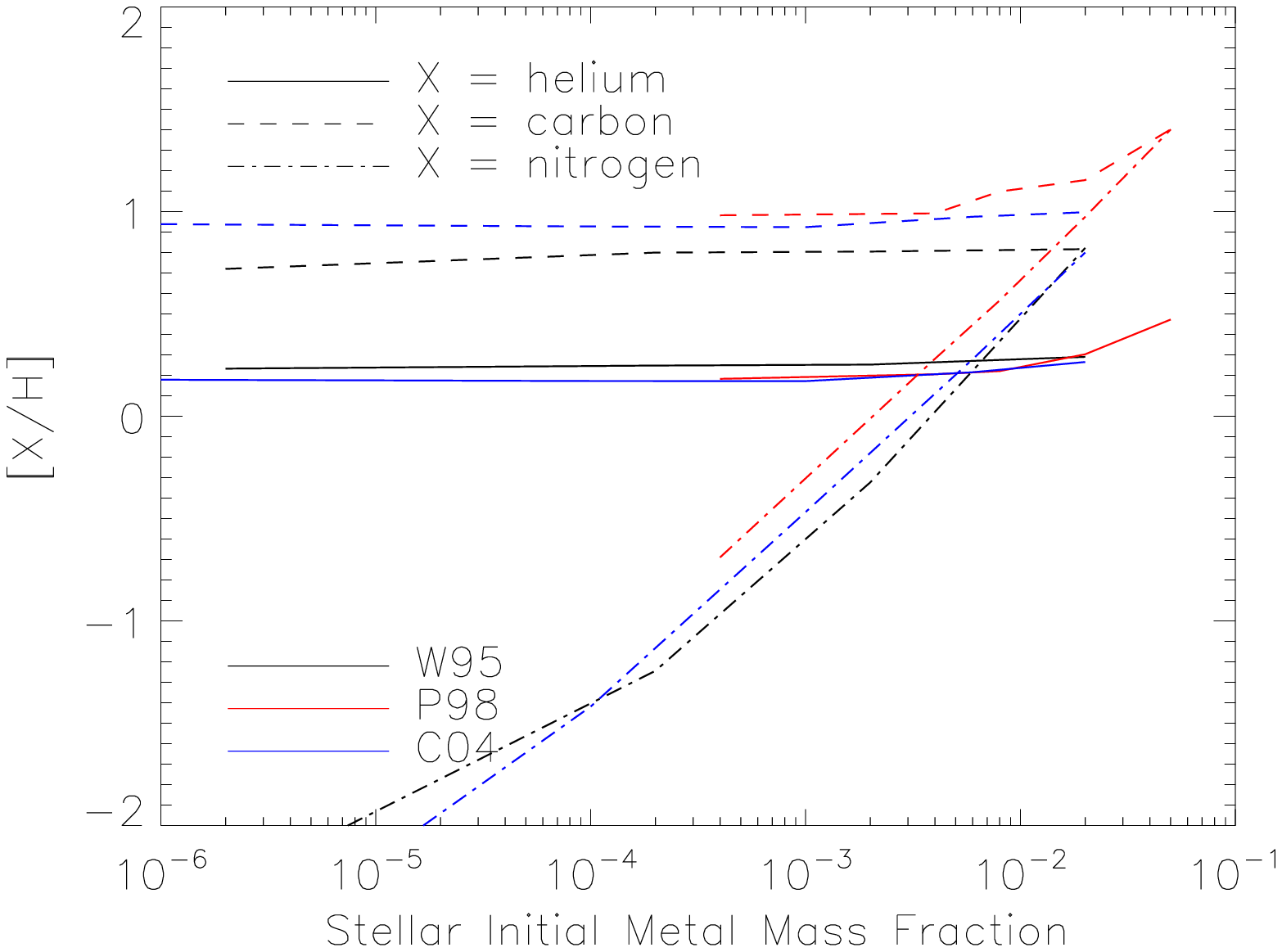}
\includegraphics[width=84mm]{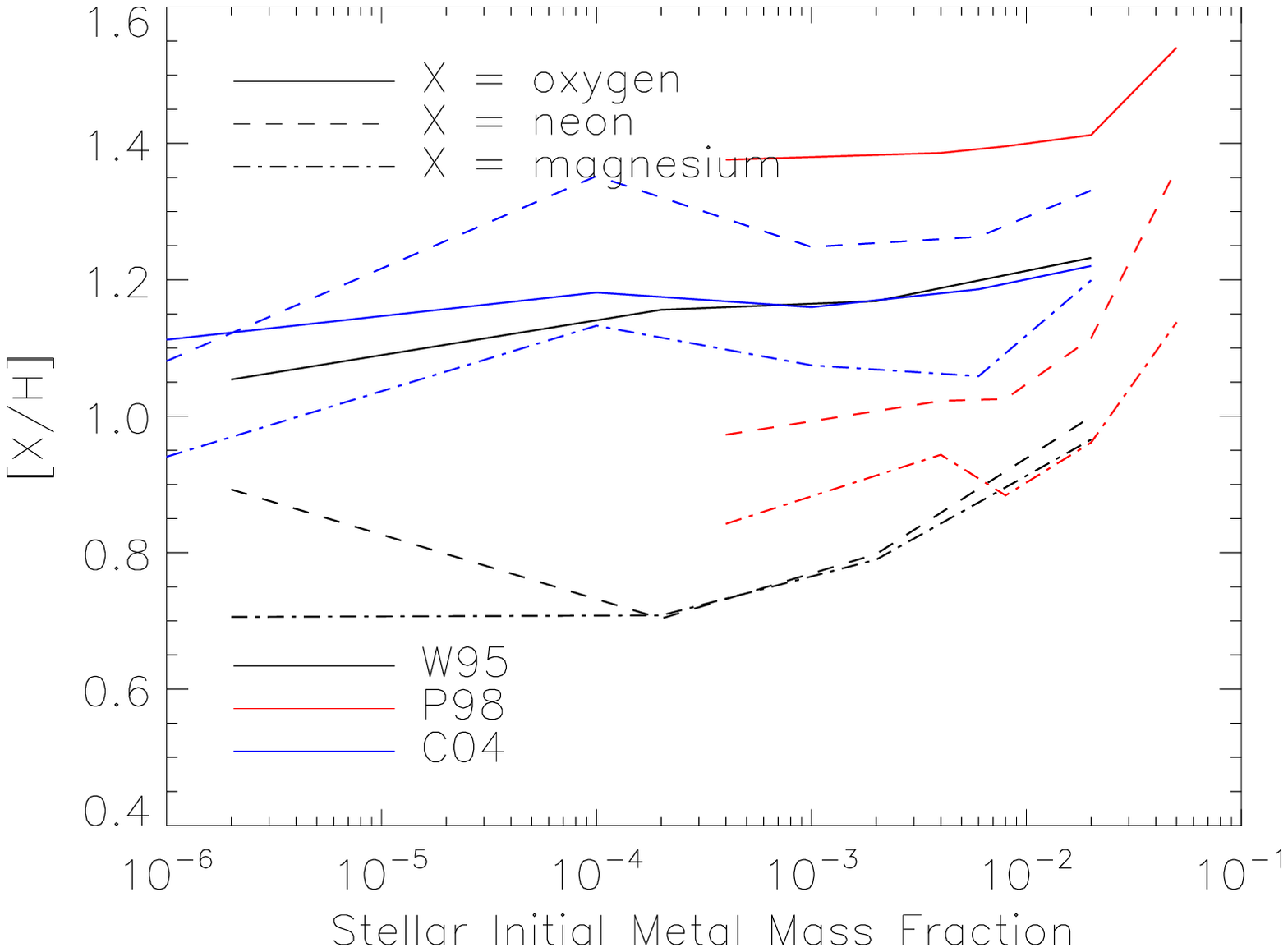}
\includegraphics[width=84mm]{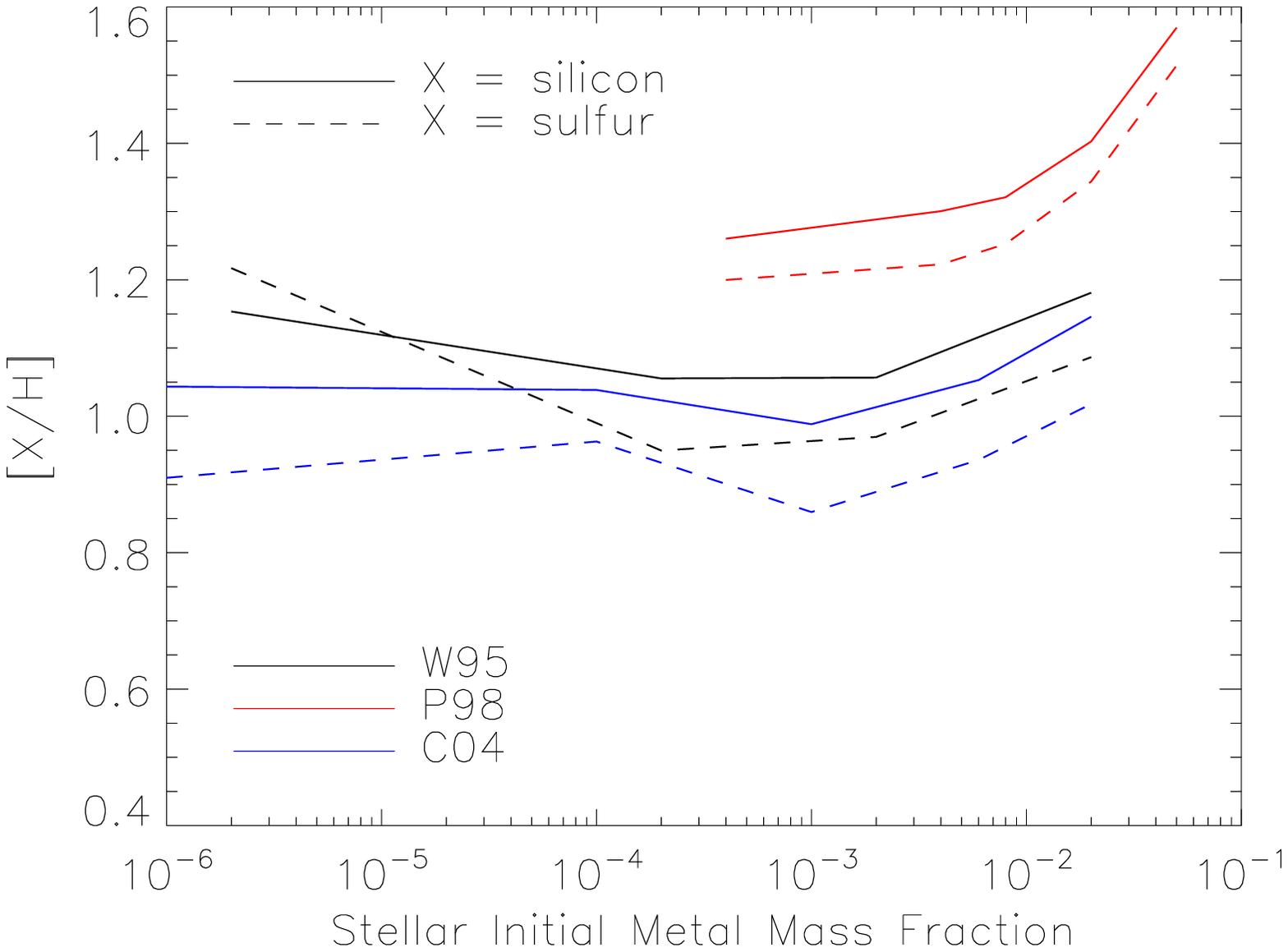}
\includegraphics[width=84mm]{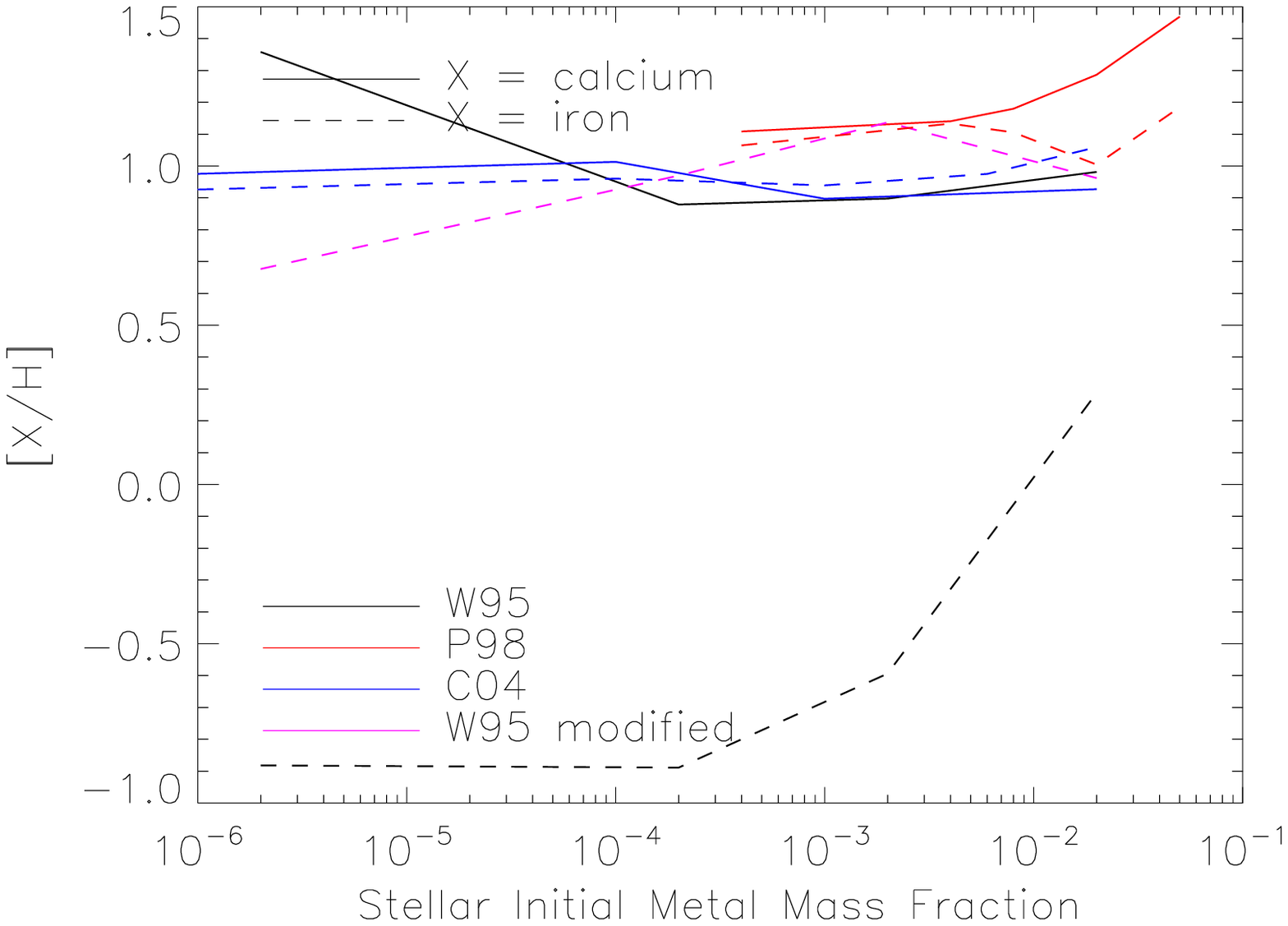}
\caption{Composition of the integrated SN type II ejecta of an SSP at
  time $t=\infty$ as a function of its initial stellar metal mass 
  fraction, assuming the yields of \protect\citet{WW95}
  (\textit{black}), \protect\citet{Peta98} (\textit{red}), or
  \protect\citet{CL04} (\textit{blue}). Shown are the abundances of
  helium, carbon, nitrogen, oxygen, neon, magnesium, silicon, sulphur,
  calcium, and iron, as indicated in the legends. The calculations
  assume a Chabrier IMF and integrate the yields over the stellar
  initial mass range $[8,40]~\Msun$. The (\textit{magenta}) dashed
  line in the lower right panel indicates the result of transferring the
  $^{56}$Ni to the Iron yield of \protect\citet{WW95}. While the
  different yield sets agree well for the lightest elements (upper
  panels), there are large differences for elements heavier than
  nitrogen. \label{fig-SNII}}
\end{figure*}

Table~\ref{tab-SNII-param} and Figure~\ref{fig-SNII} compare the SN
type II yields of \citet{WW95}, \citet{Peta98}, and
\citet{CL04}. Figure~\ref{fig-SNII} shows the abundance relative
to hydrogen, in solar units, of various elements in the ejecta as a
function of stellar metallicity. The elements shown are those
that, together with hydrogen, dominate the radiative cooling of
(photo-)ionized plasmas \citep{Wiersma2009}. These calculations are for an SSP with a Chabrier IMF in
the mass range $[8,40]~\Msun$ at time $t = \infty$. It is clear that the yields of elements that
are produced by type II SNe depend only weakly on metallicity, unless
the metallicity is supersolar. However, except
for very low metallicities, the nitrogen abundance in the ejecta is
proportional to the stellar metallicity, indicating that it is simply
passing through rather than being produced. 

The different yield sets
agree well for helium, carbon, and nitrogen, but there are large
differences for heavier elements. The difference between
\citet{Peta98} and \citet{WW95} is due to the fact that the former add
their stellar evolution to the \citet{WW95} nucleosynthesis
calculations. The difference between \citet{CL04} and \citet{WW95} is
caused mostly by the difference in the assumed mass cut. 

The yields presented in \citet{WW95} consider the state of the ejecta
$10^5\,$s after the explosion. Because a number of isotopes have rather short
decay times, it is
customary to consider -- as most recent yield sets have done -- the state of
the ejecta at a much later time ($10^8\,$s in the case of
\citealt{CL04}). This is 
especially important for $^{56}$Ni, which decays rapidly
into $^{56}$Fe. When \citet{Peta98} incorporate the \citet{WW95}
yields into their calculations, they simply add the $^{56}$Ni to the
$^{56}$Fe. We have added an additional magenta line to figure
\ref{fig-SNII} to show the iron yield from \citet{WW95} with this
adjustment. This agrees much better with the other two yield sets.

\begin{figure}
\includegraphics[width=84mm]{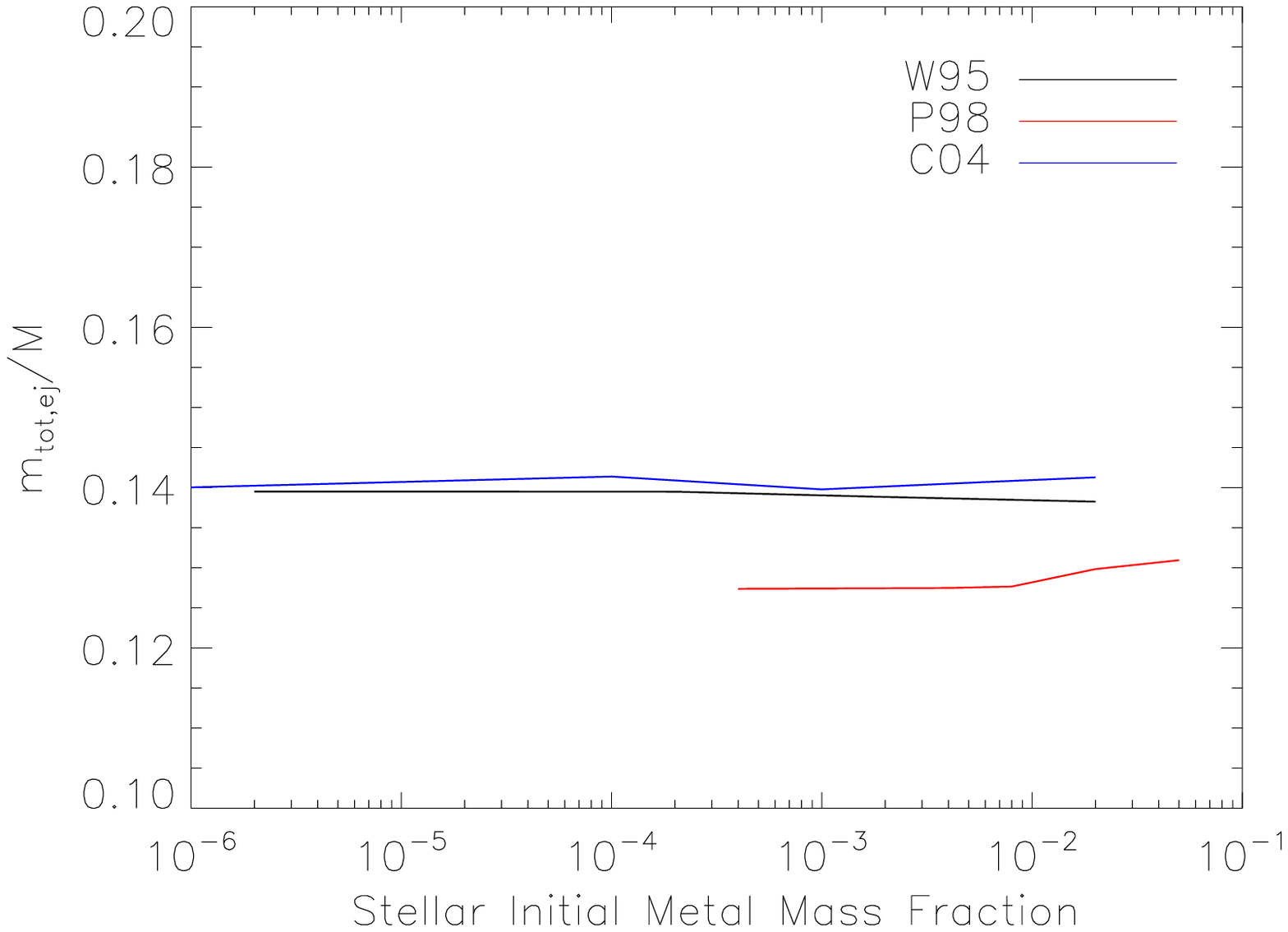}
\caption{Total fraction of the initial mass of an SSP ejected by SNe
  type II at time $t=\infty$ as a function of initial stellar
  metal mass fraction, assuming the yields of 
  \protect\citet{WW95} (\textit{black}), \protect\citet{Peta98} (\textit{red}), or
  \protect\citet{CL04} (\textit{blue}). The calculations assume 
  a Chabrier IMF and integrate the yields over the stellar initial
  mass range $[8,40]~\Msun$, but the fractions are normalized tot he
  mass range $[0.1,100]~\Msun$. The ejected mass fraction is
  insensitive to metallicity and different yields sets predict very
  similar results. \label{fig-SNII_ej}}
\end{figure}

Figure~\ref{fig-SNII_ej} shows the integrated fraction of the initial mass of an
SSP that is ejected by type II SNe at time $t=\infty$ as a function
of metallicity. These calculations again assume a Chabrier IMF and
integrate the yields over the mass range $[8,40]~\Msun$, but normalize
the mass fraction to the mass range $[0.1,100]~\Msun$. 
The ejected mass is insensitive to metallicity and the three yield
sets are in excellent agreement. 

We use the yields of \citet{Peta98} since they include mass loss from massive stars 
and because they form a self-consistent set together with the stellar
lifetimes of these authors and the AGB yields of \cite{M01}. Their 
models allow for the possibility of electron capture supernova 
for intermediate mass stars ($7-9~{\rm M}_{\odot}$). Unfortunately, nucleosynthesis calculations for these stars 
have only recently been performed \cite[e.g.,][]{Wanajo2008}. The
\citet{Peta98} yields therefore only consider the shedding 
of the envelope and the mass loss up to that stage. 
While both \citet{CL04} and \citet{WW95} evolve their models until the point of supernova 
and then begin supernova calculations, they do not include mass loss in their yield tables (since 
the goal of their studies was to investigate explosive nucleosynthesis).

Following the recommendation of L.~Portinari (private communication),
we have adjusted the massive star yields by the factors inferred from
comparisons of chemodynamical models and current
Galactic abundances. Following \citet{Peta98}, we have multiplied the
type II SN yields of C, Mg, and Fe for all masses and metallicities
by factors of 0.5, 2, and 0.5, respectively. Note that these factors
were \emph{not} included in Fig.~\ref{fig-SNII}. The adjustments to Mg and 
Fe can be justified due to  
uncertainties in the explosive nucleosynthesis
models by \citet{WW95} as discussed by a number of authors
\cite[e.g.][]{Timmes1995, Lindner1999, Carigi2000, Goswami2000, Carigi2003, Gavilan2005, Nykytyuk2006}.  These
adjustments have also been incorporated into other studies
\cite[e.g.][]{Liang2001, LPC02, Portinari2004, Nagashima2005}.
These \textit{ad hoc} adjustments reveal just how uncertain the yields are. 

\subsubsection{Type Ia Supernovae}

Despite the fact that the progenitors of type Ia SNe are still
uncertain, yield calculations can be made based on an explosion of a
Chandrasekhar mass carbon-oxygen dwarf. While most models assume the
progenitor to be a binary system, the yields usually consider just  
the explosion of the white dwarf itself, ignoring the
companion completely.

It is instructive to anticipate a few features of these calculations.
 First, since a type Ia SN explosion occurs as soon as the compact
object reaches the Chandrasekhar mass, the yields from this process will
be independent of the initial stellar mass. Second, by the time a star
gets to the white 
dwarf phase, it is almost completely composed of carbon, nitrogen and
oxygen, and while the balance between these elements may depend on initial mass or 
composition, the models assume such a dependence is insignificant. 

Several research groups have published yields from type
Ia SNe. Their calculations range from one- to 
three-dimensional calculations \citep{Teta04a}. The standard to which
most 
results are compared is the spherically symmetric calculation dubbed ``W7'' \citep{Neta84, Neta97a, Ieta99, 
  Beta00, Teta03}. We use the latest incarnation (at the time of code
development) of the W7 model, i.e., \cite{Teta03}.

\subsection{SN type Ia rates}
\label{sec:SNIa_rates}

The recipes for the release of mass by AGB stars and SNII are quite
simple because these processes are direct results of stars reaching the
ends of their lifetimes. Hence, one merely needs to combine the IMF
with the stellar lifetime function 
and the yields and integrate over the time interval specified. 
The recipe for SNIa is, however, somewhat more complex. 

Two SNIa channels are thought to be most plausible (see,
e.g., \citealt{Podsiadlowski2008} for a review). According to the 
single degenerate
scenario, SNIa result when accretion onto a
white dwarf by a binary companion pushes the former over the
Chandrasekhar mass. The double
degenerate scenario, on the other hand, involves the merger of two
white dwarfs, thus 
putting the merger product over the Chandrasekhar mass. 

The SNIa rate of an SSP as a function of its age  
can be determined using late
age stellar and binary evolution theory. The formulation that is used most often is 
given by \citet{GR83}. This formulation requires knowledge of the distribution 
of secondary masses in binaries and makes some assumptions about binary evolution. 
These are poorly constrained and may therefore be subject to large uncertainties. Hence, it is
attractive to take an alternative, more empirical approach. Guided by
observations of SNIa rates, we can simplify the prescription by
specifying a functional form for the empirical delay time function
$\xi(t)$, which gives the SNIa rate as a function of the age of an
SSP, normalized such that $\int_0^\infty \xi(t)dt =1$. The parameters
of this function can then be determined by fitting to observations of
SNIa rates  \cite[e.g.,][]{Barris2006, Forster2006}. 

The number of SNIa explosions per unit stellar mass in a time step
$\Delta t$ for a given SSP is then
\begin{equation}
N_{{\rm SNIa}}(t; t + \Delta t) = \nu \displaystyle\int^{t + \Delta t}_{t}\xi(t')dt',
\label{eq:N_SNIa_standard}
\end{equation}
where $\nu$ is the number of SNe per unit formed stellar
mass that will ever occur. While this approach is attractive since it
does not 
require us to make any assumptions regarding the progenitors, we will
employ yields that correspond to a scenario
involving at least one white dwarf so the SNIa rate should take
that into account (i.e., we should not have any SNIa before the
white dwarfs have evolved). We therefore take an approach similar to
\citet{Mannucci2006} and write:  
\begin{equation}
N_{{\rm SNIa}}(t; t + \Delta t) = a \displaystyle\int^{t + \Delta
  t}_{t}f_{\rm wd}(t')\xi(t')dt',
\label{eq:N_SNIa_us}
\end{equation}
where $a$ is a normalization parameter and $f_{\rm wd}(t)$ is the
number of stars that have evolved into white dwarfs up until time $t$
(i.e., the age of the SSP) per unit initial stellar mass:
\begin{equation}
f_{{\rm wd}}(t) = \left\{\begin{array}{ll} 0 & \textrm{if} ~M_Z(t) > m_{{\rm wdhigh}} \\
 \displaystyle\int^{m_{{\rm wdhigh}}}_{m_{{\rm SNIalow}}(t)} \Phi(M) dM & \textrm{otherwise} 
\end{array}\right.
\end{equation}
Here $m_{{\rm wdhigh}}$ and $m_{{\rm wdlow}}$ are the maximum and minimum white 
dwarf masses respectively, $M_Z(\tau)$ is the inverse\footnote{The lifetime function is invertible because it is a monotonic
function of mass for a fixed metallicity.} of the lifetime function
$\tau_Z(M)$, and
\begin{equation}
m_{\rm SNIalow}(t) = \max(M_Z(t), m_{{\rm wdlow}}).
\label{eq:N_SNIa_limits}
\end{equation}
Note that the shape of the SNIa rate as a function of time differs between
equations (\ref{eq:N_SNIa_standard}) and
(\ref{eq:N_SNIa_us}). 

It is thought that stars with main sequence masses between $3~\Msun$
and $8~\Msun$ evolve into a SNIa progenitor white dwarf. Note that
while our type II SN yields range 
to masses as low as $6~\Msun$, the models used in the yield
calculations of \citet{Peta98} for stars of $6$ and $7~\Msun$ do
not incorporate any explosive nucleosynthesis. They note that stars of
these masses could explode as electron capture SNe \textit{or} evolve
to a thermally-pulsing AGB phase, thus simply shedding their envelopes. 

\begin{figure}
\includegraphics[width=84mm]{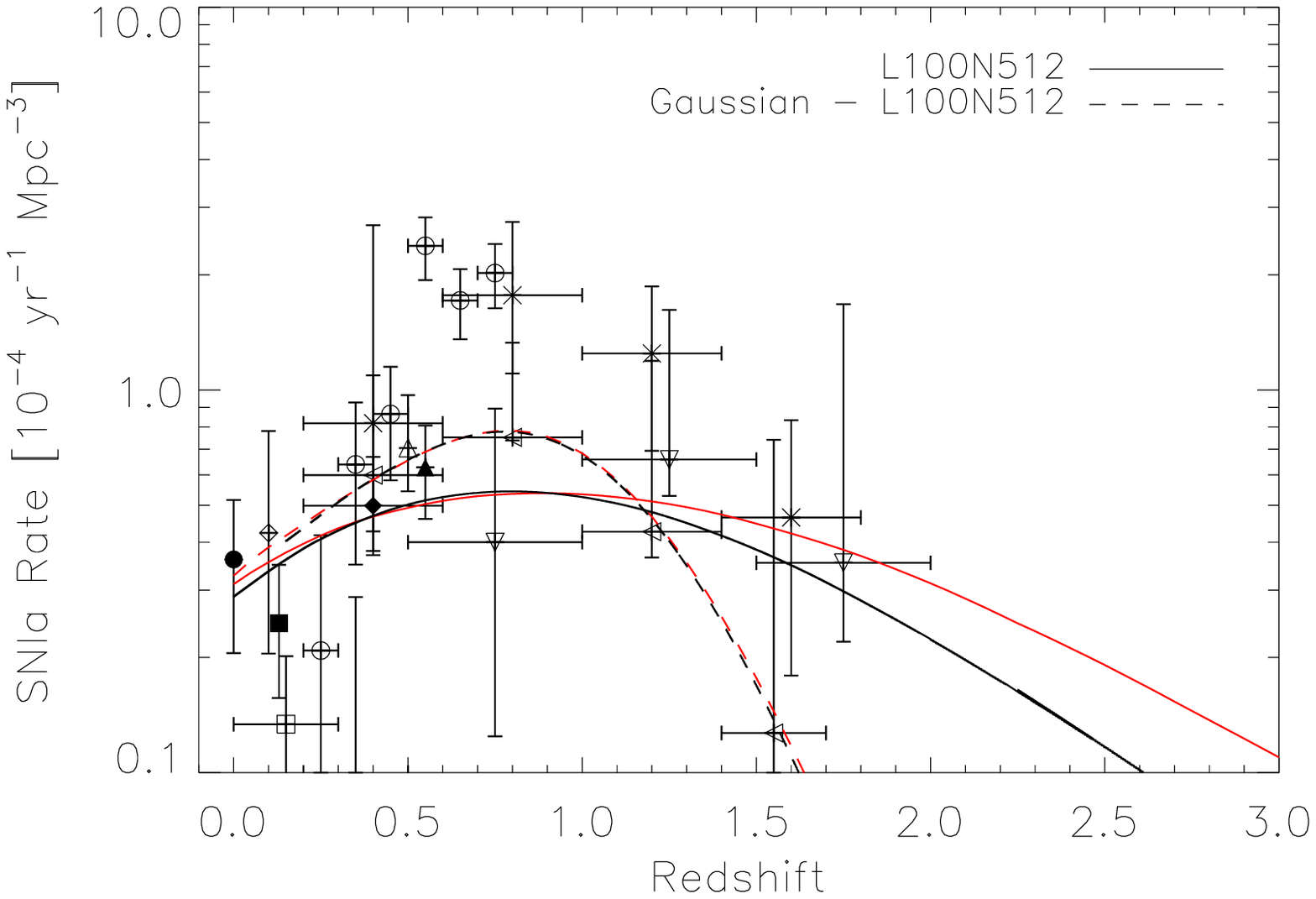}
\caption{Volumetric Type Ia supernova rate as a function of
  redshift from our simulations (\textit{black}) using either an
  e-folding (\emph{solid}) or a Gaussian (\emph{dashed}) delay function. Also shown are 
  approximate fits using a ``standard formulation'' of the cosmic SNIa rate, 
  calculated using equation (\protect\ref{eq:N_SNIa_standard}) and the 
 star formation history predicted by the simulations and a Chabrier
 IMF (\textit{red}). The data points correspond to observations
 reported by 
  \citet{Cappellaro1999} (filled circle), \citet{Madgwick2003} (open
  square), \citet{Blanc2004} (filled square), \citet{Hardin2000} (open
  diamond), \citet{Neill2006} (filled diamond), \citet{Tonry2003}
  (open black triangle), \citet{Pain2002} (filled triangle),
  \citet{Deta04} (crosses), \citet{Barris2006} (open circles), \citet{Poznanski2007}
  (upright triangles), and \citet{Kuznetsova2008}
  (left pointing triangles). \label{fig-SNIaR}}
\end{figure}

The form of the delay function $\xi$ has generated particular interest
recently \cite[e.g.,][]{Deta04}. The two types of delay functions that are most
often considered, the so-called e-folding delay and Gaussian delay
functions, are shown in figure~\ref{fig-SNIaR}. The e-folding delay
function takes the form:
\begin{equation}
\xi(t) = \displaystyle\frac{e^{-t/\tau_{\rm Ia}}}{\tau_{\rm Ia}}
\end{equation}
where $\tau_{\rm Ia}$ is the characteristic delay time. This delay  
approximates predictions made for the single degenerate scenario
via population synthesis models.

The Gaussian delay function was motivated by high
redshift observations by \citet{Deta04}, which show a marked decline
in the SNIa rate beyond $z = 1$. It takes the form:
\begin{equation}
\xi(t) = \displaystyle\frac{1}{\sqrt{2\pi \sigma^2}}
e^{-\frac{(t-\tau_{\rm Ia})^2}{2\sigma^2}}
\end{equation}
where $\sigma = 0.2\tau_{\rm Ia}$ for `narrow' distributions and $\sigma =
0.5\tau_{\rm Ia}$ for `wide' distributions. Note that the integral of this 
particular function ($\int^{\infty}_0 \xi(t)dt$ is 
only normalised to unity for $\sigma \ll \tau_{\rm Ia}$. Gaussian delay functions often
feature long characteristic times ($\tau_{\rm Ia} = 4~{\rm Gyr}$) in order to fit the
data. Strengths and weaknesses exist for both delay functions, and are
discussed in the references cited above.  Unfortunately, the choice of
distribution (including all the ensuing parameters) is rather poorly
constrained. In particular, both the cosmic star formation rate and
the type Ia rate contain large uncertainties beyond a redshift of
1. For instance, the type Ia rate quoted at $z = 1.6$ by
\citet{Deta04} is based on only two detection events. 

In addition to the cosmic SNIa rate, \cite{Mannucci2006} attempt to use 
other observations to constrain the shape of the delay function. They cite 
the dependence of the SNIa rate on galaxy colour observed in the local universe 
\citep{Mannucci2005} as well as a dependence on radio loudness observed by 
\cite{Della2005}. Their analysis suggests that the delay time function
could have two components, a ``prompt" mode and a ``tardy" mode \cite[see also][]{Scannapieco2005a}. These
modes may have physical counterparts with the two progenitor channels,
but this is still uncertain. Using such a delay function could improve our
fits to the observations marginally, but would introduce another unknown
parameter (the ratio of 
contributions of the modes). Moreover, the e-folding delay function
also includes significant prompt and late contributions. We therefore
chose to use an e-folding delay function with $\tau_{\rm Ia}=2$~Gyr, but have also
performed one simulation with a Gaussian delay function using
$\tau_{\rm Ia}=3.3$~Gyr and $\sigma=0.66$~Gyr. These parameter values were
chosen to roughly agree with the constraints mentioned above. The
coefficient $a$ appearing in equation (\ref{eq:N_SNIa_us}) was chosen
to roughly match observations of the cosmic SNIa rate.

We plot the current measurements of the cosmic SNIa rate in figure \ref{fig-SNIaR}.
We have self-consistently adjusted all data points to our
cosmology of choice\footnote{Correcting for cosmology is important
since observations are taken over a volume in co-moving space. For
instance, the highest redshift point of \citet{Poznanski2007} is greater than that
of \citet{Deta04} when using their cosmology, but when converting
to our cosmology the \citet{Deta04} point is reduced by a greater
amount since it was taken over a larger redshift range. Thus, our plot
shows the \citet{Deta04} point to be greater than the
\citet{Poznanski2007} point.}, $(h, \Omega_m, \Omega_{\Lambda}) = (0.73, 0.238,
  0.762)$. Note that the different SNIa measurements
are strongly discrepant, indicating that the statistical errors have been
underestimated and/or that some rates suffer from systematic
errors.

The solid, black curve in Figure~\ref{fig-SNIaR} shows the evolution
of the SNIa rate in our L100N512 simulation, which used the e-folding
time delay function and $a=0.01$. Also shown (dashed, black curve) is
the predicted type Ia 
rate for another simulation that is identical to L100N512 except for
the fact that it uses the Gaussian delay function and
$a=0.0069$. The e-folding and Gaussian models result in reduced
$\chi^2$ values of 2.6 and 2.2, respectively. We consider this
acceptable since the data are internally inconsistent. Note that the
predicted SNIa rates agree better with the more recent
measurements.

While preparing this publication, we encountered a bug in the code
that resulted in $m_{\rm wdlow}$ being set to zero in
equation~(\ref{eq:N_SNIa_limits}). The net result is a shift in the
effective delay time to lower 
redshifts and an increase in the effective value of $a$. Because this
error complicates the interpretation of our parameter values and
because most of the literature uses equation
(\ref{eq:N_SNIa_standard}) rather than (\ref{eq:N_SNIa_us}) we have
tried to approximately match the SNIa rates predicted by our L100N512
simulations using equation (\ref{eq:N_SNIa_standard}), taking the star
formation history predicted by the simulations as input and using the
same (Chabrier) IMF as was used in the simulation. We expect the effects of 
this bug to be nominal since our rates still pass comfortably though the 
observations. Moreover, a simulation with a Gaussian delay 
function (which we will present in a later publication) showed negligible
difference to the e-folding model in nearly all respects (e.g., star formation 
history, distribution of metals). The difference between the type Ia
rates is much greater than the difference between the bugged and unbugged 
versions.

The matching ``standard formulation'' SNIa rates 
are shown as the red curves in figure~\ref{fig-SNIaR}. The e-folding
delay function (solid, red curve) uses $\tau_{\rm Ia}=3$~Gyr while the Gaussian
delay function (dashed, red curve) uses $\tau_{\rm Ia} = 3.3~{\rm Gyr}, \sigma
= 0.66~{\rm Gyr}$ (which is identical to the original delay function used in
combination with equation \ref{eq:N_SNIa_us}). When
using equation (\ref{eq:N_SNIa_standard}) it 
is useful to parametrize the normalization in terms of $\eta$, the
fraction of white dwarfs that eventually explode as SNIa, or the  
\emph{Type Ia efficiency}, which is related to $\nu$, the number of SNIa
per unit formed stellar mass via
\begin{equation}
  \eta \int^{8}_{3} \Phi(M)dM = \nu \int^{100}_{0.1} M \Phi(M)dM.
  \label{nu}
\end{equation}
The red curves in figure~\ref{fig-SNIaR} correspond to $\eta = 2.50~\%$
for the e-folding model and $\eta = 2.56~\%$ for the Gaussian
model. Note that these efficiencies correspond to our Chabrier IMF,
whereas much of the literature quotes efficiencies for a Salpeter IMF.

We end the discussion by noting that we match another constraint in that we find iron abundances 
in our galaxies to be roughly solar at $z = 0$, as we will show
elsewhere. 

\section{Varying the size of the simulation box}
\label{sec:boxsize}

\begin{figure*}
\includegraphics[width=168mm]{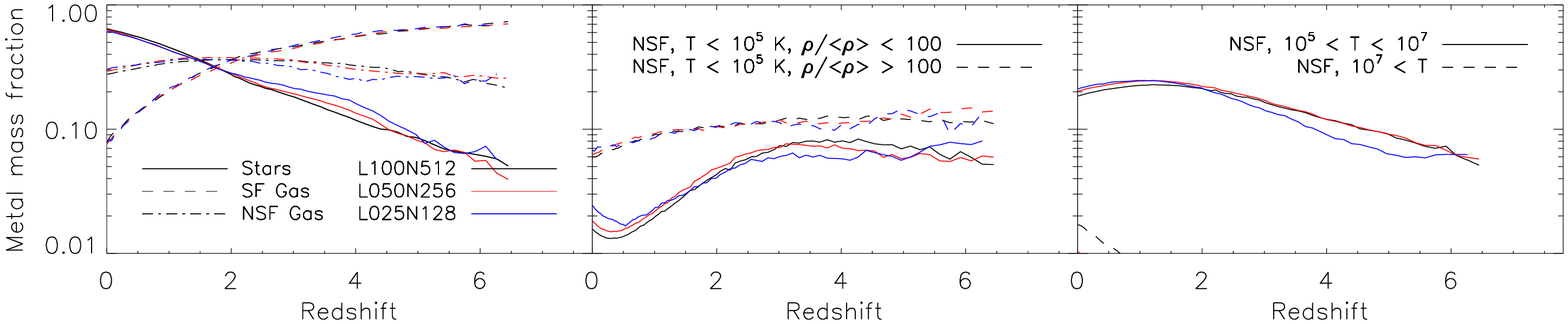}\\
\includegraphics[width=168mm]{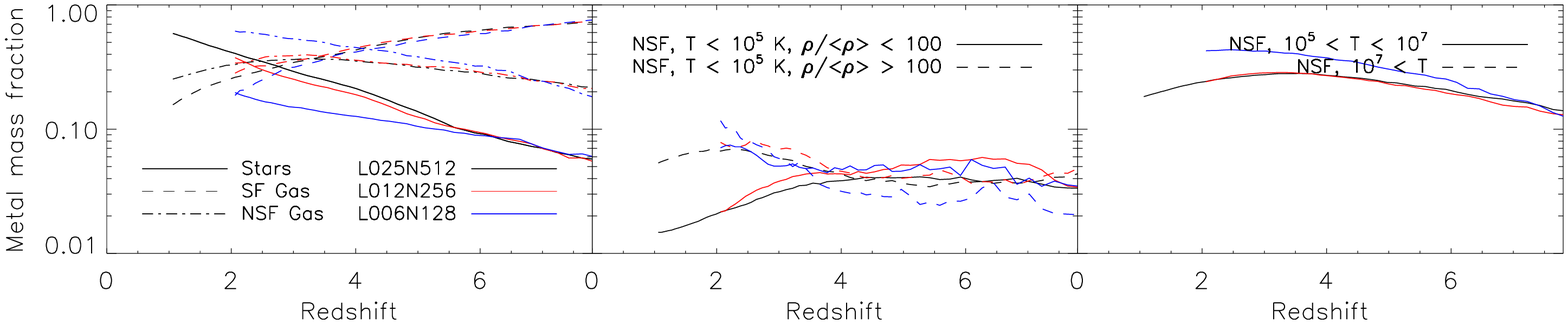}
\caption{Dependence on the simulation box size of the evolution of
  the fractions of the metal mass in various components. The curves in
  the top panels correspond to the L100N512 (\textit{black}), L050N256
  (\textit{red}), and the L025N128 (\textit{blue}) simulations. The
  curves in the bottom panels are for the L025N512 (\textit{black}),
  L012N256 (\textit{red}), and the L006N128 (\textit{blue})
  simulations. The line styles are identical to those shown in
  Fig.~\ref{fig-metevol}. The parts of the curves corresponding to redshifts for
  which the total metal mass is smaller than $10^{-6}$ of the total
  baryonic mass have been omitted because they become very
  noisy. Except for the ICM ($T>10^7\,\K$), the $50~\hMpc$ and
  $12.5~\hMpc$ boxes have nearly converged for $z<2$ and $z>2$,
  respectively.
\label{fig-box12metevol}}
\end{figure*}

\begin{figure*}
  \includegraphics[width=168mm]{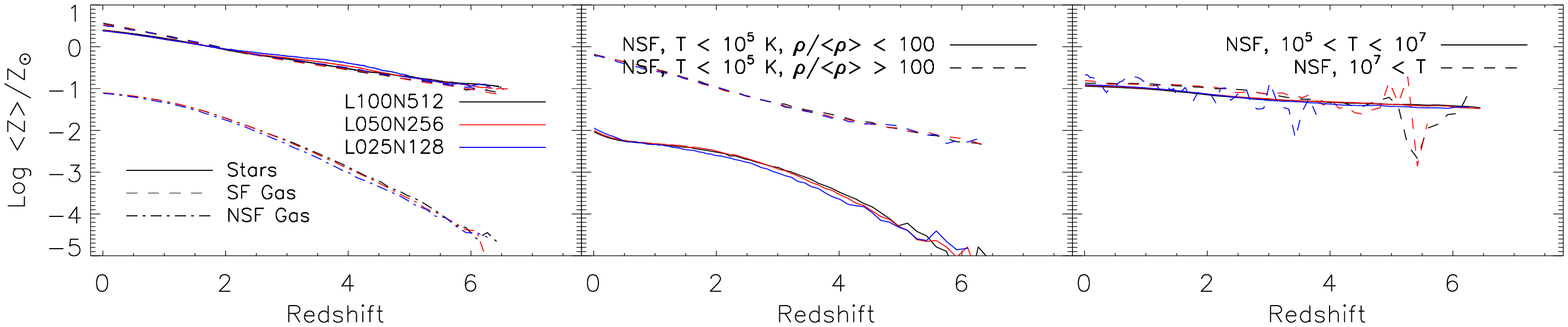}\\
  \includegraphics[width=168mm]{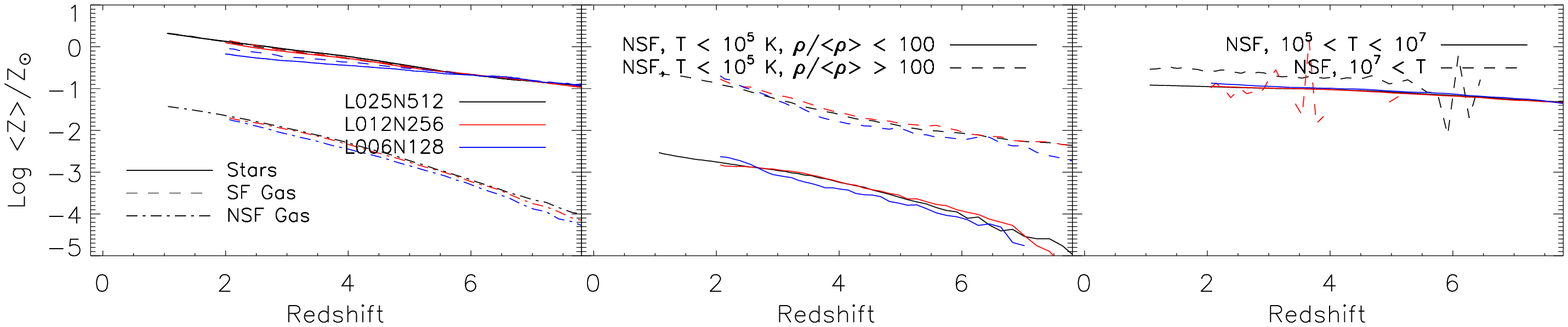}
\caption{Dependence on the simulation box size of the evolution of
  the metallicities of various components. The colors and line styles
  are identical do those used in Fig.~\protect\ref{fig-box12metevol}.
  The $25~\hMpc$ and
  $12.5~\hMpc$ boxes have converged for $z<2$ and $z>2$,
  respectively.\label{fig-box1Zevol}}
\end{figure*}

\begin{figure*}
\includegraphics[width=168mm]{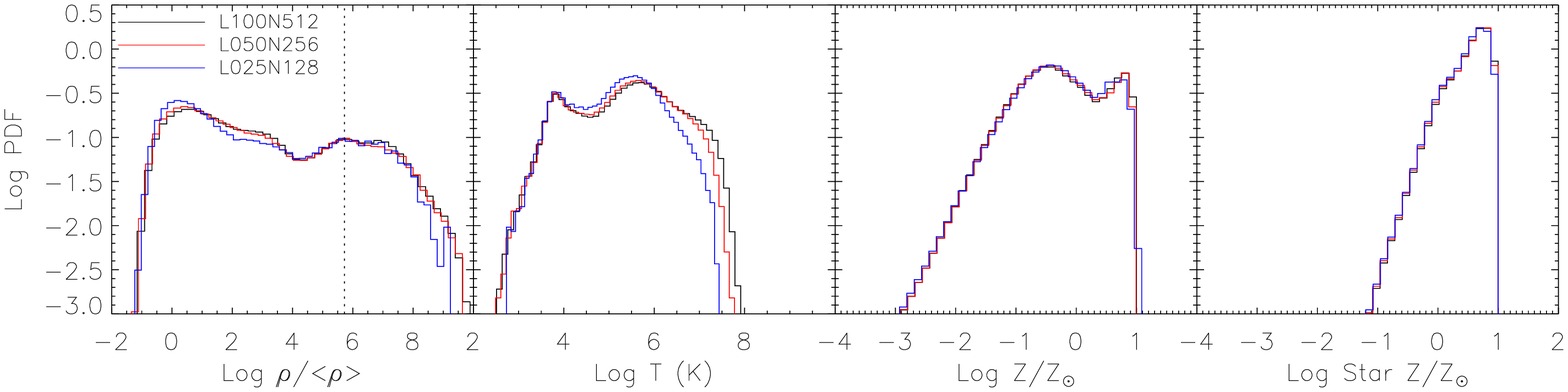}\\
\includegraphics[width=168mm]{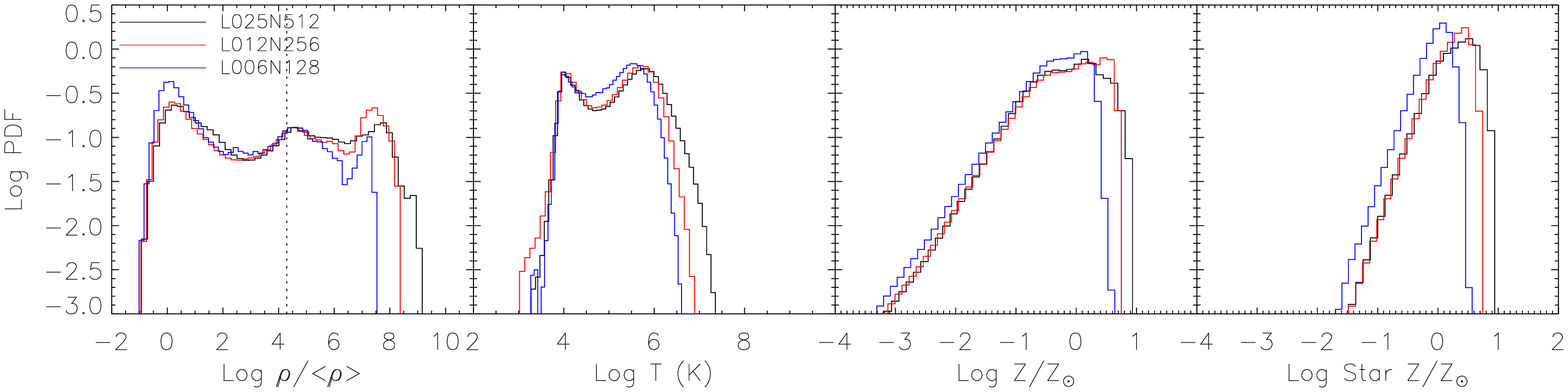}
\caption{Dependence on the size of the simulation box of the
  probability density function, weighted by metal mass, of the gas density 
  (\textit{left}), temperature 
  (\textit{centre-left}), gas metallicity
  (\textit{centre-right}), and stellar metallicity
  (\textit{right}) at $z=0$ (\textit{top}) and $z=2$
  (\textit{bottom}). The colors and line styles are identical to those 
  used in Fig.~\protect\ref{fig-box12metevol}. The vertical, dotted lines in the
  left panels indicate the threshold for star formation. Note that
  star-forming gas was 
  excluded from the temperature panels. Except for the extremes of the
  distributions, the $50~\hMpc$ and
  $12.5~\hMpc$ boxes have nearly converged for $z<2$ and $z>2$, respectively.
\label{fig-box1met1D}}
\end{figure*}

In this appendix we use the suite of cosmological simulations listed in
Table~\ref{tbl:sims} to test for convergence with the size of our
simulation box. The size of the simulation box is important because it
determines what kind of objects can form. Rare, large-scale structures
(both high- and low-density peaks) can obviously only be sampled
correctly if they are much smaller than the size of the box. Moreover,
since Fourier components of the density field only evolve
independently in the linear regime, the simulation box must be large
compared with the scale that has last gone non-linear. Because this
scale increases with time, larger boxes are required at lower
redshifts. To isolate the effect of the size of the simulation volume,
we will compare simulations that use different box sizes,
\textit{while holding the resolution constant}.

For this test we use two sets of simulations that differ in terms of
their resolution. The first set consists of L100N512, L050N256, and
L025N128 and the second set comprises L025N512, L012N256, and
L006N128. Within each set the size of the box is thus decreased by
factors of two and four, respectively. The second set uses a particle
mass that is 64 times smaller than the first set, but these higher
resolution runs were not continued down to $z=0$.

The top and bottom rows of figure~\ref{fig-box12metevol} show the
evolution of the metal mass fractions in the various components for
the low and high-resolution sets, respectively. The panels are
analogous to those shown in Fig.~\ref{fig-metevol}. In particular, the black curves
in the top panels, which correspond to L100N512, are identical to
those shown in Fig.~\ref{fig-metevol}. The top panels demonstrate that, except for
the $T>10^5\,\K$ gas (right panel), the results have already nearly
converged for the 
$25~\mpch$ box. For the WHIM (right; solid) we require a $50~\mpch$
box, although even the $25~\mpch$ box seems to be nearly converged for
$z<2$. For the ICM (right; dashed), however, even the $50~\mpch$ box
is too small. The bottom panels demonstrate that while a $12.5~\mpch$ box is
sufficient for $z>2$ (except for the ICM, which accounts for too few
metals to be visible in the plot, and the photo-ionized IGM), boxes as
small as $6.125~\mpch$ give spurious results for $z<6$. 
The fact that we require larger boxes for hot gas is not
surprising because objects with high virial temperatures are rare and
large. 

The evolution of the metallicities of the different components in the
two sets of simulations is shown in
Fig.~\ref{fig-box1Zevol}. Interestingly, except for the ICM, the
results appear to have converged already for our smallest simulation
boxes (although this is not quite true for the stellar metallicity in
L006N128). Together with Fig.~\ref{fig-box12metevol} this suggests that it
is easier to obtain convergence for the metallicity than for the metal
mass fraction. This in turn implies that the lack of convergence of
the metal mass fractions for the smallest boxes was caused by
non-convergence of the baryonic mass fraction rather than the
metallicity. 

However, part of the difference in appearance can be accounted
for by the fact that the $y$-axis spans six decades in
Fig.~\ref{fig-box1Zevol}, but only two decades in
Fig.~\ref{fig-box12metevol}. Closer inspection reveals that
while this explanation may be viable for the stars, it does not hold
for the gaseous components. Furthermore, the axis ranges differ for a
reason: they show the range of interest. While we generally do not
care whether the metal mass fraction in a particular component is
$10^{-3}$ or $10^{-2}$ when drawing up a census of metals (both
fractions are negligible), it is very interesting to know whether the
diffuse IGM has a metallicity of $10^{-3}$ or $10^{-2}$ solar because
the difference is measurable and because it has important consequences
for enrichment scenarios. 

Finally, we plot the probability density functions of the gas density,
temperature and metallicity as well as the stellar metallicity, all
weighted by metal mass, in Fig.~\ref{fig-box1met1D}. The results
confirm the conclusions drawn from the previous two figures. This
figure illustrates clearly that differences first show up at the
extremes of the distribution, which is not surprising since those
correspond to rare objects.

Summarizing, a $50~\mpch$ box is sufficient to obtain a converged
picture of the cosmic metal mass fractions and metallicities of
most components down to $z=0$. Somewhat
smaller boxes may suffice if one is only interested in higher
redshifts (except for the ICM, $12.5~\hMpc$ is sufficient for $z>2$)
or in low-mass objects. We caution the reader that it is 
possible that larger box sizes are needed if other aspects of the
cosmic metal distribution are investigated (e.g., clustering strengths
of metal absorption lines).

\section{Varying the resolution}
\label{sec:resolution}

While the size of the simulation box mainly determines the types
of structures that can form, 
numerical resolution can even have a large effect on common
objects. For example, simulations that do not resolve the Jeans scales
may underestimate the fraction of the mass in collapsed structures and
hence the total amount of metals produced. Moreover, processes like
metal mixing (see \S\ref{sec:zsm}) may depend on resolution. To
isolate the effect of numerical resolution, it is important to hold the 
simulation box size constant (or to use a box sufficiently large
for the result to have converged with respect to the size of the
simulation box). 

Before showing the results of the convergence tests, it is useful to
consider what to expect.
The temperature of substantially overdense\footnote{Gas with very low
overdensities can have temperatures substantially below $10^4\,\K$
due to adiabatic expansion, but the Jeans scales corresponding to
these low densities are nevertheless large.} gas does not drop much below
$10^4~\K$ in our simulations (see Fig.~\ref{fig-h3massweight}). In
reality, gas at interstellar densities 
($n_{\rm H}\ga 10^{-1}\,\cm^{-3}$) is sufficiently dense and
self-shielded to form a cold ($T\ll 
10^4\,\K$), interstellar phase, allowing it to
form stars \citep{Schaye2004}. However, our
simulations impose an effective equation of state for gas with
densities exceeding our star formation threshold of $n_{\rm H} =
10^{-1}\,\cm^{-3}$. For
our equation of state, $P\propto \rho^{4/3}$, the Jeans mass is
independent of the density. Hence, provided we resolve the Jeans mass
at the star formation threshold, we resolve it everywhere. The Jeans
mass is given by
\begin{equation}
M_{\rm J} \approx 1 \times 10^7\,\hMsun f_{\rm g}^{3/2} \left
({n_{\rm H} \over   10^{-1}\,\cm^{-3}}\right )^{-1/2} \left ({T \over
  10^4\,\K}\right )^{3/2},
\end{equation}
where $f_{\rm g}$ is the local gas fraction. Hence, we do not expect
convergence unless the gas particle mass $m_{\rm g} \ll
10^7\,M_{\odot}$. To achieve convergence, a simulation will, however,
also need to resolve the Jeans length $L_{\rm J}$, which implies that
the maximum, proper gravitational softening, $\epsilon_{\rm prop}$, must be
small compared with the Jeans Length 
\begin{equation}
L_{\rm J} \approx 1.5~\hkpc ~f_{\rm g}^{1/2} \left
({n_{\rm H} \over 10^{-1}\,\cm^{-3}}\right )^{-1/2} \left ({T \over
  10^4\,\K}\right )^{1/2}.
\end{equation}
Note that since $L_{\rm J}$ scales as $L_{\rm J}\propto \rho^{-1/3}$
for our equation of state, the softening scale will always exceed
$L_{\rm J}$ for sufficiently large densities. However, since the Jeans
mass does not decrease with density, runaway collapse is not expected
for star-forming gas. Comparing the above equations with the particle
mass and softening scales listed in Table~\ref{tbl:sims}, 
we see that while our highest resolution simulation (L025N512)
marginally resolves the Jeans scales for $f_{\rm g}\approx 1$, this is not
the case for the simulations that go down to $z=0$, although L050N512
has $m_{\rm g}\approx M_{\rm J}$ and $\epsilon_{\rm prop} \approx
L_{\rm J}$ and is therefore not far off. 

\begin{figure*}
  \includegraphics[width=168mm]{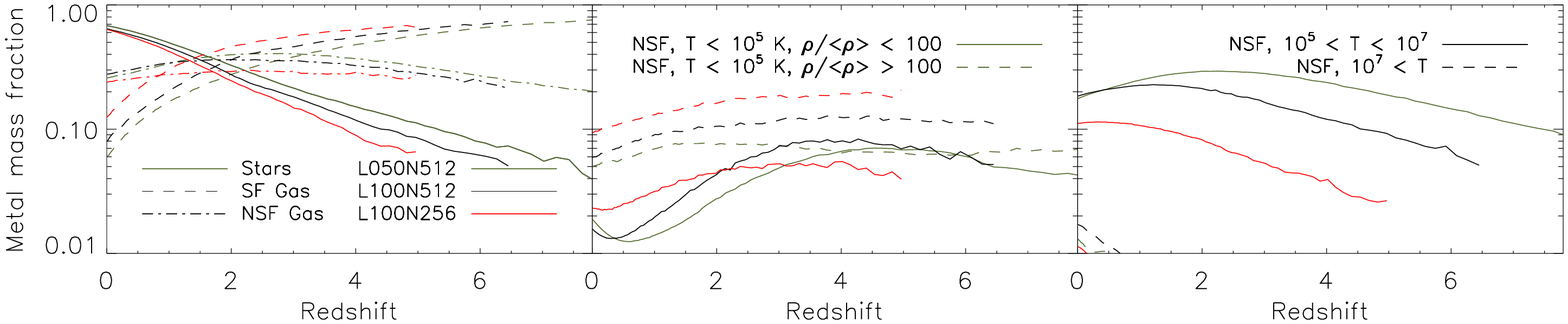}\\
  \includegraphics[width=168mm]{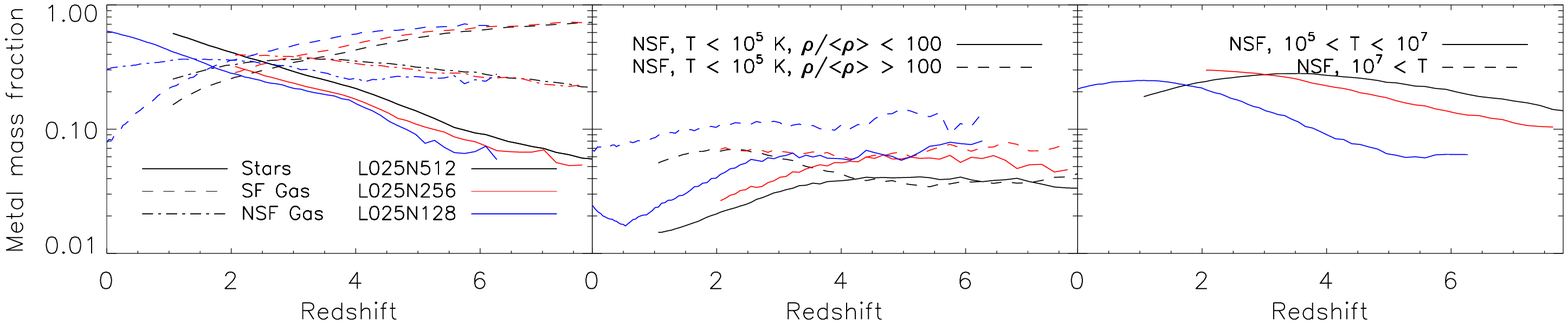}
\caption{Dependence on the numerical resolution of the evolution of
  the fractions of the metal mass in various components. The curves in
  the top panels correspond to the L050N512 (\textit{dark green}), L100N512
  (\textit{black}), and the L050N256 
  (\textit{red}) simulations. The
  curves in the bottom panels are for the L025N512 (\textit{black}),
  L025N256 (\textit{red}), and the L025N128 (\textit{blue})
  simulations. The line styles are identical to those shown in
  Fig.~\ref{fig-metevol}. The parts of the curves corresponding to redshifts for
  which the total metal is smaller than $10^{-6}$ of the total
  baryonic mass have been omitted because they become very
  noisy. Convergence is relatively poor for
  $z>2$ in the top row and for $z>4$ in the bottom row. \label{fig-resmetevol}}
\end{figure*}

\begin{figure*}
  \includegraphics[width=168mm]{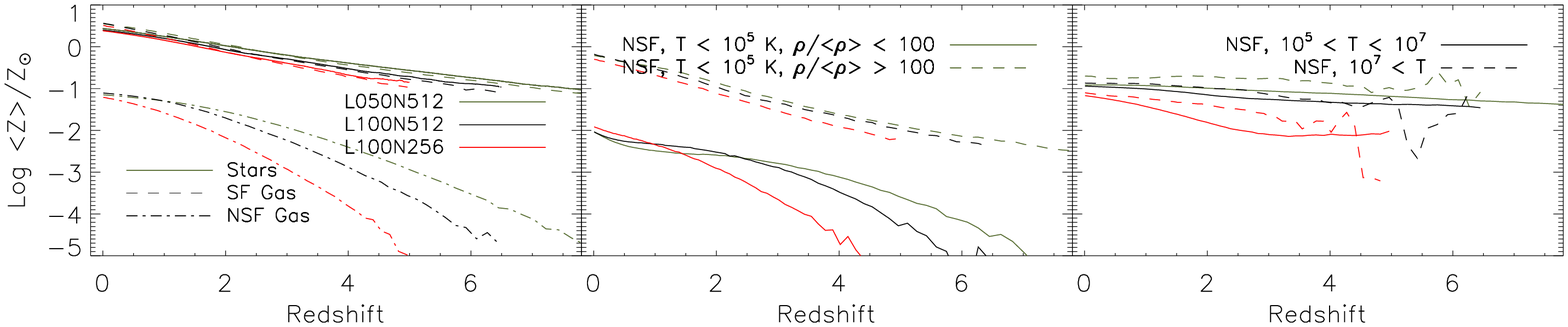}\\
  \includegraphics[width=168mm]{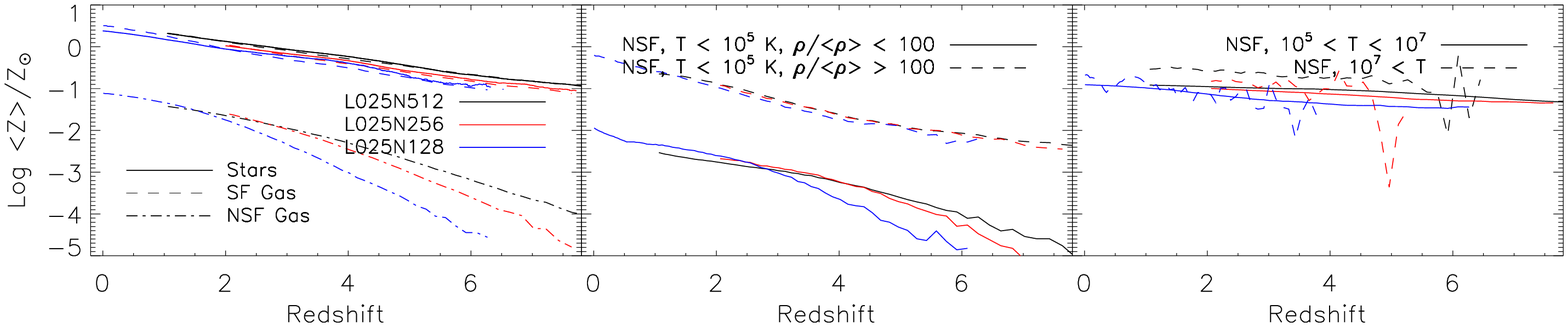}
\caption{Dependence on numerical resolution of the evolution of
  the metallicities of various components. The colors and line styles
  are identical to those used in Fig.~\protect\ref{fig-resmetevol}.
  Convergence is good for $z<2$ in the top row and for $2<z<4$ in the
  bottom row. \label{fig-resZevol}}
\end{figure*}

However, resolving the Jeans scales post-reionization may not
even be sufficient, given that the simulations start at $z \gg z_{\rm
  reion}$. Prior to reionization the gas temperature can be much lower
in which case we have no hope of resolving the Jeans scales at the
threshold for star formation. On the other hand, a photo-dissociating
UV background may well prevent the collapse of haloes with virial
temperatures smaller than $10^4\,\K$. In any case, we expect the
duration of  this period to be comparatively short. Our simulations
assume $z_{\rm 
  reion} = 9$ and include a photo-dissociating background for $z>9$
(see \S\ref{sec:sim}). Hence, in our simulations the formation of
haloes with $T_{\rm vir} \ll 10^4\,\K$ will be efficiently
suppressed. 

We test for convergence with resolution using two sets of
simulations, which use different box sizes. The first set consists of
L100N256, L100N512, and L050N512 and the second set comprises
L025N128, L025N256, and L025N512. Within each set the mass (spatial)
resolution is thus increased by factors of 8 and 64 (2 and 4),
respectively. Note that L050N512, the highest resolution run of the
first set, uses a $50~\hMpc$ box whereas the other runs of this set
used a $100~\hMpc$ box. However, as we have shown in
appendix \ref{sec:boxsize}, $50~\hMpc$ is sufficiently large to give
converged results for all phases but the ICM. We do not show L100N128
because most of its predictions are completely unreliable due to its
extremely low resolution. The second set of simulations uses a box
size of only $25~\hMpc$, but these higher resolution runs were not
continued down to $z=0$. 

\begin{figure*}
\includegraphics[width=168mm]{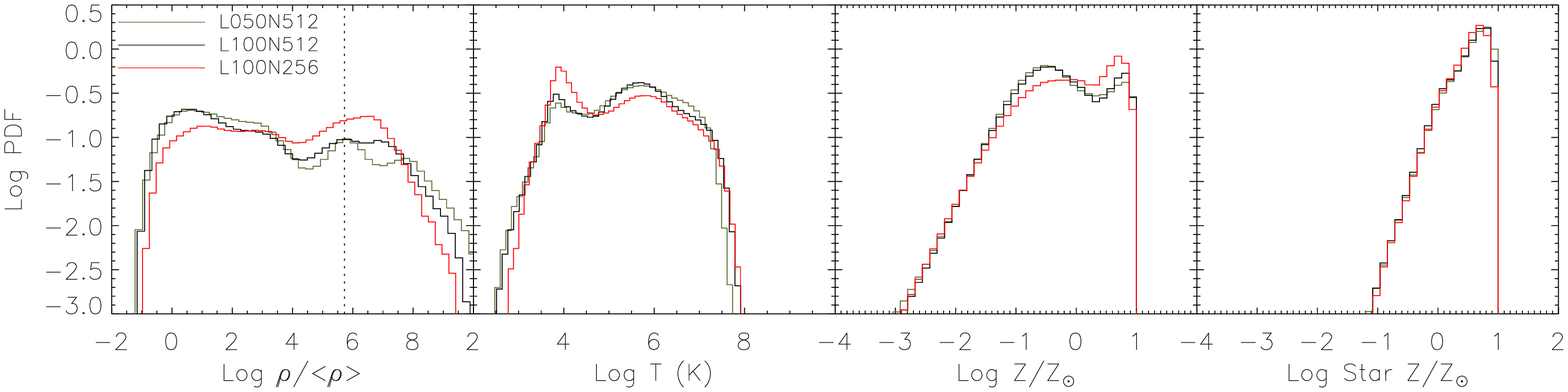}\\
 \includegraphics[width=168mm]{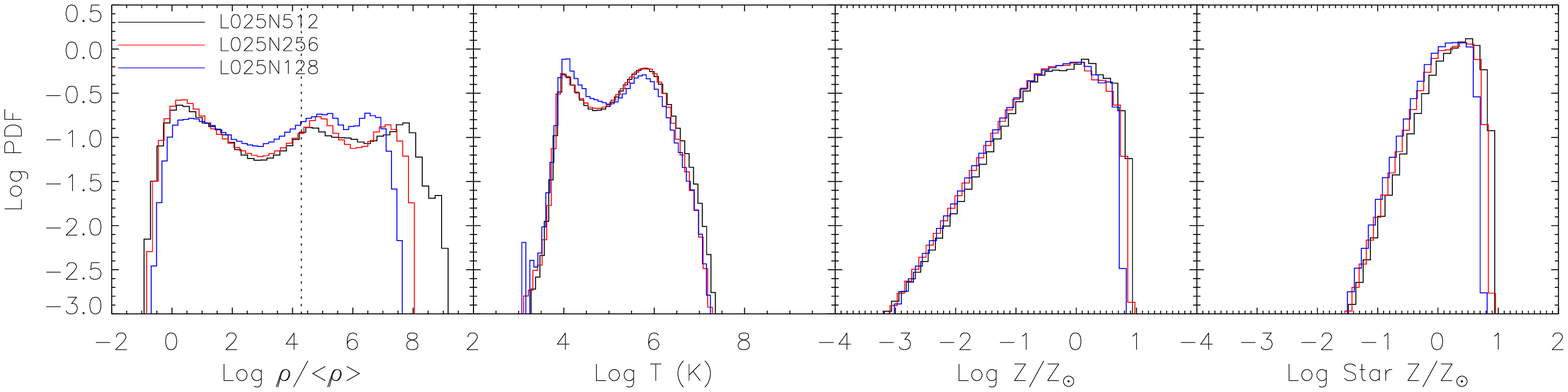}
\caption{Dependence on numerical resolution  of the
  probability density function, weighted by metal mass, of the gas density 
  (\textit{left}), temperature 
  (\textit{centre-left}), gas metallicity
  (\textit{centre-right}), and stellar metallicity
  (\textit{right}) at $z=0$ (\textit{top}) and $z=2$
  (\textit{bottom}). The colors and line styles are identical to those 
  used in Fig.~\protect\ref{fig-resmetevol}. The vertical, dotted lines in the
  left panels indicate the threshold for star formation. Note that
  star-forming gas was 
  excluded from the temperature panels. The highest resolution
  simulations have nearly converged.
\label{fig-resmet1D}}
\end{figure*}

Figures~\ref{fig-resmetevol} -- \ref{fig-resmet1D} are analogous to
figures~\ref{fig-box12metevol} -- \ref{fig-box1met1D} but show the
effect of varying the resolution rather than the size of the
simulation box. The top row of Fig.~\ref{fig-resmetevol} shows no
evidence for full convergence, but the difference between L100N512
(black) and L050N512 (dark green) is small for $z<1$. From the bottom row of
panels we can see that L025N512, while certainly not fully converged,
is similar to L025N256 for $z<4$. As the resolution is increased, the
fractions of the metal 
mass in stars and the WHIM tend to increase, while the fractions in
non-star-forming gas with $T<10^5\,\K$ tend to decrease. 

It is not surprising that
convergence is better at lower redshift, because the typical mass of
star-forming haloes increases with decreasing redshift which means it
will be sampled with more particles per halo. In particular, lower
resolution simulations will start forming stars later and it takes
some time for the resultant underestimate of the metal mass to become
negligible.

Figure~\ref{fig-resZevol} shows that it is much easier to get
convergence for the metallicity of the different components than it is
to get converged predictions for the metal mass fractions. As for
convergence with box size, this partly reflects the fact that we do
not require as much precision for the metallicities since the dynamic
range of interest is much larger than it is for the metal mass
fractions. The metallicities are nearly converged in the
L050N512 run for $z<2$ and in the L025N512 for $z<4$. Finally,
Fig.~\ref{fig-resmet1D} confirms the finding from Fig.~\ref{fig-resmetevol} that
increasing the resolution mainly moves metals from cold-warm halo and
ISM gas to the WHIM. In addition, Fig.~\ref{fig-resmet1D} shows that
the high-density cut-off to the gas 
distribution is mainly set by resolution, which is expected because
many particles and small softening lengths are needed to sample the
tail of the density distribution. The 
main conclusion we draw from this figure is, however, that the
convergence is sufficiently good to draw interesting conclusions,
particularly for the $25~\hMpc$ box at $z=2$ and the $50~\hMpc$ box at
$z=0$.

Thus, the results of the convergence tests mostly agree with the
expectations based on comparisons of the numerical resolution with
the Jeans scales. The resolution of the L100N512 run ($m_g\approx
9\times 10^7\,\hMsun$) is sufficient to
draw qualitative conclusions regarding the mass fractions in various
phases and for $z < 2$ the L050N512 ($m_g\approx 1\times 10^7\,\hMsun$)
may be close to convergence. Run
L100N512 already has sufficient resolution to obtain interesting,
quantitative predictions for the metallicities of the different phases
at $z<2$. For $z>2$ we need higher resolution. While the mass
fractions are still not fully converged for L025N512 ($m_g\approx
1\times 10^6\,\hMsun$), its predictions for the metallicities are
robust for $z<4$.

\end{document}